\documentclass[aip,
               amsmath,
               amssymb,
               booktabs,
               reprint,
               floatfix,
               superscriptaddress,]{revtex4-1}

\usepackage{helvet}
\usepackage[normalem]{ulem}

\usepackage[medium]{titlesec}
\titleformat*{\section}{\bfseries}
\titleformat*{\subsection}{\bfseries}
\renewcommand{\thesection}{\Roman{section}} 
\renewcommand{\thesubsection}{\Alph{subsection}}
\renewcommand{\thesubsubsection}{\arabic{subsubsection}}
\titleformat{\subsubsection}{\normalsize\itshape\bfseries}{\thesubsection.\thesubsubsection}{1em}{}

\usepackage{multirow}
\usepackage{siunitx}
\usepackage[hidelinks,colorlinks=true,allcolors=blue]{hyperref}
\makeatletter
\newcommand\footnoteref[1]{\protected@xdef\@thefnmark{\ref{#1}}\@footnotemark}
\makeatother

\usepackage{graphicx}
\usepackage[figurename=Figure,labelsep=period,labelfont=bf,font=small,justification=Justified]{caption}
\usepackage{subcaption}
\usepackage{changepage}
\usepackage{color}
\definecolor{pblue}{RGB}{0,0,255}
\definecolor{pgreen}{RGB}{0,128,0}
\definecolor{pred}{RGB}{255,0,0}

\begin{document}
\title{\Large \textbf{Characterization of reduced-order turbulence models in the L-mode pedestal-forming region in JET}}

\author{G.~Snoep}
\email[E-mail:]{g.snoep@differ.nl}
\affiliation{Dutch Institute for Fundamental Energy Research, 5612 AJ Eindhoven, Netherlands}
\affiliation{Department of Applied Physics and Science Education, Eindhoven University of Technology, 5612 AZ Eindhoven, Netherlands}

\author{C.~Bourdelle}
\affiliation{CEA, IRFM, F-13108 Saint-Paul-lez-Durance, France}

\author{J.~Citrin}
\altaffiliation{Present affiliation: Google DeepMind, London N1C4AG, UK}
\affiliation{Dutch Institute for Fundamental Energy Research, 5612 AJ Eindhoven, Netherlands}
\affiliation{Department of Applied Physics and Science Education, Eindhoven University of Technology, 5612 AZ Eindhoven, Netherlands}

\author{A.~Ho}
\affiliation{MIT - Plasma Science and Fusion Center, Cambridge, Massachusetts 02139, USA}

\author{M.J.~Pueschel}
\affiliation{Dutch Institute for Fundamental Energy Research, 5612 AJ Eindhoven, Netherlands}
\affiliation{Department of Applied Physics and Science Education, Eindhoven University of Technology, 5612 AZ Eindhoven, Netherlands}

\author{P.~Vincenzi}
\affiliation{Consorzio RFX, 35127 Padova, Italy}
\affiliation{Institute for Plasma Science and Technology, National Research Council, 35127 Padova, Italy}

\author{E.R.~Solano}
\affiliation{Laboratorio Nacional de Fusion, CIEMAT, Madrid, Spain}

\author{M.~Sertoli}
\affiliation{Tokamak Energy Ltd, Oxfordshire, OX14 4SD, United Kingdom of Great Britain and Northern Ireland}

\author{E.~Delabie}
\affiliation{Oak Ridge National Laboratory, Oak Ridge, TN37831-6169, United States of America}

\author{JET Contributors}
\altaffiliation{See the author list of “Overview of T and D-T results in JET with ITER-like wall” by C.F. Maggi \textit{et al.~}to be published in Nuclear Fusion Special Issue: Overview and Summary Papers from the 29th Fusion Energy Conference (London, UK, 16-21 October 2023)}
\noaffiliation


\begin{abstract}
  \begin{adjustwidth}{-1.5cm}{0.5cm}
  \noindent 
  Linear instability characterization of seven JET discharges just prior to the L-H transition is performed at $\rho_{\text{tor}} \in [0.85,0.9,0.95]$ with the gyrokinetic GENE code. The discharges cover both the low- and high-density branches of the L-H transition at two different triangularities. Sensitivities to driving gradients, normalized electron collisonality $\nu_e^*$, hydrogen isotope mass, magnetic geometry and finite-$\beta$ effects are all characterized. At $\rho_{\text{tor}}=0.85$ and $0.9$, trapped-electron modes (TEMs) propagating in both the electron- or ion-drift direction are observed at the lowest densities. At higher density ion-temperature-gradient (ITG) modes are dominant, some of which exhibit trapped-ion drive and unconventional ballooning structures. At $\rho_{\text{tor}}=0.95$, the low-density cases are similar to inner radii, while at higher densities subdominant modes are destabilized by higher collisionalities. The electron collisonality $\nu_e^*$ is scanned around the experimental values at all three radii and for the seven discharges studied. The experimental collisionality range corresponds to a region of minimum linear drive between ITG-TEM mode branches at lower collisionalities and resistive mode branches at higher collisionalities. Moreover, the quasilinear particle flux is directed inward only in the collisionality domain where the linear drive is minimized at $\rho_{\text{tor}}=0.85$ for all densities and $0.9$ only for the highest densities. Model fidelity reduction is performed on the GENE simulations to evaluate the impact of various assumptions and simplifications made by the state-of-the-art quasilinear models QuaLiKiz and TGLF. QuaLiKiz is found to be inadequate beyond $\rho_{\text{tor}}=0.85$, while TGLF-SAT2 agrees well with linear spectra and the quasilinear heat fluxes from GENE up to and including $\rho_{\text{tor}}=0.9$.
  \end{adjustwidth}
\end{abstract}
\maketitle

\section{INTRODUCTION} \label{intro}
To aid in the development and optimization of tokamak plasma scenarios, fast, self-consistent, full-radius modeling of the transport of conserved quantities, such as energy, particles and momentum, is required.
During a tokamak discharge, the transport in the plasma undergoes a series of transitions as the discharge progresses from breakdown, through the low-confinement phase (L-mode), and into the desired final high-confinement phase (H-mode).
In recent years substantial progress has been made in integrated transport modeling of all these discharge phases \cite{Ho2021,Kiefer2021,Meneghini2021,Angioni2022,Bonanomi2022,Luda2023,Slendebroek2023,Fajardo2024} with the JINTRAC \cite{Romanelli2014}, ASTRA \cite{Fable2013} and OMFIT \cite{OMFIT2015} integrated modeling frameworks.
In these frameworks, models that encompass all relevant physics, such as current diffusion, impurities, magnetohydrodynamic equilibrium, radiation losses and heat/particle/momentum sources and transport fluxes, are integrated in a coupled transport simulation in a self-consistent manner.
The computation of the transport fluxes due to turbulent microinstabilities is typically the bottleneck in such simulations.

Although the availability of ever more powerful high-performance computing has led to significant advances in full-physics turbulent transport simulations \cite{Candy2009a,DiSiena2022,Fernandez2022,Howard2024}, the computational cost currently limits their widespread application in scenario analysis and development.
Therefore, reduced-order turbulent transport models have been developed over the past two decades.
These models strike a balance between the level of physics detail and speed, ensuring tractability.
At present, the two principal physics-based, reduced-order turbulence models employed in integrated modeling frameworks are QuaLiKiz \cite{Bourdelle2007,Bourdelle2016,Citrin2017} and TGLF \cite{Staebler2007,Staebler2010}.
These models are quasilinear (QL), meaning both solve for linear eigenmodes of the system and then apply an ad-hoc saturation rule to get the correct nonlinearly saturated fluxes.
Several QL saturation rules have been developed \cite{Citrin2017,Staebler2010,Staebler2017,Staebler2021a,Staebler2021b,Dudding2022}, and this continues to be an active area of research. 
Recent findings suggest the current QL saturation rules may need further refinement, due to an increase in Kubo number for ions near the plasma edge in L-mode \cite{Ashourvan2024}. 

For plasma control applications and true predict-first discharge optimization, real-time capable machine learning surrogate models \cite{Plassche2020,Ho2021,Meneghini2021,Citrin2023} are needed for use in so-called pulse design simulators like FENIX \cite{Janky2019}, RAPTOR \cite{Felici2012,Felici2018} and TORAX \cite{Citrin2024}.
Before considerable computing resources are spent to create further machine learning surrogates of the current state-of-the-art reduced-order models, large-scale validation against experiments across relevant parameters regimes is warranted.

Until recently, the outer boundary conditions in integrated modeling simulations were set at around 85-90\% of the plasma minor radius, using data from experimental measurements to inform the values.
This was done to avoid the need for self-consistent modeling of phenomena near the plasma boundary (the separatrix), as it is not trivial to predict the particle sources in this region in a tractable manner \cite{Simpson2021}.
As an ad-hoc solution, the (neutral) particle source at the simulation boundary is often allowed to vary based on feedback on the simulated line-averaged density.
Recently, in the context of preparations for the first operational phase of ITER, the predictive capabilities of the reduced-order turbulent transport models have been investigated closer to the separatrix within the integrated modeling frameworks \cite{Kiefer2021,Angioni2022,Angioni2023a,Slendebroek2023}.
This is particularly important for L-mode plasmas, where microturbulence is the dominant transport mechanism all the way to the plasma boundary.
In Refs.~\cite{Angioni2022} and \cite{Slendebroek2023}, TGLF-SAT2 \cite{Staebler2021a,Staebler2021b} was used to predict density and temperature profiles for large databases of L- and H-mode discharges, respectively. 
The parametric dependencies of global, zero-dimensional quantities were found to be captured within $\leq$ 20\% error.
However, the accuracy of the predicted density and temperature profiles in the pedestal-forming region varied, and it is difficult to validate the results using flux-driven simulations due to the large uncertainties on the local particle sources.
Therefore, both Ref.~\cite{Angioni2022} and Ref.~\cite{Slendebroek2023} state the need for a detailed verification of the reduced-order models near the edge of L-mode plasmas.

The nature of the turbulence regime(s) in the L-mode edge prior the formation of a pedestal and entry into H-mode has been studied extensively. 
Compared to the inner core, the plasma resistivity is higher, due to the lower temperatures near the separatrix, and this gives rise to different classes of turbulent microinstabilities, such as resistive drift waves (RDW) and/or ballooning modes (RBM).
The key role of resistivity near the separatrix was already identified in the late 1990s and early 2000s, when Scott \textit{et al.}~\cite{Scott1997,Scott2003,Scott2005} and Rogers \textit{et al.}~\cite{Rogers1998} compared fluid turbulence models to experimental scrape-off layer (SOL) data in L- and H-modes, see for example Ref.~\cite{LaBombard2005}. 
Nevertheless, this continues to be an active area of study, comparing experimental trends to high-fidelity simulations using modern gyrokinetic \cite{Bourdelle2012,Bourdelle2014,Bonanomi2019,Bonanomi2021} and nonlinear (gyro)fluid codes \cite{Stegmeir2019,DeDominici2019,Giacomin2020}. 

Often two density branches for the L-to-H transition are observed during experiments as the input heating power is ramped up: (1) the low-density branch, for which the threshold power crossing the separatrix necessary to trigger the pedestal formation $P_{L-H}$ decreases with increasing line-averaged density $\bar{n}_e$, and (2) the high-density branch, where $P_{L-H}$ increases with higher $\bar{n}_e$ \cite{Ryter1996}. 
This suggests that the transport regime near the plasma edge in L-mode changes based on local conditions.
Several qualitative explanations based on changes in the underlying turbulent transport, which is stabilized by the mean flow shear, have been proposed \cite{Singh2014,Bourdelle2015,Bilato2020,Manz2020,Bourdelle2020,Eich2021}.
Previous linear stability analyses with both gyrokinetic and (gyro)fluid codes near the edge of L-mode discharges prior to H-mode entry \cite{Bourdelle2014,Bourdelle2015,DeDominici2019} found a competition between microinstabilities typically dominant in the core, such as trapped electron modes (TEM)  and ion temperature gradient (ITG) modes, at low collisionalities and resistive ballooning modes \cite{Bourdelle2014} and/or resistive drift waves \cite{DeDominici2019} at higher collisionalities. 
In such studies, the experimental collisionality was found to lead to microinstabilities just prior to or at the minimum of the growth rate as a function of collisionality \cite{Bourdelle2014,DeDominici2019,Bonanomi2021}.

In this context, we conducted a verification of reduced-order turbulent transport models in the pedestal-forming region in the L-mode phase just prior to H-mode entry, focusing on verification of the implementations of linear solvers of such models across the different density branches near the plasma edge in L-mode.
Such a comparison is necessary to guide potential future developments of these reduced models and provide further guidance on their validity to present-day users of integrated modeling frameworks.
Validation of nonlinear fluxes based on different saturation rules and the impact of $E \times B$ shear flow stabilization are left for future studies.
For this verification study we used a set of seven JET-ILW (ITER-like wall) discharges that were part of an in-depth study of the L-H transition threshold power $P_{\text{L-H}}$ as a function of line-averaged density $\bar{n}_e$ \cite{Vincenzi2022}. 
We characterized the linear instabilities in all seven discharges at three radii in the L-mode pedestal-forming region: $\rho_{\text{tor}}$ = 0.85, 0.9 and 0.95 with local gyrokinetic simulations using the GENE code \cite{Jenko2000,Goerler2011}.
We show that experimentally observed sensitivities of the L-H power threshold to the line-averaged density, electron collisionality, isotope mass, plasma shaping and dilution due to impurities are all directly reflected in the characteristics of the linear instabilities dominant in the simulations of the JET discharges considered here.
We then verify to what degree the two main fast, reduced-order turbulent transport models, QuaLiKiz and TGLF, cover the observed instability characteristics.

This paper is structured as follows: in Sec.~\ref{sec:data}, we outline the methodology used to prepare the input data for the simulations performed in this work from the experimental measurements.
This is followed in Sec.~\ref{sec:gene} by a linear gyrokinetic stability analysis and sensitivity study based on GENE simulations. 
Details on the model setup and the dominant linear instabilities in the seven discharges at the three radii considered are presented.
In Sec.~\ref{sec:ver}, the impact of several physics fidelity reductions made by the quasilinear models are investigated with (modified) linear GENE simulations, followed by a comparison with QuaLiKiz and TGLF.
Finally, in Sec.~\ref{sec:disc} we summarize our findings on the characteristics of the linear instabilities on both density branches of $P_{L-H}$ and discuss the applicability of the reduced-order models near the separatrix in L-mode.

\section{EXPERIMENTAL DATA}
\label{sec:data}
\begin{figure}[!b]
   \centering
   \includegraphics[width=0.49\textwidth]{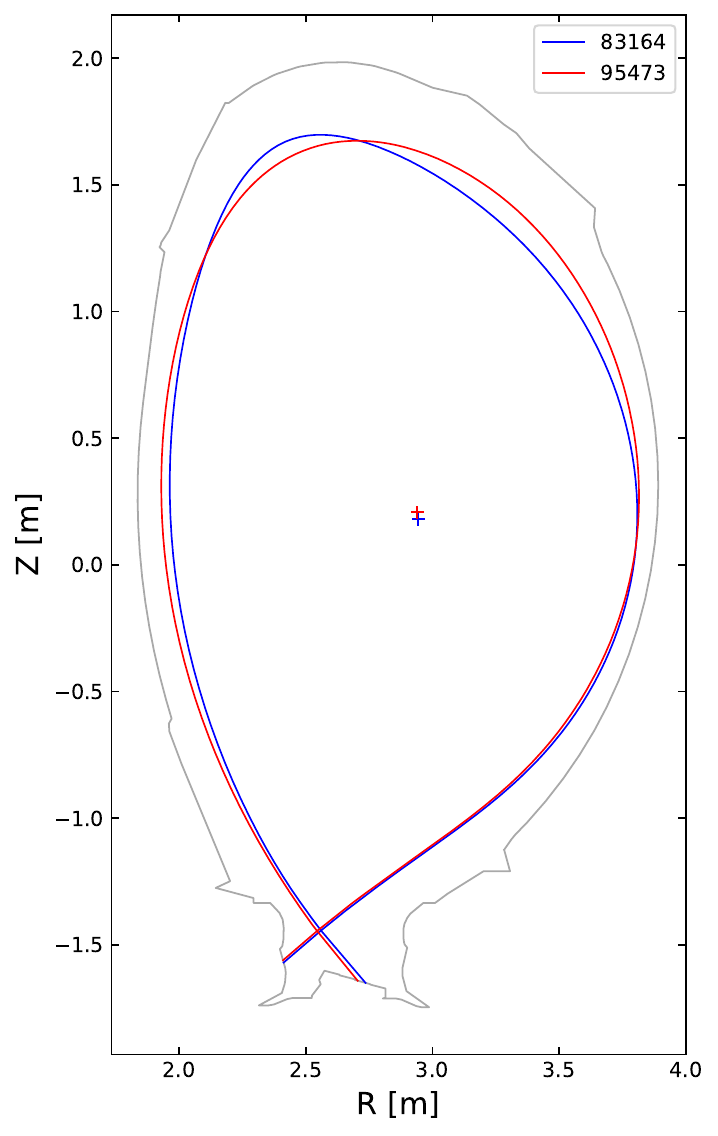}
   \caption{The separatrices of two studied discharges with distinct upper triangularities and horizontal target divertor configuration inside the JET-ILW vessel (grey). As representative examples, the two plasma shapes for discharges \#83164 (blue) and \#95473 (red) are shown. The positions of the magnetic axis are also indicated (plus symbols).}
   \label{fig:separatrix}
\end{figure}
\begin{figure}[!h]
   \centering
   \includegraphics[width=0.49\textwidth]{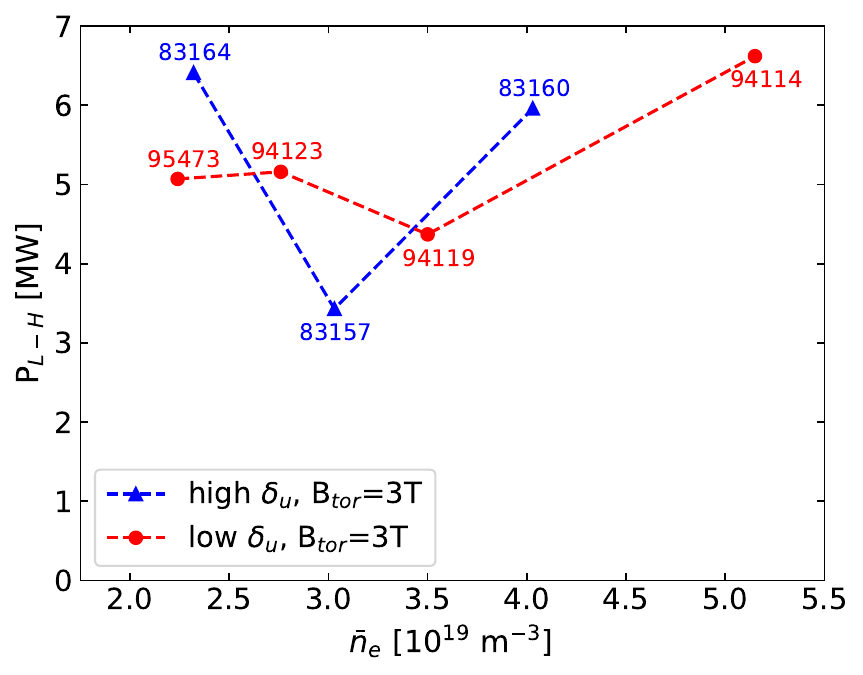}
   \caption{L-H threshold power P$_{\text{L-H}}$ as a function of line-averaged density $\bar{n}_e$ for two sets of JET-ILW NBI-heated deuterium discharges as reported in Ref.~\cite{Vincenzi2022}. The two sets of discharges differ in upper triangularity $\delta_u$ (high in blue triangles, low in red circles).}
   \label{fig:PLH_vs_ne}
\end{figure}

In this work, we make use of experimental data from seven deuterium discharges from the JET tokamak (major radius $R_0$ = 2.96 m, minor radius $a$ = 1 m) with the ITER-like wall (ILW), \textit{i.e.}~full tungsten divertor and a beryllium first wall.
These discharges were heated using only neutral beam injection (NBI), and their line-averaged densities $\bar{n}_e$ span a range of between \SI[]{2E19}[]{\per\cubic\meter} and \SI[]{5.5E19}[]{\per\cubic\meter}.
The toroidal field strength and divertor configuration were held constant across the discharges, at B$_{\text{tor}}$ = \SI{3.0}{\tesla} and outer horizontal target configuration (VH(C)), respectively.
The plasma current was $I_p$ = \SI{2.5}{\mega\ampere} for five out of the seven discharges, while for the high-triangularity discharges \#83160 and \#83164 $I_p$ = \SI{2.75}{\mega\ampere}.
Two plasma shapes with distinct upper triangularities $\delta_u$ were applied, shown in Figure \ref{fig:separatrix}, with three of the seven discharges having an increased $\delta_u$ relative to the other four.
The NBI heating was scanned stepwise in each discharge to identify the threshold power $P_{L-H}$ crossing the separatrix at the time of the low-to-high confinement mode (L-H) transition $t_{L-H}$, see Vincenzi \textit{et al.}~for more details \cite{Vincenzi2022}.
$P_{L-H}$ does not scale linearly with the line-averaged density $\bar{n}_e$, as can be seen in Figure \ref{fig:PLH_vs_ne}, but instead exhibits a minimum that separates the so-called low- and high-density branches of the L-H transition.
For these discharges $P_{L-H}$ varied by about a factor 2, between \SI{3.4}{\mega\watt} and \SI{6.6}{\mega\watt}.
More experimental points around the minima are reported in Ref.~\cite{Vincenzi2022}.
Furthermore, modification of the plasma shape also appears to impact $P_{L-H}$, with larger $\delta_u$ resulting in a higher power threshold, consistent with previous observations \cite{Maggi2014}.

\begin{figure*}[!t]
   \centering
   \includegraphics[width=\textwidth]{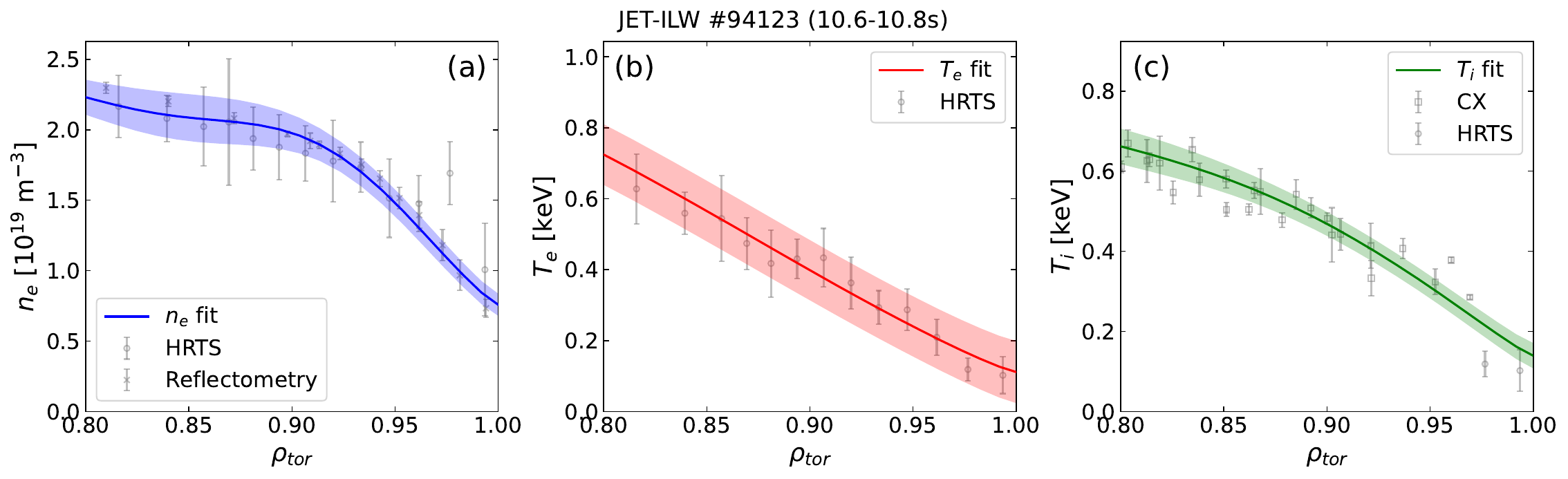}
   \caption{Gaussian process regression fits of time-averaged experimental measurements (grey markers) of the (a) electron density $n_e$ (blue), (b) electron temperature $T_e$ (red) and (c) ion temperature $T_i$ (green) of the low-density branch, low-triangularity JET-ILW discharge \#94123 (10.6-10.8s) for $\rho_{\text{tor}}\in[0.8,1]$.
   High-resolution Thomson scattering (HRTS) and reflectometry data were used for $n_e$, HRTS for $T_e$, and impurity charge exchange (CX) supplemented with some $T_e$ HRTS data near the separatrix for $T_i$. 
   Propagated uncertainties of the fits are indicated with shaded 1$\sigma$ confidence intervals.}
   \label{fig:gpr_fits}
\end{figure*}

\subsection{Data treatment}
\label{sec:data:fit}
\textbf{Profile fitting:} A Gaussian process regression (GPR) method was used to fit the density, temperature and rotation profiles for these JET-ILW discharges \cite{Ho2019}.
This GPR method self-consistently propagates the typically large uncertainties of measurements in the pedestal-forming region to provide a $1 \sigma$ confidence bound on both the radial profiles \textit{and} their normalized logarithmic gradients, to which plasma microturbulence is highly sensitive.
All profiles were fit on experimental measurements and pressure-constrained EFIT++ \cite{Lao1985,Appel2018,Szepesi2021} magnetic equilibria time-averaged over a \SI[]{200}[]{\milli\second} boxcar window just prior to the observed $t_{L-H}$ in the discharge.
For a few discharges the window was shifted by \SI[]{50}[]{\milli\second} away from $t_{L-H}$ to avoid time-averaging over a sawtooth crash.
For this study, $t_{L-H}$ was defined as the onset of dithering identified in the $D_\alpha$ signal while supplied heating power was scanned \cite{Solano2018}.
The normalized poloidal flux, $\psi_N = \psi_{\text{pol}}/\psi_{\text{pol,LCFS}}$, was used as the radial coordinate for all GPR profile fitting. 
Adjustments, such as line-of-sight corrections, made in the original treatment of the data using ProfileMaker were preserved \cite{Vincenzi2022}.
In Figure \ref{fig:gpr_fits} the GPR fits of $n_e$, $T_e$ and $T_i$ for $\rho_{\text{tor}} \in [0.8,1]$, where $\rho_{\text{tor}}$ is the square root of the normalized toroidal flux, of the low-density, low-triangularity discharge JET-ILW \#94123 for 10.6-10.8s, are shown as an example. 
Fits to the experimental measurements for the other six discharges can be found in Appendix \ref{apdix:profiles}.

The electron density $n_e$ profiles were fit using data from both the high-resolution Thomson scattering (HRTS) \cite{Pasqualotto2004,Frassinetti2012} and density profile reflectometry \cite{Sirinelli2010} diagnostics.
The electron temperature $T_e$ profiles were fit using data from both the HRTS and electron cyclotron emission (ECE) \cite{Giroud2008} diagnostics.
Due to an alignment mismatch between the ECE channels measuring in the outer part of the plasma and HRTS measurements at the time of fitting, the ECE data was cut off below \SI[]{800}[]{\electronvolt}, effectively eliminating it as a constraint on the profiles in our region of interest. 
This alignment was corrected after the profiles were fitted and a large part of our analysis was already performed.
Although adding the ECE data improves the constraining of the profile uncertainty in the region of interest, the nominal fit values were typically affected by $5-20$\% \footnotemark[1]\footnotetext[1]{Adding the ECE measurements reduces the $T_e$ GPR fit uncertainty estimates by $\sim$2$\times$ in the region of interest.
For the low-triangularity discharges (\#9xxxx) the nominal fit values were typically affected by $\leq 5$\%.
However, for the high-triangularity discharges (\#83xxx) the impact was larger, with the fitted values of $T_e$ affected by 15-20\% in our region of interest.
In particular for \#83164, the original nominal fit of the electron temperature near the edge might have been too low.
Nevertheless, the uncertainties of the fits of $T_e$ used in this work were much larger than such changes (20-60\%) and sensitivity analyses show that none of the reported trends are affected by such differences. 
Therefore, the analyses are presented on the basis of the original GPR fit profiles.}, well within the uncertainty of the fits of the original data treatment.
The ion temperature $T_i$ profiles were fit using data from both the core main-ion \cite{Hawkes2018} and edge impurity \cite{Delabie2016} charge exchange (CX) measurements.
CX measurements were supplemented with HRTS $T_e$ data when either main-ion CX data was unavailable (high $\delta_u$ set), or impurity CX measurements close to the separatrix were too noisy and assuming $T_i=T_e$ was deemed suitable due to high density (high $\bar{n}_e$ branch).
The core rotation profiles (not shown) were also fit using data derived from the core CX measurements when available (only for five out of seven discharges).

\textbf{Impurities:} Impurity composition and density profile data were derived from measurements with the methodology of Sertoli \textit{et al.}~\cite{Sertoli2018,Sertoli2019}.
This resulted in density profiles for a low-, medium- and high-Z impurity species; specifically beryllium, nickel and tungsten.
The deuterium density profiles were computed from plasma quasi-neutrality.
In our region of interest, the effective ion charge $Z_{\text{eff}}\approx1.2-1.4$ \footnotemark[2]\footnotetext[2]{The values of the line-averaged $Z_{\text{eff}}$, which were used as part of the impurity composition calculations, were revised after a large part of this work was already performed.
The updated values were typically 25-40\% larger than the values used in the computation of the impurity compositions used in this work.
Therefore, the local values of $Z_{\text{eff}}$ quoted here are underestimating the actual experimental values.
This also affects the resulting impurity densities and the main-ion dilution used in this work.
A sensitivity analysis to characterize the impact of $Z_{\text{eff}}$ and the resulting dilution was conducted.}, see Table \ref{tab:params}. 
In the low-density branch discharges, all three impurity species contribute significantly to $Z_{\text{eff}}$, while in the high density branch discharges beryllium is the dominant impurity.
For two discharges (high $\delta_u$, mid and high $\bar{n}_e$) this analysis could not be performed, and the composition and profiles were extrapolated from the available data.

\textbf{Magnetic equilibria:} Safety factor $q$ and magnetic shear $\hat{s}$ profiles self-consistent with the GPR fit profiles were obtained from current diffusion simulations using the fixed-boundary Grad-Shafranov solver ESCO coupled to the JETTO transport solver in the JINTRAC suite \cite{Romanelli2014}.
The pressure-constrained EFIT++ $q$-profile was used as an initial condition, while the total plasma current from the experiment and the EFIT++ last-closed flux-surface (LCFS) were used as boundary conditions.
Source profiles related to NBI injection were taken from ASCOT simulations from Ref.~\cite{Vincenzi2022}.
Current was allowed to diffuse in the simulations until the radial location at which $q=1$ matched the sawtooth inversion radius $\rho_{\text{inv}}$ determined from core ECE $T_e$ measurements. 
A high resolution (451 grid points) and a larger-than-usual number of spline knots (210) were required in ESCO to get numerically converged solutions near the edge of the plasma.

These self-consistent equilibria were also used as the numerical magnetic geometry in the gyrokinetic simulations in the next section.
Plasma shape parameters required for an analytical parameterization of the local magnetic equilibrium, such as $s$-$\alpha$ or Miller geometry \cite{Miller1998}, were extracted with the MEGPy Python package \cite{Snoep2023,Megpy}.

\textbf{Processing:} All data was processed and automatically converted into simulation input for all the codes utilized in this work with the GyroKit \cite{Gyrokit} Python package, which was developed during this work.

\begin{figure}[!h]
   \centering
   \includegraphics[width=0.475\textwidth]{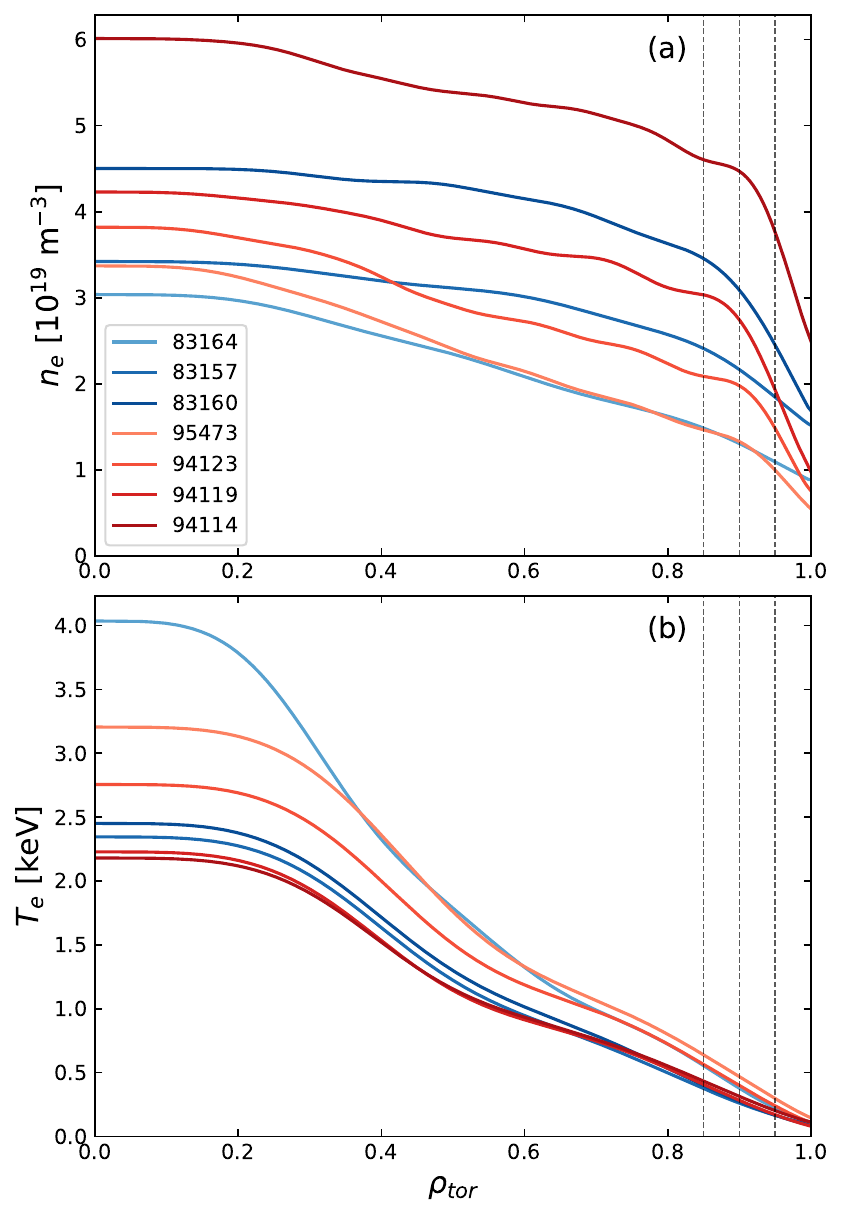}
   \caption{Nominal Gaussian process regression fits of time-averaged measurements of (a) the electron density $n_e$ and (b) the electron temperature $T_e$ as functions of $\rho_{\text{tor}}$ for both the high- (shades of blue) and low-triangularity (shades of red) JET-ILW discharges. Radial locations of interest in the pedestal-forming region $\rho_{\text{tor}} \in [0.85,0.90,0.95]$ are indicated (vertical dashed).}
   \label{fig:neTe_profiles}
   \vspace*{-2em}
\end{figure}

\subsection{Profile observations}
\label{sec:data:obs}
The nominal electron density and temperature GPR fits of all discharges are shown in Figure \ref{fig:neTe_profiles}(a) and (b), with the high- and low-triangularity sets in shades of blue and red, respectively.
The vertical dashed lines indicate the three radial positions of interest in the pedestal-forming region.
The electron density profiles in Figure \ref{fig:neTe_profiles}(a) show similar trends for both plasma geometries in the plasma center, although the low-triangularity discharges have higher density peaking.
In the pedestal-forming region, where the differences in plasma shaping are significant, the density profiles start diverging.
The low-triangularity discharges have lower electron densities $n_e$ towards the separatrix and larger normalized logarithmic density gradients $a/L_{n_e}$ relative to the corresponding high-triangularity discharges, also see Table \ref{tab:params}.
These differences are significant in comparison to the uncertainties of the fits, and are particularly striking at low line-averaged densities, which could indicate local difference in particle transport.
However, the main-chamber wall-clearance and the divertor geometry differed slightly, and the gas fueling flows and injection positions varied between the low and high-triangularity discharge sets as well.
These factors can influence the particle sources at the separatrix \cite{Lomanowski2023,Salmi2023}, in addition to shaping-related changes in transport.
Furthermore, recent improvements to the data processing methodology of the reflectometry \cite{Morales2024} suggest that the uncertainty of the radial positioning of measurements near the separatrix might have been large, which is not accounted for here.
This makes it challenging to draw any definitive conclusions.

For the electron temperature profiles in Figure \ref{fig:neTe_profiles}(b) we again see very similar profile trends in the plasma center of all discharges.
In the pedestal-forming region, a clear bifurcation between the low- and high-density branches can be observed, with the low-density branch exhibiting higher electron temperatures than the high-density branch.
The high-triangularity discharges consistently have lower $T_e$ in the pedestal-forming region compared to the adjacent low-triangularity discharges, but showing the opposite trend further inwards.
However, compared to the other normalized logarithmic gradient scale lengths, $a/L_{T_e}$ changes little in the pedestal-forming region across the ranges of $\bar{n}_e$ and $\delta_u$, varying consistently $<20\%$ (see Table \ref{tab:params}).

\begin{table*}[]
   \caption{Nominal experimental plasma parameters for seven JET-ILW deuterium discharges at radial positions $\rho_{\text{tor}}\in[0.85,0.90,0.95]$, used as input for linear gyrokinetic simulations. Here $\delta_u$ is the upper triangularity; $n_e$ and $T_e$ are the electron density in $10^{19}$\SI{}{\per\meter\cubed} and temperature in \SI{}{\electronvolt}, respectively; $a/L_{n_e}$, $a/L_{T_e}$ and $a/L_{T_i}$ are the normalized logarithmic density and temperature gradients; $Z_{\text{eff}}$ is the local effective ion charge; $q$, $\hat{s}$ and $\alpha$ are the safety factor, the local magnetic shear and the normalized pressure gradient, respectively; $\beta_e$ is normalized electron pressure in \%, $\nu^*_e$ is the normalized electron collision frequency and $\hat{\nu}_{ei} = \nu_{ei}/(c_s/a)$ is the normalized electron-ion collision frequency.}
   \label{tab:params}
   \setlength{\tabcolsep}{3.75pt}
   \setlength{\extrarowheight}{0.75pt}
   \centering
   \begin{tabular}{c c c c c c c c c c c c c c c c}
      \hline\hline\\[-2.25ex]
      $\rho_{\text{tor}}$ & \# & $\delta_u$ & $n_e$ & $T_e$ & $\frac{T_i}{T_e}$ & $a/L_{n_e}$ & $a/L_{T_e}$ & $a/L_{T_i}$ & $Z_{\text{eff}}$ & $q$ & $\hat{s}$ & $\alpha$ & $\beta_e$ & $\nu^*_e$ & $\hat{\nu}_{ei}$\\[1ex]
      \hline
      \multirow{7}{*}{0.85}
         & \textcolor{pblue}{83164} & 0.21 & 1.5 & 555 & 1.1 & 2.6 & 7.7 & 4.6 & 1.4 & 2.6 & 3.0 & 0.13 & 0.034 & 0.35 & 0.52\\
         & \textcolor{pblue}{83157} & 0.21 & 2.4 & 374 & 1.1 & 2.0 & 7.8 & 6.2 & 1.2 & 2.9 & 3.0 & 0.18 & 0.037 & 1.19 & 1.59\\
         & \textcolor{pblue}{83160} & 0.21 & 3.5 & 398 & 1.2 & 1.7 & 8.0 & 7.0 & 1.2 & 2.6 & 2.9 & 0.24 & 0.056 & 1.29 & 1.87\\
         & \textcolor{pred}{95473} & 0.11 & 1.5 & 640 & 0.9 & 1.8 & 6.4 & 2.4 & 1.2 & 2.9 & 2.8 & 0.12 & 0.039 & 0.25 & 0.36\\
         & \textcolor{pred}{94123} & 0.11 & 2.1 & 565 & 1.0 & 0.9 & 7.0 & 3.8 & 1.3 & 2.9 & 2.7 & 0.16 & 0.049 & 0.48 & 0.66\\
         & \textcolor{pred}{94119} & 0.10 & 3.0 & 408 & 0.9 & 0.8 & 7.6 & 6.1 & 1.2 & 2.9 & 2.7 & 0.19 & 0.052 & 1.19 & 1.68\\
         & \textcolor{pred}{94114} & 0.11 & 4.6 & 434 & 1.0 & 0.8 & 6.7 & 5.1 & 1.2 & 2.9 & 2.5 & 0.28 & 0.084 & 1.58 & 2.16\\
      \hline
      \multirow{7}{*}{0.90}
         & \textcolor{pblue}{83164} & 0.25 & 1.3 & 378 & 1.3 & 4.0 & 11.9 & 5.9 & 1.3 & 3.0 & 3.9 & 0.16 & 0.020 & 0.68 & 0.93\\
         & \textcolor{pblue}{83157} & 0.25 & 2.2 & 261 & 1.1 & 3.5 & 10.7 & 9.5 & 1.2 & 3.4 & 3.9 & 0.23 & 0.023 & 2.28 & 2.84\\
         & \textcolor{pblue}{83160} & 0.25 & 3.1 & 273 & 1.2 & 4.2 & 11.2 & 10.9 & 1.2 & 3.0 & 3.8 & 0.33 & 0.034 & 2.59 & 3.50\\
         & \textcolor{pred}{95473} & 0.13 & 1.3 & 467 & 1.1 & 4.1 & 9.6 & 5.7 & 1.2 & 3.3 & 3.5 & 0.22 & 0.026 & 0.45 & 0.59\\
         & \textcolor{pred}{94123} & 0.13 & 2.0 & 398 & 1.2 & 3.0 & 10.5 & 7.4 & 1.3 & 3.3 & 3.4 & 0.29 & 0.033 & 0.93 & 1.21\\
         & \textcolor{pred}{94119} & 0.12 & 2.7 & 283 & 1.0 & 5.3 & 11.1 & 9.5 & 1.2 & 3.3 & 3.4 & 0.33 & 0.033 & 2.35 & 3.12\\
         & \textcolor{pred}{94114} & 0.13 & 4.5 & 314 & 1.0 & 1.6 & 9.6 & 8.4 & 1.2 & 3.3 & 3.2 & 0.42 & 0.059 & 3.01 & 3.90\\
      \hline
      \multirow{7}{*}{0.95}
         & \textcolor{pblue}{83164} & 0.30 & 1.1 & 221 & 1.7 & 6.0 & 17.9 & 7.5 & 1.2 & 3.7 & 6.0 & 0.19 & 0.010 & 1.66 & 2.01\\
         & \textcolor{pblue}{83157} & 0.29 & 1.8 & 165 & 1.1 & 5.5 & 15.0 & 16.2 & 1.2 & 4.0 & 5.9 & 0.29 & 0.012 & 5.34 & 5.80\\
         & \textcolor{pblue}{83160} & 0.29 & 2.5 & 166 & 1.2 & 8.9 & 16.3 & 18.7 & 1.2 & 3.6 & 5.9 & 0.39 & 0.017 & 6.14 & 7.38\\
         & \textcolor{pred}{95473} & 0.15 & 1.0 & 296 & 1.2 & 12.4 & 15.6 & 15.9 & 1.2 & 3.9 & 5.1 & 0.36 & 0.012 & 0.91 & 1.09\\
         & \textcolor{pred}{94123} & 0.15 & 1.5 & 240 & 1.3 & 13.7 & 17.3 & 15.6 & 1.3 & 3.9 & 5.2 & 0.48 & 0.015 & 2.13 & 2.51\\
         & \textcolor{pred}{94119} & 0.15 & 1.9 & 170 & 1.0 & 14.7 & 17.2 & 16.3 & 1.3 & 3.9 & 5.2 & 0.40 & 0.014 & 5.22 & 6.22\\
         & \textcolor{pred}{94114} & 0.15 & 3.8 & 202 & 1.1 & 8.5 & 14.2 & 14.8 & 1.2 & 3.9 & 4.9 & 0.69 & 0.032 & 6.49 & 7.81\\
      \hline
   \end{tabular}
\end{table*}

\section{LINEAR PHYSICS PRIOR TO $t_{L-H}$}
\label{sec:gene}
To characterize the linear instabilities found on both density branches of $P_{\text{L-H}}$ just prior to $t_{L-H}$, local linear gyrokinetic simulations were performed with the GENE code \cite{Jenko2000}.
This characterization was conducted with two objectives in mind: (1) to develop a comprehensive dataset that can be used for the verification of the reduced-order models in Sec.~\ref{sec:ver}, and (2) to investigate the potential correlation between the behavior of $P_{\text{L-H}}$ as a function of $\bar{n}_e$ and the dominant linear instability regimes.

First, the general setup of the linear gyrokinetic simulations presented in this section is described. 
Second, the linear stability of both density branches is covered in detail, including the effect of the normalized logarithmic density and temperature gradients $a/L_{n_e}$, $a/L_{T_e}$ and $a/L_{T_i}$, changes in the mode structures and spectral dependence in the radial wavenumber $k_x$.
Third, it is shown that the impact of collisionality on the dominant linear instabilities can account for the differences in behavior of the low and high $\bar{n}_e$ branches.
Fourth, the sensitivity of the linear instabilities to the main ion (isotope) mass is examined.
Finally, a brief overview of the sensitivity of the linear stability analysis to electromagnetic effects, toroidal rotation (gradient), magnetic geometry and impurity composition is reported.

\subsection{GENE setup and convergence}
Nominal plasma parameters from the GPR fits for all seven JET discharges at three radial locations, $\rho_{\text{tor}}$ = 0.85, 0.9 and 0.95, were used for all GENE simulations, see Table \ref{tab:params}. 
For the magnetic geometry, numerical equilibria (num.~eq.) generated with the fixed-boundary code ESCO, as described in section \ref{sec:data:fit}, were mapped onto the GENE grid using the TRACER interface.

Initial value simulations were set up with three or four kinetic species: electrons, deuterium, beryllium and, when included, a mass-weighted composite of the nickel and tungsten species.
This second impurity species was only added when the relative contribution of the medium- and high-$Z$ impurities on $Z_{\text{eff}}$ was larger than 5\%, typically only on the low $\bar{n}_e$ branch.
$Z_{\text{eff}}$ was only used in the evaluation of the impurity composition and was not set as a direct input to GENE.
Due to the substantial collisionalities considered here, the high-fidelity Sugama collision operator \cite{Sugama2009,Crandall2020} was used.
The impact of finite-Larmor radius effects on the collision operator was evaluated for a few cases, but they were subsequently ignored given that ion-scale growth rates were typically affected by less than 3\%, in accordance with Ref.~\cite{Belli2017}.
Electromagnetic (EM) effects (\textit{i.e.}~finite $\beta$ and parallel magnetic fluctuations $\delta B_{||}$), were included for all simulations presented unless stated otherwise.
$\beta$ was evaluated self-consistently from the numerical equilibria, and the full drift velocity was considered in the evaluation of the pressure gradient term.
Plasma rotation was not included in most of the simulations, due to missing measurements for two out of seven discharges, but was later tested as a sensitivity.
$E \times B$ shear was also not included, as in linear simulations this leads to Floquet modes, significantly complicating the analysis \cite{Dagnelie2019}.

Eigenvalue simulations were run for a small subset of cases with the \textit{`harmonic'} SLEPc solver option.

GENE uses a field-aligned coordinate system ${x,y,z,v_{||},\mu}$, where $x$, $y$ and $z$ are the radial, the binormal and field-aligned coordinates respectively, $v_{||}$ is the parallel velocity and $\mu$ is the magnetic moment.
Typical grid settings for the linear simulations after convergence checks were [$n_x$,$n_z$,$n_{v_{||}}$,$n_{\mu}$] $\approx$ [9-17,48-128,64,18].
Following stronger ballooning at higher wavenumbers $n_x$ was decreased from 17 to 9 for $k_y\rho_s \geq 0.4$, where $\rho_s$ is the ion sound gyroradius, and $n_z \in [48,80,128]$ for $\rho_{\text{tor}} \in [0.85,0.9,0.95]$, respectively.
The binormal wavenumber was scanned for $0.05 \leq k_y\rho_s \leq 100$, focussing on $k_y\rho_s \leq 1$.

\subsection{Linear stability analysis with GENE}
\label{sec:gene:lsa}
In this section, the observed mode characteristics of the dominant linear instabilities at $\rho_{\text{tor}} \in [0.85,0.9,0.95]$ are described in detail.
In this analysis, we limit ourselves to the ion-scale ($k_y \rho_s \leq 1$) instabilities, as these are expected to dominate the turbulent transport in the experiments.
For details on the electron-scale spectra see Appendix \ref{apdix:elecs}.

The linear growth rate $\gamma$, frequency $\omega$, heat flux ratio q$_i$/q$_e$ and convective heat flux ratio $\tfrac{3}{2} \text{T}_e \Gamma_e$/q$_e$ spectra of the most unstable microinstabilities in the GENE simulations are shown as a function of $k_y$ at the three radial locations considered in Figure \ref{fig:lin_rho=0.85}, \ref{fig:lin_rho=0.9} and \ref{fig:lin_rho=0.95}.
The growth rates and frequencies are normalized to $c_s/a$, where $c_s = \sqrt{T_e/m_i}$, with $T_e$ the local electron temperature, $m_i$ the deuterium ion mass and $a$ the plasma minor radius of the last closed flux-surface.
The discharges are consistently grouped by suspected density branch based on their $P_{L-H}$ and $\bar{n}_e$, with discharges on the low $\bar{n}_e$ branch in the left column and on the high $\bar{n}_e$ branch on the right.
To easily distinguish between low- and high-triangularity discharges, the shades of red and blue introduced in Figure \ref{fig:neTe_profiles} are used respectively across all figures when data from both are shown.

Cross-correlation phases between fluctuations in the electrostatic potential $\tilde{\phi}$ and both the electron density $\tilde{n}_e$ and perpendicular temperature $\tilde{T}_{e,\perp}$ at all three radial locations are shown for a representative discharge in Figure \ref{fig:cphases}.
The cross-phases are weighted by the product of the integrated absolute values of the two compared quantities and split between the passing and trapped parts of the perturbed distribution function in the simulations, in accordance with the methodology described in Ref.~\cite{Dannert2005}.

The mode structures of the electrostatic potential $\phi$ in ballooning representation \cite{Candy2004b} are shown in Figure \ref{fig:phi_gene} for two discharges, one low and one high $\bar{n}_e$ branch discharge as a function of both $k_y$ and $\rho_{\text{tor}}$.

References to these characteristics, are made throughout the following sections.
 
\subsubsection{Mode characteristics at $\rho_{\text{tor}}=0.85$}
\label{sec:gene:lsa:85}
\begin{figure}[!t]
   \centering
   \hspace{-2em}
   \includegraphics[width=0.51\textwidth]{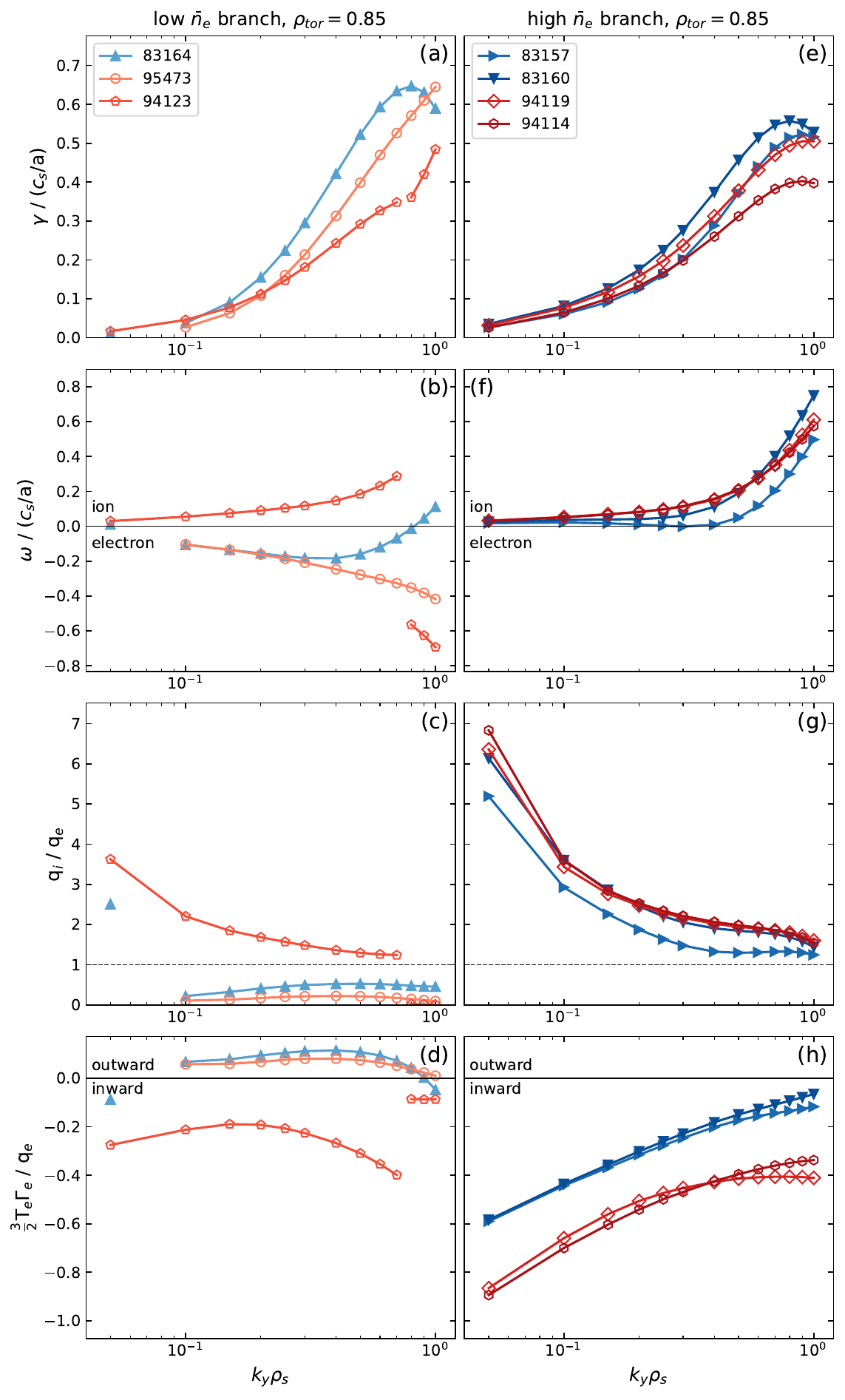}
   \caption{Linear GENE ion-scale spectra as function of binormal wavenumber $k_y$ at $\rho_{\text{tor}} = 0.85$.
   The linear (a, e) growth rate $\gamma$, (b, f) frequency $\omega$, (c, g) heat flux ratio q$_i$/q$_e$ and (d, h) convective heat flux ratio $\frac{3}{2}$T$_e\Gamma_e$/q$_e$ are shown for both the low- and high-density branches on the left and right, respectively.
   Positive/negative frequencies indicate mode propagation in the ion(/electron) diamagnetic drift direction, while positive/negative convective heat flux ratios indicate outward/inward particle fluxes.}
   \label{fig:lin_rho=0.85}
   \vspace*{-2em}
\end{figure}

\textbf{Low $\bar{n}_e$ branch:}
For \#95473 (light red circles), the dominant modes for all wavenumbers are propagating in the electron diamagnetic drift direction ($\omega < 0$), as can be seen in Figure \ref{fig:lin_rho=0.85}(b).
These instabilities have trapped-electron-mode (TEM) characteristics, as q$_i$/q$_e < 1$ and weighted cross-phases between fluctuations in the electrostatic potential $\tilde{\phi}$ and perpendicular electron temperature $\tilde{T}_{e,\perp}$, see Figure \ref{fig:cphases}(b), had the largest correlation amplitude for the trapped electrons \cite{Dannert2005}.
Although $\eta_e > 1$, where $\eta_e = (\partial\ln{T_e}/\partial r) / (\partial\ln{n_e}/\partial r) = L_{n_e}/L_{T_e}$, sensitivity checks showed these modes experience drive by both $a/L_{n_e}$ and $a/L_{T_e}$, suggesting these modes are a mix of both density-gradient-driven ($\nabla n$) and temperature-gradient-driven ($\nabla T$) TEM.
The mode structures of $\phi$ are extended along the field lines, \textit{i.e.}~had significant amplitude over more than one poloidal turn, as can be seen in Figure \ref{fig:phi_gene}(a).

For \#83164 (light blue triangles), Figure \ref{fig:lin_rho=0.85}(a-d) show that the low-wavenumber instabilities have similar TEM characteristics.
However, with increasing $k_y$, sensitivity checks showed $a/L_{T_i}$ also starts driving these modes, already at negative mode frequencies, as the mode propagation direction gradually shifts towards the ion diamagnetic direction.
Such characteristics match the transition from collisionless TEM (CTEM) to ubiquitous TEM (UTEM) \cite{Coppi1977}, with the $\omega$ cross-over happening at the peak of $\gamma$ \cite{Faber2015,Shen2019}.
The UTEM $\phi$ mode structures are first slightly extended along the field lines, similar to their TEM counterparts, and they become increasingly ballooned around the outboard midplane over the course of the propagation direction shift, see \textit{e.g.}~the magenta dashed mode structures in Figure \ref{fig:phi_gene}(g) and (h).

For \#94123 (red pentagons), the dominant ion-scale instabilities show ion-temperature-gradient (ITG) mode characteristics in Figure \ref{fig:lin_rho=0.85}(a-d), as $\omega > 0$ and q$_i$/q$_e > 1$.
Furthermore, the strongest cross-phase correlation was found between $\tilde{\phi}$ and $\tilde{T}_i$ and the mode structures were strongly ballooned at the outboard midplane, similar to those for $\rho_{\text{tor}}=0.85$ in Figure \ref{fig:phi_gene}(e-h).
This ITG is also dominant at $k_y\rho_s = 0.05$ for \#83164.
The difference in dominant modes for \#94123 coincides with an $a/L_{n_e}$ 2-3$\times$ smaller than that of the other two low $\bar{n}_e$ discharges at this radial location, see Table \ref{tab:params}.
This provides further evidence that the TEMs observed in the other discharges are a hybrid of $\nabla n$- and $\nabla T$-TEM.
At high wavenumbers, a sharp transition to instabilities with a mix of TEM and electron-temperature-gradient (ETG) mode characteristics occurs, see Appendix \ref{apdix:elecs}.

\textbf{High $\bar{n}_e$ branch:}
The dominant instabilities for the ion-scale wavenumbers for all four discharges also show ITG mode characteristics in Figure \ref{fig:lin_rho=0.85}(e-h), as $\omega > 0$ and q$_i$/q$_e > 1$.
Sensitivity checks (at $k_y\rho_s=0.2$) showed that these modes are indeed only driven by normalized logarithmic ion temperature gradient $a/L_{T_i}$ and that they are stabilized equally by both normalized logarithmic electron density and temperature gradients $a/L_{n_e}$ and $a/L_{T_e}$.
Although the growth rates, frequencies and q$_i$/q$_e$ ratios of all discharges show little variation as function of both $\bar{n}_e$ and $\delta_u$, the convective heat flux ratios in Figure \ref{fig:lin_rho=0.85}(h) show a clear difference between the low- (in red) and high-triangularity (in blue) discharges.
The particle fluxes are directed inward (as the total heat fluxes are always directed outward), with the low-triangularity discharges having relatively larger fluxes compared to the high-triangularity ones.
This correlates with stronger density profile peaking on axis for the low-triangularity discharges.
Gradient sensitivity scans showed no change in the particle flux direction for both $a/L_{n_e}$ and $a/L_{T_i}$, but did show a change from inward to outward when $a/L_{T_e}$ was reduced beyond the uncertainties of the fits.
This matches previous findings on the particle flux direction sensitivity of ITG modes \cite{Angioni2005b,Fable2010}.

\subsubsection{Mode characteristics at $\rho_{\text{tor}}=0.9$}
\label{sec:gene:lsa:90}
\begin{figure}[!b]
   \centering
   \hspace{-2em}
   \includegraphics[width=0.51\textwidth]{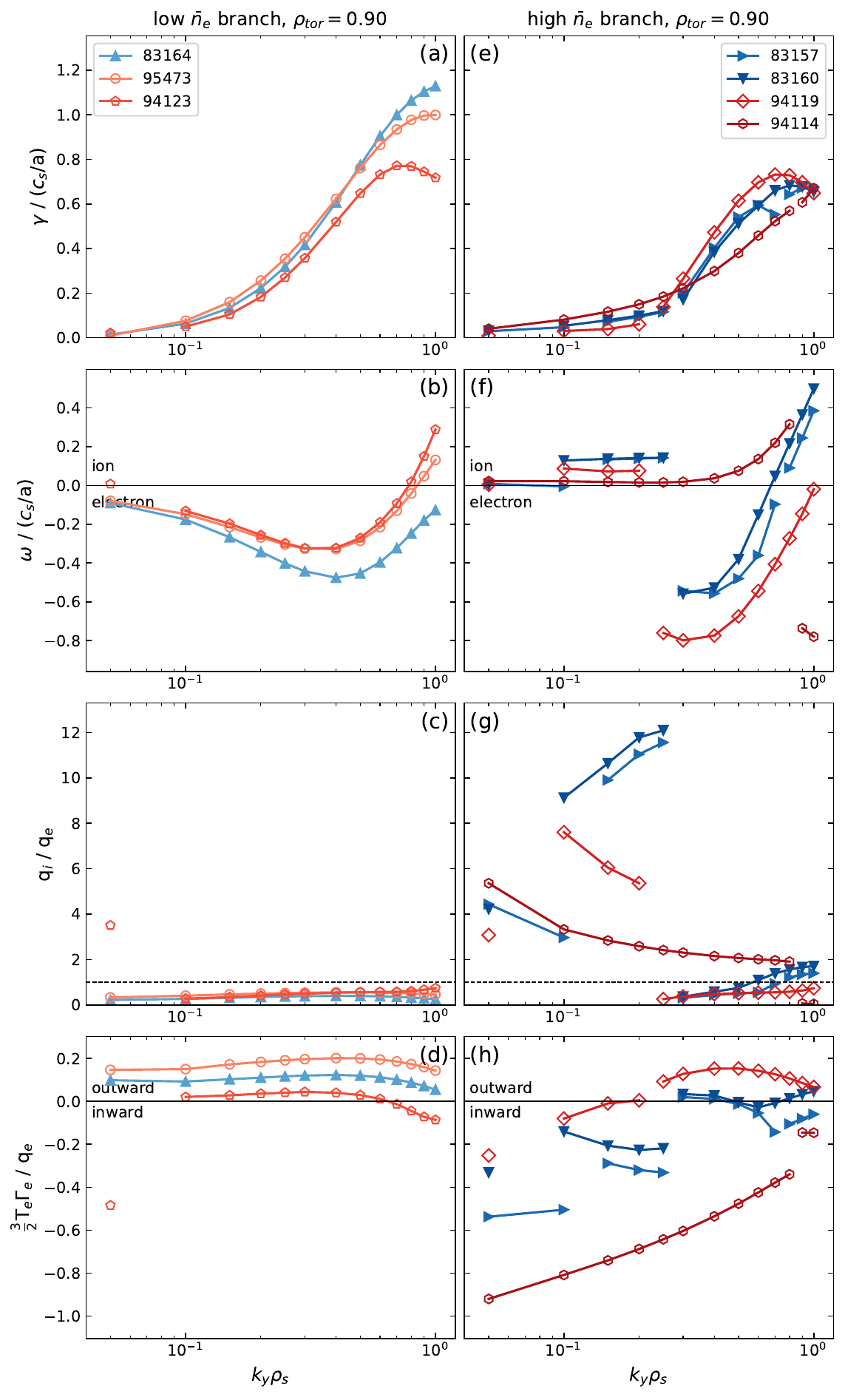}
   \caption{Linear GENE ion-scale spectra as a function of binormal wavenumber $k_y$ at $\rho_{\text{tor}} = 0.9$.
   The linear (a, e) growth rate $\gamma$, (b, f) frequency $\omega$, (c, g) heat flux ratio q$_i$/q$_e$ and (d, h) convective heat flux ratio $\frac{3}{2}$T$_e\Gamma_e$/q$_e$ are shown for both the low- and high-density branches on the left and right, respectively.
   Positive/negative frequencies indicate mode propagation in the ion(/electron) diamagnetic drift direction, while positive/negative convective heat flux ratio values indicate outward/inward particle fluxes.}
   \label{fig:lin_rho=0.9}
   \vspace*{-2em}
\end{figure}

\textbf{Low $\bar{n}_e$ branch:}
As Figure \ref{fig:lin_rho=0.9}(a-d) shows, all three discharges have similar CTEM-UTEM spectra as seen for \#83164 at $\rho_{\text{tor}}=0.85$.
The low-$k_y$ mode structures of $\phi$ are still somewhat extended along the field line, but less so compared to those at $\rho_{\text{tor}}=0.85$, as can be seen in Figure \ref{fig:phi_gene}(a) and (b).
However, for \#83164 the change in propagation direction of the UTEM does not occur here and instead a smooth transition to TEM-ETG hybrid instabilities happens for $k_y\rho_s>1$, as can be seen in Appendix \ref{apdix:elecs}.

\textbf{High $\bar{n}_e$ branch:}
For low wavenumbers $k_y\rho_s<0.25$, \#83157, \#94119 and \#83160 are dominated by instabilities propagating in the ion-direction, as can be seen in Figure \ref{fig:lin_rho=0.9}(f).
These instabilities have odd (tearing) parity, see for example Figure \ref{fig:phi_gene}(e) and (f).
Their mode structures remain odd-parity in electrostatic ($\beta = 0$) simulations, thus they are not microtearing modes.
Unconventional ballooning structures of the dominant instabilities have been reported previously at the steep normalized logarithmic gradients found in the pedestal(-forming) region \cite{Xie2015,Han2017}.
Among such modes are higher-order harmonics ($\ell>0$, where $\ell$ is the mode quantum number) of common even-parity modes \cite{Pueschel2019}.
The odd-parity modes reported here fit with a first-order ($\ell$=1) excitation of an ion-direction mode. 
Examples of similarly ballooned ion modes are the unconventional ITG\cite{Han2017}/tearing ITG (TITG)\cite{Pueschel2019}, and the trapped ion-temperature-gradient mode \cite{Li1996}.
Gradient sensitivity scans confirmed that these $\ell$=1 modes are also purely driven by $a/L_{T_i}$ and stabilized by both $a/L_{n_e}$ and $a/L_{T_e}$, similarly to the typical ITG modes we observed at $\rho_{\text{tor}}=0.85$.
As can be seen in Figure \ref{fig:lin_rho=0.9}(g) the heat flux ratio q$_i$/q$_e \gg 1$ for these $\ell$=1 ion modes, 2-3$\times$ higher than for the $\ell=0$ ITG.
The weighted cross-correlation phases between $\tilde{\phi}$ and $\tilde{T}_i$ of both passing and trapped ions for these modes had similar amplitudes for both deuterium and beryllium.
Furthermore, the cross-phase angle $\alpha_{\tilde{\phi}\times\tilde{T}_{i,\perp}} \approx \alpha_{\tilde{\phi}\times\tilde{T}_{i,\perp,\text{trap}}} \approx \pi/2$.
This indicates that the trapped-ion populations drive, or at least contribute to, the destabilization of these modes, in contrast to typical ITGs.
For the trapped beryllium, the cross-phase angle $\alpha_{\tilde{\phi}\times\tilde{n}_i} \approx \pi/2$.
Nevertheless, the overall effect of the impurities on the growth rate was found to be slightly stabilizing, similar to that for typical ITG, based on simulations without impurity species. 
Dilution by the impurities was found to reduce the growth rates by 5\% at $k_y\rho_s=0.2$.
Thus, impurities are not required to destabilize these $\ell$=1 ion modes.
Finally, further sensitivity scans indicated the magnetic shear $\hat{s}$ stabilizes these modes at the relatively large $\hat{s}$ seen in the pedestal-forming region.

For larger wavenumbers $k_y\rho_s \geq 0.25$, \#83157, \#94119 and \#83160 become dominated by similar TEM-UTEM mode branches as seen in the discharges on the low $\bar{n}_e$ branch. 
Yet, there are some differences in characteristics.
The UTEM propagation direction change occurs slightly after the peak in growth rate, as can be seen in Figure \ref{fig:lin_rho=0.9}(e) and (f), while for the high-triangularity discharges this is paired with q$_i$/q$_e > 1$, as shown in Figure \ref{fig:lin_rho=0.9}(g).
Gradient sensitivity checks showed that these modes are still driven by both $a/L_{T_e}$ and $a/L_{T_i}$, but slightly stabilized by $a/L_{n_e}$.
Furthermore, weighted cross phases showed that trapped particles are still contributing the most to the instability, making it unlikely this marks a transition to ITG.

The highest density discharge \#94114 (dark red hexagons) remains dominated by the ITG modes seen on the high-density branch at $\rho_{\text{tor}}=0.85$, see Figure \ref{fig:lin_rho=0.9}(e-h).
This regime difference is again correlated with an $a/L_{n_e}$ smaller by a factor of 2-3 relative to the other three discharges, as can be seen in Table \ref{tab:params}.

\subsubsection{Mode characteristics at $\rho_{\text{tor}}=0.95$}
\label{sec:gene:lsa:95}
\begin{figure}[!t]
   \centering
   \hspace{-2em}
   \includegraphics[width=0.51\textwidth]{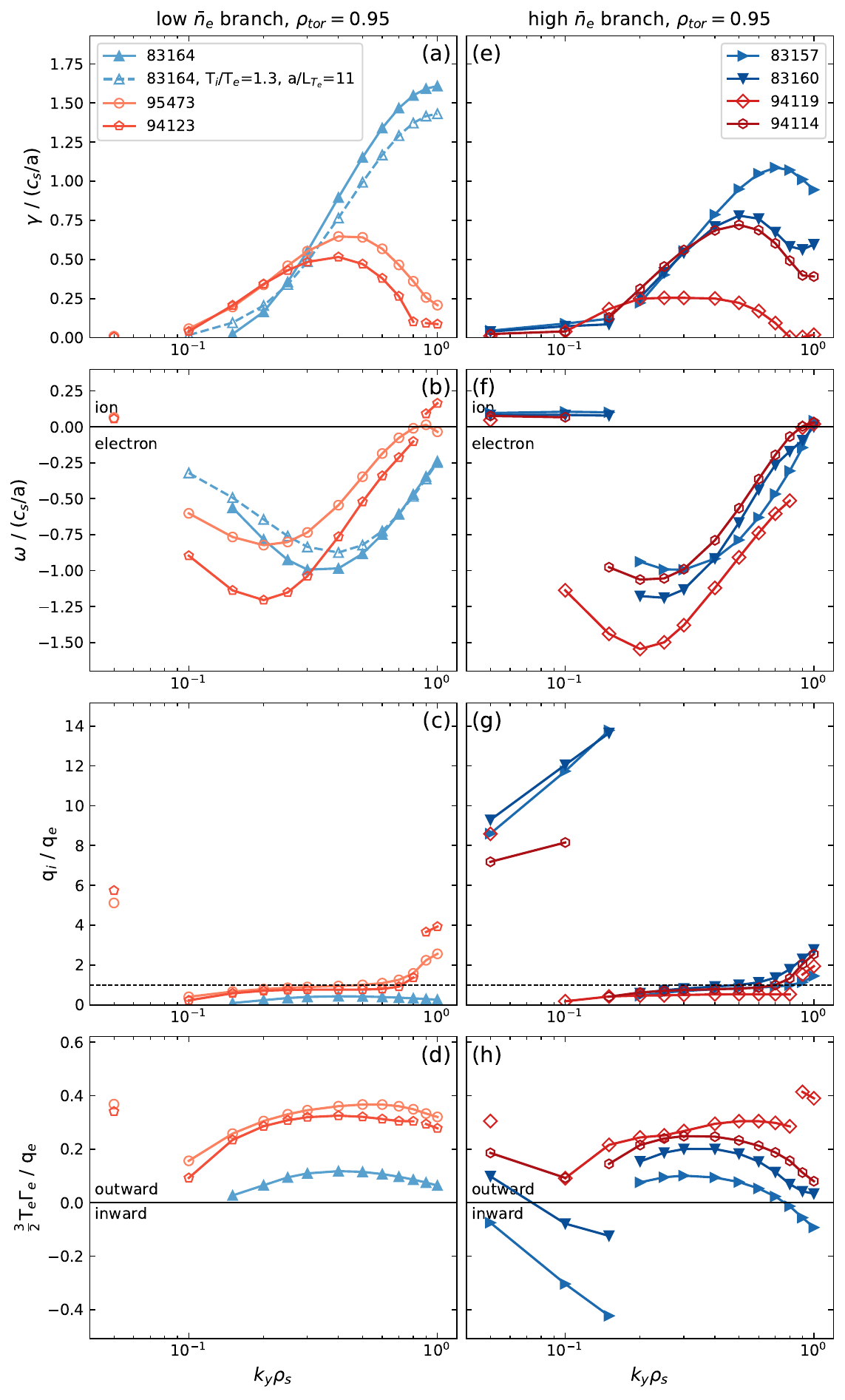}
   \caption{Linear GENE ion-scale spectra as a function of binormal wavenumber $k_y$ at $\rho_{\text{tor}} = 0.95$.
   The linear (a, e) growth rate $\gamma$, (b, f) frequency $\omega$, (c, g) heat flux ratio q$_i$/q$_e$ and (d, h) convective heat flux ratio $\frac{3}{2}$T$_e\Gamma_e$/q$_e$ are shown for both the low- and high-density branches on the left and right, respectively.
   Positive/negative frequencies indicate mode propagation in the ion(/electron) diamagnetic drift direction, while positive/negative convective heat flux ratio values indicate outward/inward particle fluxes.}
   \label{fig:lin_rho=0.95}
   \vspace*{-2em}
\end{figure}

\textbf{Low $\bar{n}_e$ branch:}
The ion-scale spectra of all three discharges show modes propagating in the electron direction, as can be seen in Figure \ref{fig:lin_rho=0.95}(a-d).
The mode frequency and flux ratio spectra show similarities to the TEM-UTEM regimes at previous radii, but several mode characteristics are different.
Compared to previous radii, there are also more pronounced differences between the low- and high-triangularity discharges.
\begin{figure}[!t]
   \centering
   \includegraphics[width=0.475\textwidth]{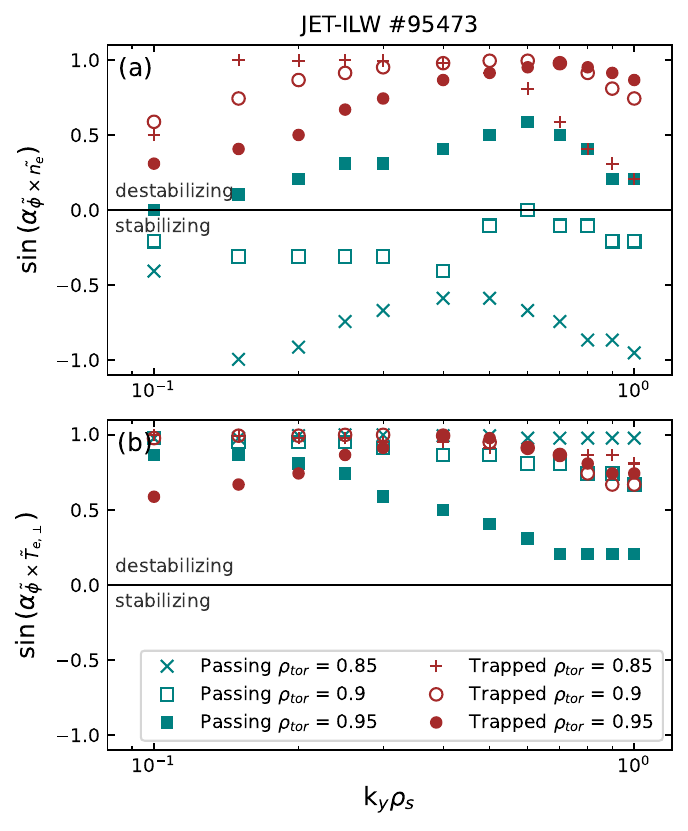}
   \caption{Sine of the mean cross-phase angles $\alpha$ between fluctuations in the electrostatic potential $\tilde{\phi}$ and (a) the electron density $\tilde{n}_e$ and (b) perpendicular electron temperature $\tilde{T}_{e,\perp}$ as function of $k_y$ for \#95473 at $\rho_{\text{tor}}$ = 0.85 (crosses), $\rho_{\text{tor}}$ = 0.9 (open symbols) and 0.95 (solid symbols). 
   Passing (x, squares) and trapped (+, circles) parts of the perturbed distribution function are separately considered. 
   The cross phases are weighted by the product of the integrated absolute values of the two compared quantities. 
   Positive/negative values indicate destabilizing/stabilizing contributions to the mode growth.}
   \label{fig:cphases}
\end{figure}
\begin{figure*}
   \centering
   \includegraphics[width=0.9\textwidth]{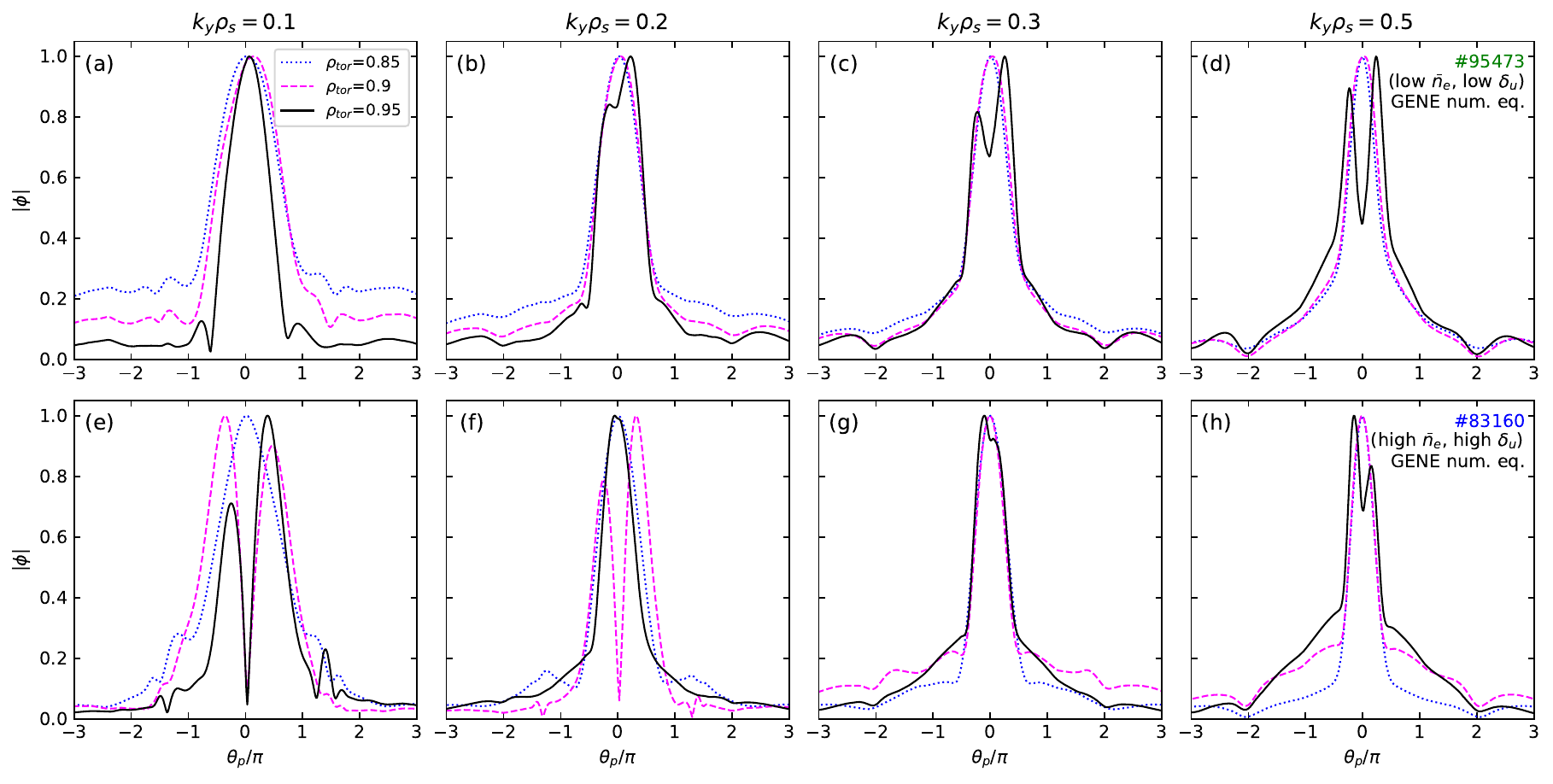}
   \caption{Linear GENE (num.~eq.) ballooning representation of the normalized electrostatic potential $\phi$ as a function of ballooning angle $\theta_{p}$.
   Mode structures for $k_y\rho_s$ = 0.1, 0.2, 0.3 and 0.5 are shown for \#95473 (a, b, c, d on top row) and \#83160 (e, f, g, h on bottom row) at $\rho_{\text{tor}}$ = 0.85 (blue dotted), 0.90 (magenta dashed) and 0.95 (black solid).
   A truncated view of the simulated parallel domain $\theta_{p} \leq \pm17\pi$, centered on the outboard midplane, is shown for clarity.}
   \label{fig:phi_gene}
\end{figure*}

\begin{figure*}
   \vspace{-1em}
   \centering
   \begin{minipage}[t]{0.48\textwidth}
      \centering
      \includegraphics[width=0.85\textwidth]{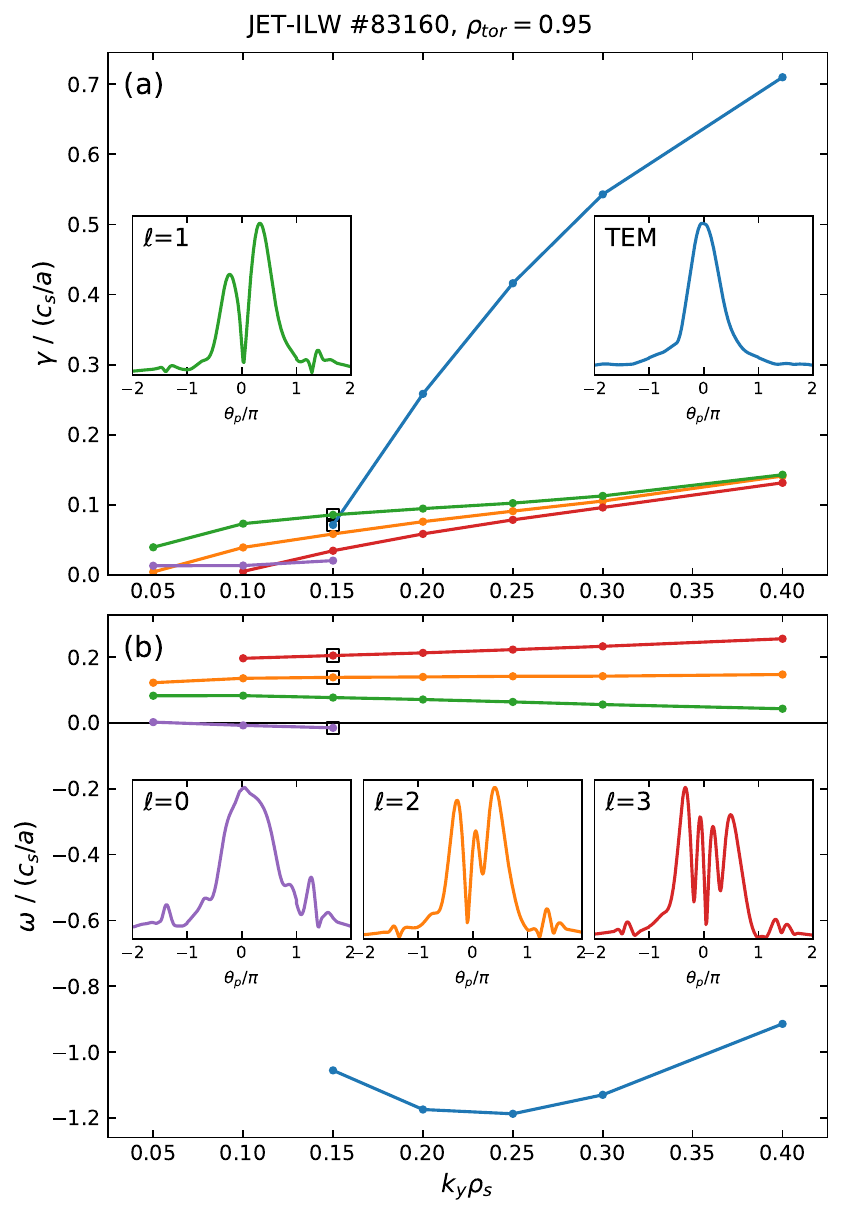}
      \caption{Linear (a) growth rate $\gamma$ and (b) frequency $\omega$ spectra from GENE eigenvalue simulations for \#83160 at $\rho_{\text{tor}}=0.95$.
      Insets show a truncated view (two poloidal turns centered on the outboard midplane) of multiple eigenmodes at $k_y\rho_s = 0.15$.}
      \label{fig:83160_ev}
   \end{minipage}
   \hfil
   \begin{minipage}[t]{0.48\textwidth}
      \centering
      \includegraphics[width=0.85\textwidth]{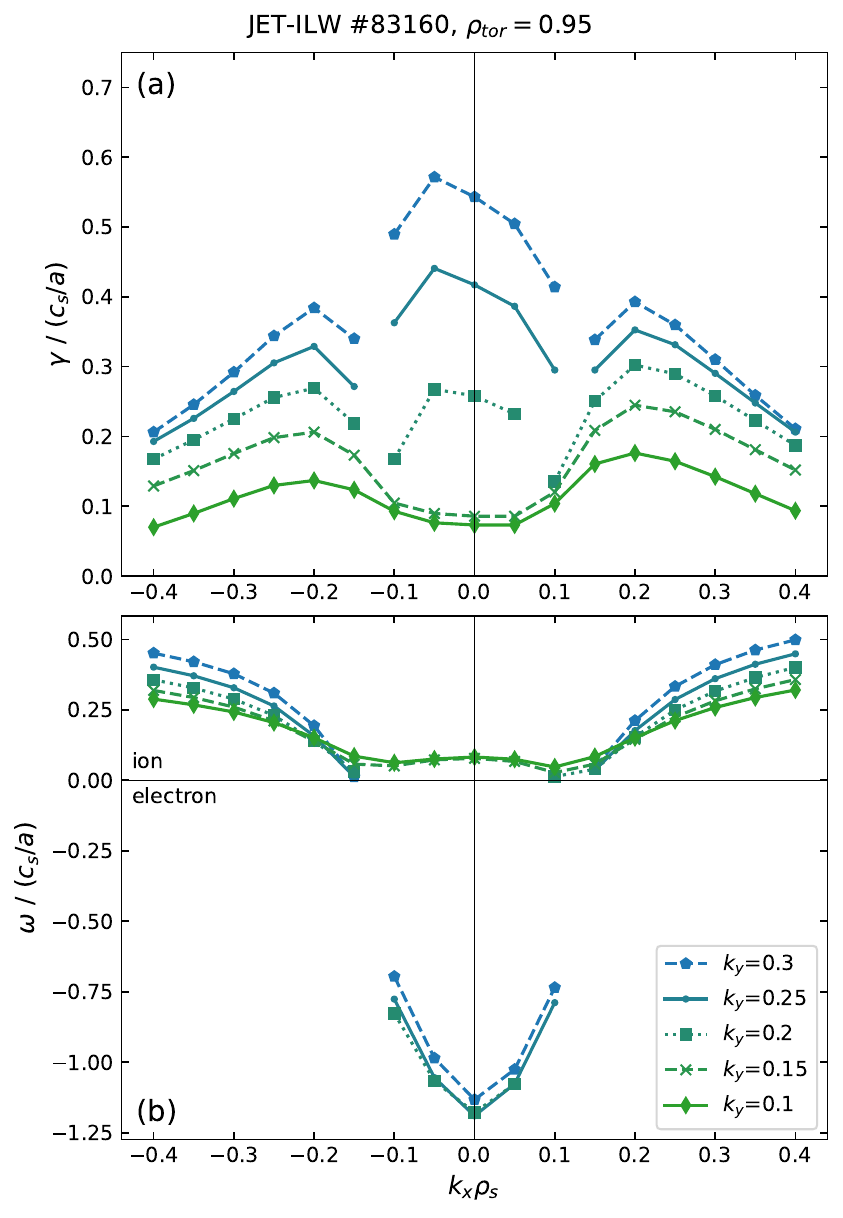}
      \caption{Linear (a) growth rate $\gamma$ and (b) frequency $\omega$ spectra of the most unstable modes as functions of radial wavenumber $k_x$ from GENE initial-value simulations for \#83160 at $\rho_{\text{tor}}=0.95$.}
      \label{fig:83160_kx}
   \end{minipage}
 \end{figure*}

The mode structures of $\phi$ for $k_y\rho_s < 0.4$ are less extended along the field lines than at the previous two radii, as illustrated in Figure \ref{fig:phi_gene}(a) and (b).
This coincides with a reduction in the contribution of trapped electrons to the mode destabilization compared to the instabilities dominant at $\rho_{\text{tor}}=0.9$.
An example of this is shown for \#95473 in Figure \ref{fig:cphases}(a) and (b), where the weighted cross-phase angles $\alpha$ between $\tilde{\phi}$ and both $\tilde{n}_e$ and $\tilde{T}_{e,\perp}$ show a clear reduction in the contribution from the trapped part of the perturbed distribution function for electrons to the instability when comparing $\rho_{\text{tor}}=$ 0.9 and 0.95 (open vs solid circles).
For the low-triangularity discharges, the contribution of passing electrons also changes from stabilizing to destabilizing from $\rho_{\text{tor}}=$ 0.9 to 0.95, while for the high-triangularity discharges the passing electrons become near-adiabatic ($\alpha_{\phi \times n_{e,p}} \approx 0$).
Previous studies have demonstrated a similar increase in the importance of the passing electron dynamics near the edge of tokamak plasmas, attributing this to electron drift waves becoming dominant \cite{Scott2005,Bonanomi2019}.
However, the mode frequencies of the low-$k_y$ modes found here are on average at least 5$\times$ smaller than the electron diamagnetic drift frequency $\omega_{*e}$, the characteristic frequency of electron drift waves \cite{Conner1994}.
Furthermore, gradient sensitivity scans showed that these modes are driven by both $a/L_{T_e}$ and (to a lesser extent) $a/L_{T_i}$, while their response to $a/L_{n_e}$ exhibited a non-monotonic dependence on the other two gradients.
This is consistent with the sensitivities found in Ref.~\cite{Bonanomi2019}.
Yet, a similar combination of $a/L_{T_e}$ drive and non-monotonic sensitivity to $a/L_{n_e}$ at high normalized gradient values was found in a study of TEMs in Ref.~\cite{Ernst2009}.
Thus, this combination of characteristics indicates that these instabilities have a hybrid nature \cite{Rewoldt2005}, combining both TEM, ITG and (electron) drift-wave characteristics.

The ballooning structures of $\phi$ for the dominant instabilities no longer peak at the midplane and become asymmetric, examples of which are shown in Figure \ref{fig:phi_gene}(b), (c) and (d).
These modes are even-parity and can thus be seen as (superpositions of) even-$\ell$ states.
Sensitivity scans for radial wavenumber $k_x$ showed that the peak in growth rate of these modes lies at finite $k_x$, see for example $k_y\rho_s=0.3$ in Figure \ref{fig:83160_kx}(a). 
Scans in the gradients showed that the dip in mode amplitude at the midplane increased with increasing $a/L_{n_e}$.
This also explains why it is more prominent for the low-triangularity discharges, as can be seen by comparing Figure \ref{fig:phi_gene}(d) and (h).

Towards $k_y\rho_s = 1$, a change in mode propagation direction still occurs, but only after the peak in the growth rate spectra, as shown in Figure \ref{fig:lin_rho=0.95}(a) and (b).
This is similar to the changes for UTEMs in the high $\bar{n}_e$ branch discharges at $\rho_{\text{tor}}=0.9$, suggesting that it may be related to the increase in collisionality.

Growth rates differ significantly between \#83164 and the two low-triangularity discharges, as can be seen in Figure \ref{fig:lin_rho=0.95}(a).
This is correlated with $a/L_{n_e}$ and $a/L_{T_i}$ for \#83164 being smaller by a factor two, see Table \ref{tab:params}. 
Gradient sensitivity checks confirmed that increasing both of these gradients for \#83164 results in growth rates and mode frequencies comparable to those of the low-triangularity discharges, with most of the differences arising from increased $a/L_{n_e}$.
However, even at equal $a/L_{n_e}$ and $a/L_{T_i}$, the linear spectra demonstrated increased stabilization for $k_y\rho_s \leq 0.2$ in \#83164 compared to the low-triangularity discharges. 
Further sensitivity checks indicated this can be attributed to differences in $T_i/T_e$ and magnetic geometry.
The ratio of ion-to-electron temperature $T_i/T_e$ is considerably higher for \#83164 than for the other discharges at this radial location, see Table \ref{tab:params} \footnotemark[1].
This was found to stabilize the modes at $k_y\rho_s \leq 0.2$, as shown in Figure \ref{fig:lin_rho=0.95}(a) and (b).
Nevertheless, additional simulations where the numerical equilibria for \#83164 and \#95473 were swapped revealed that the differences in magnetic geometry can also account for this additional stabilization.

\textbf{High $\bar{n}_e$ branch:}
The spectra of all four high-density discharges share similarities with those on the low $\bar{n}_e$ branch, as can be seen in Figure \ref{fig:lin_rho=0.95}.
For $k_y\rho_s \geq 0.2$, the spectra are dominated by similar electron-direction modes as on the low $\bar{n}_e$ branch.
Their cross-phase angles exhibited the same reduction for trapped and increase for passing electrons as for the low density branch discharges.
Gradient sensitivity scans also showed a similar mix of drive by both $a/L_{T_e}$ and $a/L_{T_i}$ and non-monotonic behavior for $a/L_{n_e}$.
In particular, \#94119, the low-triangularity minimum in $\bar{n}_e$, could also be considered on the low $\bar{n}_e$ branch here, because as can be seen in Figure \ref{fig:lin_rho=0.95}(e) and (f) that it has similar spectra as \#95473 and \#94123.
This is in agreement with Ref. \cite{Solano2023}.
This correlates with a similarly high ratio between $a/L_{n_e}$ and $a/L_{T_i}$ for \#94119, as can be seen in Table \ref{tab:params}.

For the lowest wavenumbers, the $\ell$=1 ion modes are dominant again, as can be seen in Figure \ref{fig:lin_rho=0.95}(e-h), but they are less prevalent than at $\rho_{\text{tor}}=0.9$.
These ion modes are more stabilized in the low-triangularity discharges because of a combination of the local balance of gradient drives ($a/L_{n_e}$, which acts stabilizing, and $a/L_{T_i}$, which acts destabilizing) and the impact of the magnetic geometry, which stabilizes the hybrid electron-direction modes more in the high-triangularity discharges.
Although these $\ell$=1 modes are less dominant throughout the $k_y$ spectrum at $\rho_{\text{tor}}=0.9$, Figure \ref{fig:83160_kx} shows that at finite $k_x$, their growth rates for $k_y\rho_s<0.3$ can be comparable to or higher than the dominant mode at $k_x\rho_s=0$.
The finite-$k_x$ growth rate spectra also show some asymmetries, with modes on the positive $k_x$ side having around 20\% higher growth rates.

Similar to $\rho_{\text{tor}}=0.9$, linear eigenvalue scans at $\rho_{\text{tor}}=0.95$ showed that in addition to the dominant $\ell$=1 excitation, there exist several unstable subdominant modes.
Higher-order excitations $\ell$ $>$1 and at low wavenumbers even the ground-state $\ell$=0 have comparable growth rates.
An example of this is shown for \#83160 in Figure \ref{fig:83160_ev}, where $\ell$=0, 1, 2 and 3 modes are unstable, in addition to the dominant hybrid TEM-like modes, over a portion of the spectrum.
Again, such behavior was not observed for the low $\bar{n}_e$ branch discharges at nominal gradients. 
This suggests that these subdominant modes are driven unstable by increasing density or collisionality.

\subsection{Effect of collisionality}
\label{sec:gene:coll}
\begin{figure*}[!p]
   \centering
   \includegraphics[width=\textwidth]{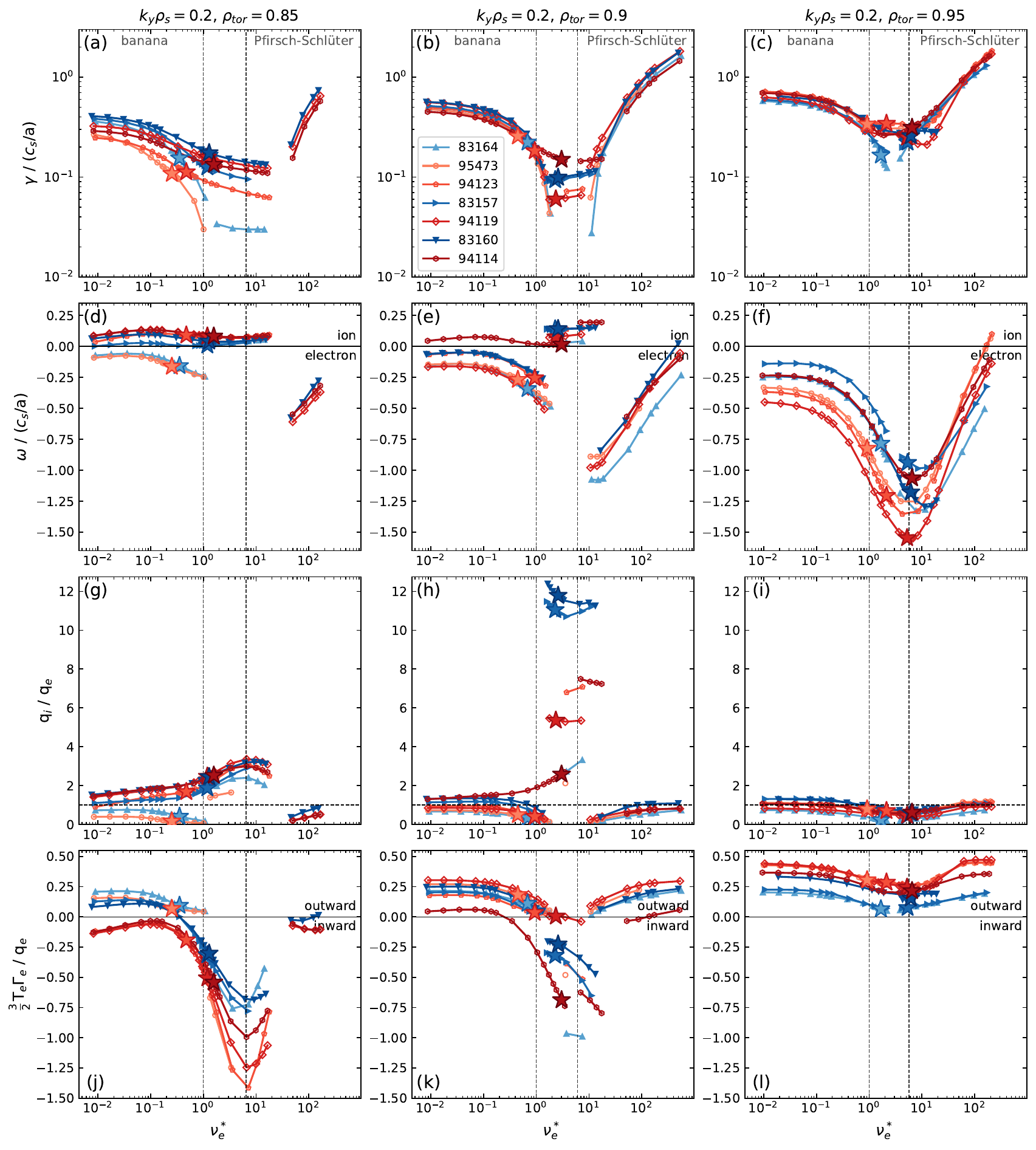}
   \caption{From top to bottom: the linear (a, b, c) growth rate $\gamma$, (d, e, f) mode frequency $\omega$, (g, h, i) heat flux ratio q$_i$/q$_e$ and (j, k, l) convective heat flux ratio $\tfrac{3}{2}$T$_e\Gamma_e/$q$_e$ as functions of normalized electron collision frequency $\nu_e^*$ for $k_y\rho_s=0.2$ at three radial positions, from left to right: $\rho_{\text{tor}} \in [0.85,0.9,0.95]$. 
   Both high- (shades of blue) and low-triangularity (shades of red) discharges are shown. 
   The normalized electron collision frequency determines the transition from core-like, collisionless instabilities to collisional/resistive instabilities found near the plasma edge.}
   \label{fig:col_scan}
\end{figure*}
The collisionality in the GENE simulations was scanned over several orders of magnitude for all discharges and radii to investigate which, if any, of the dominant mode branches have resistive characteristics.
In this work, the term resistive will be used to refer to any modification or destabilization of a mode due to finite collisionality, irrespective of whether magnetic fluctuations are involved; \textit{i.e.~}resistive mode branches by this definition can also be identified in the electrostatic limit.
In these scans, the collisionality was varied independently, while the density and temperatures were kept at the nominal experimental values.
In Figure \ref{fig:col_scan}, the resulting linear growth rates, frequencies, heat flux and convective heat flux ratios for $k_y\rho_s=0.2$ in all discharges at all three radii are shown as function of the dimensionless electron collisionality $\nu_e^* = \tfrac{4}{3\sqrt{\pi}} \tfrac{q R}{\epsilon^{3/2} v_{th,e}} \nu_{ei}$, where $\nu_{ei}$ is the electron-ion collision frequency, $q$ is the safety factor, $\epsilon=r/R$, $r$ is the minor radius of the flux-surface of interest and $v_{th,e} = \sqrt{T_e/m_e}$ is the thermal electron velocity.
The corresponding values at $k_y\rho_s=0.2$ from Sec.~\ref{sec:gene:lsa} are indicated with stars at the experimental collisionality.
The propagated uncertainty in the experimental collisionality was on average 38\%, 52\% and 84\% for $\rho_{\text{tor}} \in [0.85,0.9,0.95]$, respectively.
The wavenumber $k_y\rho_s=0.2$ was chosen to demonstrate the sensitivity of the dominant linear instabilities to collisionality at wavenumbers that typically contribute substantially to the saturated nonlinear transport.
The vertical dashed lines at $\nu_e^*=1$ and $\nu_e^*=\epsilon^{-3/2}$ indicate the boundaries between the three neoclassical collisional transport regimes, from left to right: banana, plateau and Pfirsch-Schl\"{u}ter, assuming a circular plasma \cite{Hinton1976}.

At all three radii, the growth rates exhibit a gradual decrease with increasing collisionality until they reach a minimum, after which they increase again, as can be seen in Figure \ref{fig:col_scan}(a-c).
This is consistent with previous work that found similar transitions from core-like TEM/ITG to resistive mode branches, \textit{i.e.}~instabilities that grow with increasing collisionality and hence resistivity \cite{Bourdelle2014,Bourdelle2015,DeDominici2019,Bonanomi2021}.
Additionally, we observe here that this behavior correlates with the three neoclassical collisionality regimes for the electrons.
Near the neoclassical banana-plateau boundary, where trapped electrons no longer complete their banana orbits, we find that all TEM mode branches are stabilized and/or become subdominant.
The ITG mode branches, which are not driven by trapped electrons, do not show the same correlation, but both their frequencies and heat flux ratios have inflection points near this boundary.
The minimum in growth rate is then typically found in the neoclassical plateau regime, although this again does not hold for the ITG branches at $\rho_{\text{tor}}=0.85$.
Resistive mode branches start becoming dominant close to the boundary of the Pfirsch-Schl\"{u}ter regime as we approach the edge of the plasma.
The collisionality threshold for the dominance of the resistive mode branches decreases by about an order of magnitude from $\rho_{\text{tor}}=0.85$ to 0.95, as can be seen by comparing Figure \ref{fig:col_scan}(a) and (c).
This correlation between the dominant linear gyrokinetic mode boundaries and the neoclassical collisionality regimes is observed for $k_y\rho_s < 0.4$, but it is weaker at higher $k_y$ values, where the UTEM branches become dominant.
Furthermore, the plasma is strongly shaped at these radii, which may cause the regime boundaries to deviate slightly from the circular approximations displayed in Figure \ref{fig:col_scan}.

A detailed description of how $\nu_e^*$ affects the mode characteristics of the dominant instabilities identified at the three radial locations follows.

At $\rho_{\text{tor}} = 0.85$, the dominant TEM and ITG mode branches identified in Sec.~\ref{sec:gene:lsa:85} are all stabilized by increasing collisionality, as can be seen in Figure \ref{fig:col_scan}(a) and (d).
The minima in growth rate as a function of collisionality occur at values that are at least an order of magnitude higher than the experimental collisionalities for all discharges, and thus outside of their collisionality uncertainty ranges.
The growth rate and frequency spectra indicate that both the collisionality and the balance of local gradient drives determine which mode branch is dominant.
For the high-triangularity discharges, the particle fluxes show a direction reversal with collisionality, as shown in Figure \ref{fig:col_scan}(j), while similar ITG branches for the low-triangularity discharges do not share this feature.
The particle flux depends on a balance of large terms, meaning that a small shift in parameters, including collisionality, can alter the flux direction \cite{Angioni2005b,Fable2010,Mordijck2015}.
\begin{figure}
   \centering
   \includegraphics[width=0.475\textwidth]{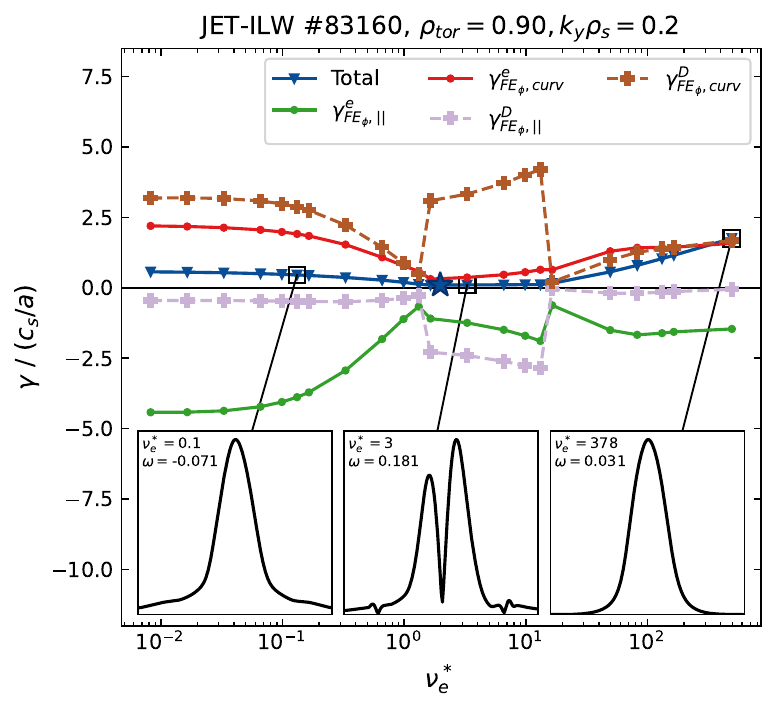}
   \caption{Linear growth rates $\gamma$ as function of normalized electron collisionality $\nu_e^*$ for $k_y\rho_s=0.2$ in \#83160 at $\rho_{\text{tor}}=0.9$.
   The growth rate is split into contributions by terms in the linear gyrokinetic equation to the electrostatic-potential part of the free-energy balance $\mathrm{FE}_{\phi}$.
   Only terms with a non-negligible contribution, here the parallel velocity ($||$) and curvature (curv) terms for the electrons (e) and deuterium (D) ions, are shown.
   Positive/negative values indicate destabilizing/stabilizing contributions.
   Insets show a truncated view (two poloidal turns centered on the outboard midplane) of the ballooning representation of $\phi$ for three representative examples of the three dominant mode branches.}
   \label{fig:fetime}
   \vspace{-1em}
\end{figure}
\begin{figure}
   \centering
   \includegraphics[width=0.45\textwidth]{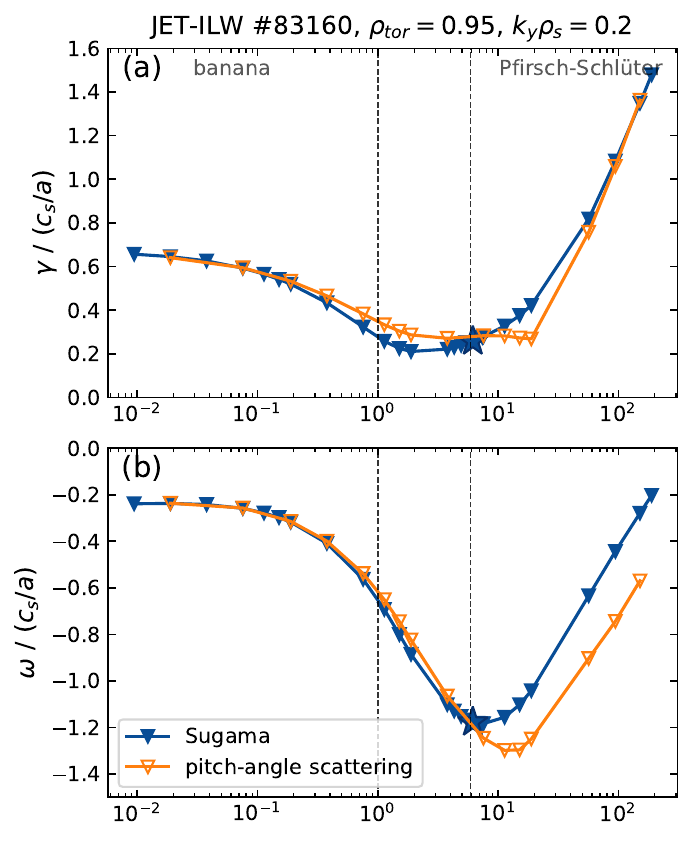}
   \includegraphics[width=0.45\textwidth]{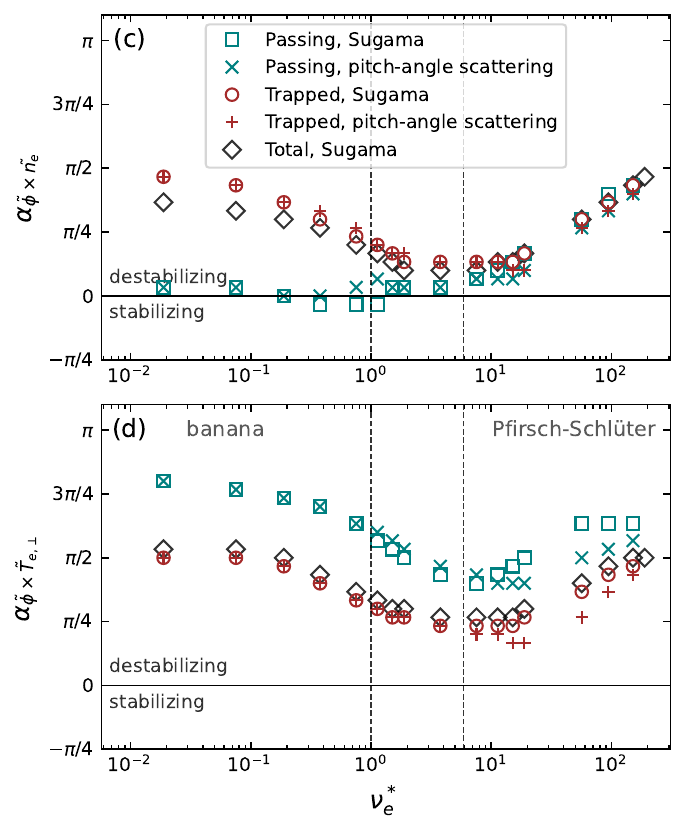}
   \caption{The effect of the collision operator on the linear (a) growth rate $\gamma$, (b) frequency $\omega$ and weighted cross-phase angles between (c) $\tilde{\phi}$ and $\tilde{n_e}$ and (d) $\tilde{\phi}$ and $\tilde{T_{e,\perp}}$ of the dominant mode at $k_y\rho_s=0.2$ as a function of normalized electron collisionality $\nu_e^*$ for \#83160 at $\rho_{\text{tor}}=0.95$ in GENE simulations with numerical equilibrium.
   Results with the Sugama operator are compared with the pitch-angle scattering operator.
   Passing (squares, x) and trapped (circles, +) parts of the perturbed distribution function are  considered separately.
   For Sugama, the cross-phases for the total distribution function are also shown (diamonds).}
   \vspace{-1em}
   \label{fig:coll_op}
\end{figure}
\begin{figure*}
   \includegraphics[width=\textwidth]{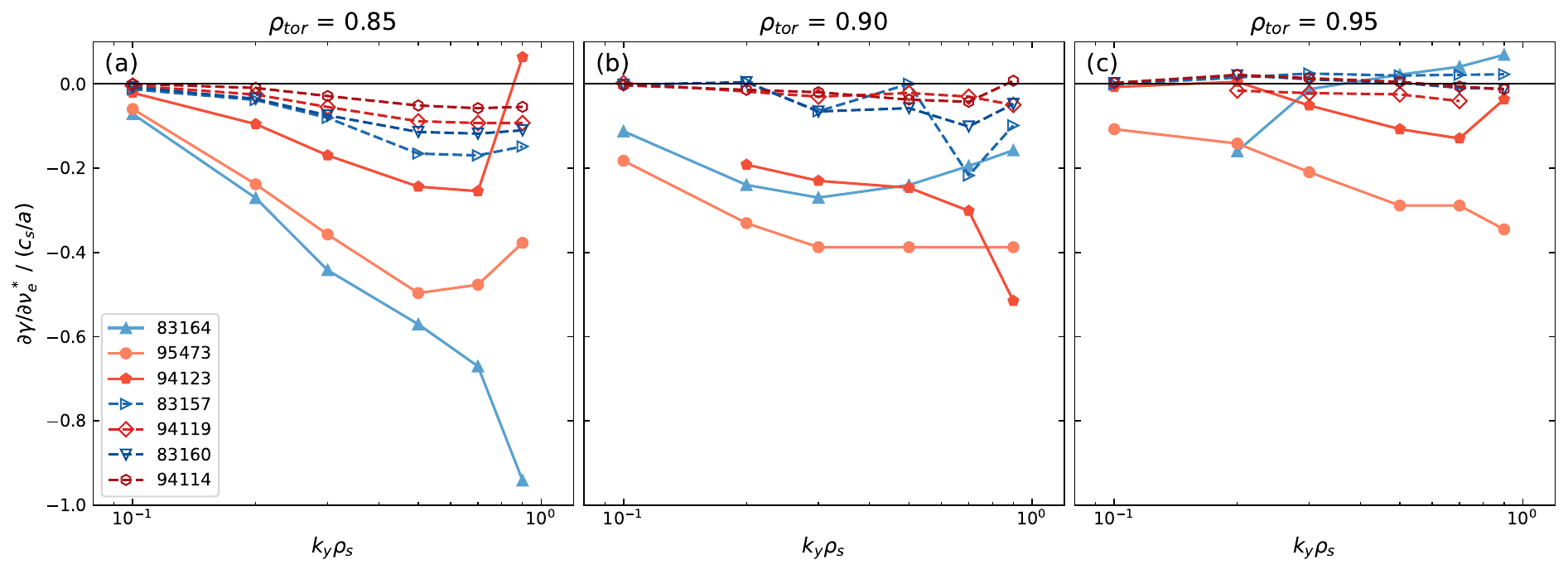}
   \caption{The derivative of the linear growth rate $\gamma$ with respect to the normalized electron collisionality $\nu_e^*$, from GENE simulations with numerical equilibria and Sugama collision operator for all seven discharges, as a function of $k_y$ at (a) $\rho_{\text{tor}}$ = 0.85, (b) $\rho_{\text{tor}}$ = 0.9 and (c) $\rho_{\text{tor}}$ = 0.95.}
   \label{fig:col_dgamma}
   \vspace*{-1em}
\end{figure*}

At $\rho_{\text{tor}} = 0.9$, the dominant TEMs observed in the low $\bar{n}_e$ branch discharges in Sec.~\ref{sec:gene:lsa:90} are once again stabilized by collisionality, as shown in Figure \ref{fig:col_scan}(b) and (e).
The $\ell$=1 ion-direction modes that are dominant for the high-density branch discharges exhibit a somewhat resistive character: their growth rates increase by approximately a decade for each order of magnitude increase in collisionality, in contrast to the conventional ITG modes.
Using the contributions of the different terms in the linear gyrokinetic equation to the electrostatic-potential part of the free energy we can estimate how much each term and each species contributes to the total growth rate of the mode \cite{Navarro2011b,Manas2015,Whelan2018}.
From this splitting of the growth rate, shown in Figure \ref{fig:fetime} for \#83160, it is clear that most of the growth of the $\ell$=1 ion modes comes from the curvature term for the deuterium ions in the linear gyrokinetic equation.
This corroborates the finding that trapped ions can contribute to the growth of this type of mode, as presented in Sec.~\ref{sec:gene:lsa:90}.
The parallel flow terms for both the electrons and ions stabilize these type of modes.
The resistive mode branches once again become dominant only outside of the experimental uncertainties for the collisionality.
Discharge \#94114, which had the lowest $a/L_{n_e}$ here, is near the mode boundary between the ITG and the $\ell$=1 ion-mode branches, as can be seen in Figure \ref{fig:col_scan}(e).

At $\rho_{\text{tor}} = 0.95$, the modes observed on both density branches in Sec.~\ref{sec:gene:lsa:95} show no discontinuities in the mode frequencies as a function of collisionality, see Figure \ref{fig:col_scan}(c) and (f).
This supports labeling them as hybrid modes.
For most discharges, the growth rates are close to the minima at the nominal $\nu_e^*$, with the dominant modes in the low and high $\bar{n}_e$ discharges on the collisionless and resistive branches, respectively.
An exception to this is \#94119, which had the highest $a/L_{n_e}$, where the minimum is found at higher collisionality than both the experimental collisionality and the minima of the other discharges.
However, the experimental $\nu_e^*$ values at this radial position have large uncertainties, making it possible that, experimentally, the collisionless modes are also dominant for (some of) the high $\bar{n}_e$ branch discharges.
Cross-phase angles between $\tilde{\phi}$ and the trapped part of the distribution function for the electrons indicate that the dominant modes for both density branches experience similar levels of drive from trapped electrons at the experimental values of $\nu_e^*$, see for example Figure \ref{fig:coll_op}(c) and (d) (open circles).
Thus, although this drive is less pronounced than for the dominant mode branches at the other radii, as previously demonstrated in Figure \ref{fig:cphases} in Sec.~\ref{sec:gene:lsa:95}, the modes are still (partially) driven by trapped electrons.
Going from the banana to the Pfirsch-Schl\"{u}ter regime, the high frequency of collisions increasingly interrupts the ability of the passing electrons to react to changes in $\phi$.
This causes their response to become increasingly non-adiabatic, as indicated by the increase in the cross-phase angle between $\tilde{\phi}$ and $\tilde{n}_{e,\text{p}}$.
This is consistent with previous findings in Ref.~\cite{Bonanomi2019}.
Thus, at the nominal collisionality levels of the low $\bar{n}_e$ branch discharges, the character of the hybrid modes is still more TEM-like, while for the high $\bar{n}_e$ branch discharges in the Pfirsch-Schl\"{u}ter regime, it becomes increasingly dominated by resistive drift-wave characteristics.
This is also supported by the cross-phases between $\tilde{\phi}$ and $\tilde{T}_{e,\perp}$, which show the closest correlation for the trapped electrons on the collisionless branches and then shift towards the passing electrons at high collisionality.
The cross-phases for the parallel temperature fluctuations $\tilde{T}_{e,||}$ were smaller in absolute magnitude and showed the strongest correlation for the passing electrons at $\pi$/4 across the entire $\nu_e^*$ range.
Given that the dominant cross-phase for $T_e$ was found to be between $\tilde{\phi}$ and $\tilde{T}_{e,\perp}$, the TEM aspect of these hybrid modes is of a $\nabla B$-driven type (as the $\nabla B$ drift scales with the magnetic moment $\mu$).
Near the plateau-Pfirsch-Schl\"{u}ter boundary, the mode characteristics are strongly influenced by collisional energy scattering and multi-species collisions, as can be seen in Figure \ref{fig:coll_op}(a).
When switching to an electron pitch-angle scattering (PAS) collision operator, which does not include these additional collisional processes, the resistive behavior of the dominant instabilities disappears.
This is correlated with a reduction in the trapped-particle cross-phases, as can be seen in Figure \ref{fig:coll_op}(c) and (d), which suggests a dissipative TEM (DTEM) drive \cite{Wang2015b} outside of the banana regime.
Further study of the exact collisional mechanism responsible for the resistive behavior of the dominant modes near the plateau-Pfirsch-Schl\"{u}ter boundary is left for future work.
For $\nu_e^* \geq 20$, the mode structures of $\phi$ start to become increasingly ballooned, see for example the third inset in Figure \ref{fig:fetime}.
The cross-phase angles between $\tilde{\phi}$ and both $\tilde{n}_e$ and $\tilde{T}_e$ shift towards $\pi$/2, for both passing and trapped electrons, as shown in Figure \ref{fig:coll_op}(c) and (d).
These mode characteristics are compatible with interchange/resistive ballooning modes (RBMs) \cite{Scott2005}.
However, contrary to what would be expected for RBMs, these modes are only weakly destabilized by the normalized plasma pressure $\beta$, as an order of magnitude increase in $\beta$ increased the growth rate by less than $10$\%. 
Thus, beyond the minimum in growth rate at $\rho_{\text{tor}} = 0.95$, first hybrid resistive (drift-wave) mode branches become dominant, which then transform into interchange/RBM as collisionality increases.
This is consistent with observations in Ref.~\cite{DeDominici2019}.

The aforementioned trends for $k_y\rho_s=0.2$ are applicable to the full ion-scale spectra at all three radii.
This can be seen in Figure \ref{fig:col_dgamma}, where the derivative of the linear growth rate $\gamma$ with respect to the normalized electron collision frequency $\nu_e^*$ is plotted as a function of $k_y$.
At low collisionality (\textit{i.e.}~at $\rho_{\text{tor}}=0.85$), all the dominant ion-scale instabilities are stabilized with increasing collisionality.
As a function of increasing density, the dominant mode branches become gradually less stabilized by collisionality, until at some point the whole spectrum becomes resistive.
The minimum in growth rate typically shifts to lower $\nu_e^*$ with increasing $k_y$.
\begin{figure}[!h]
   \centering
   \includegraphics[width=0.475\textwidth]{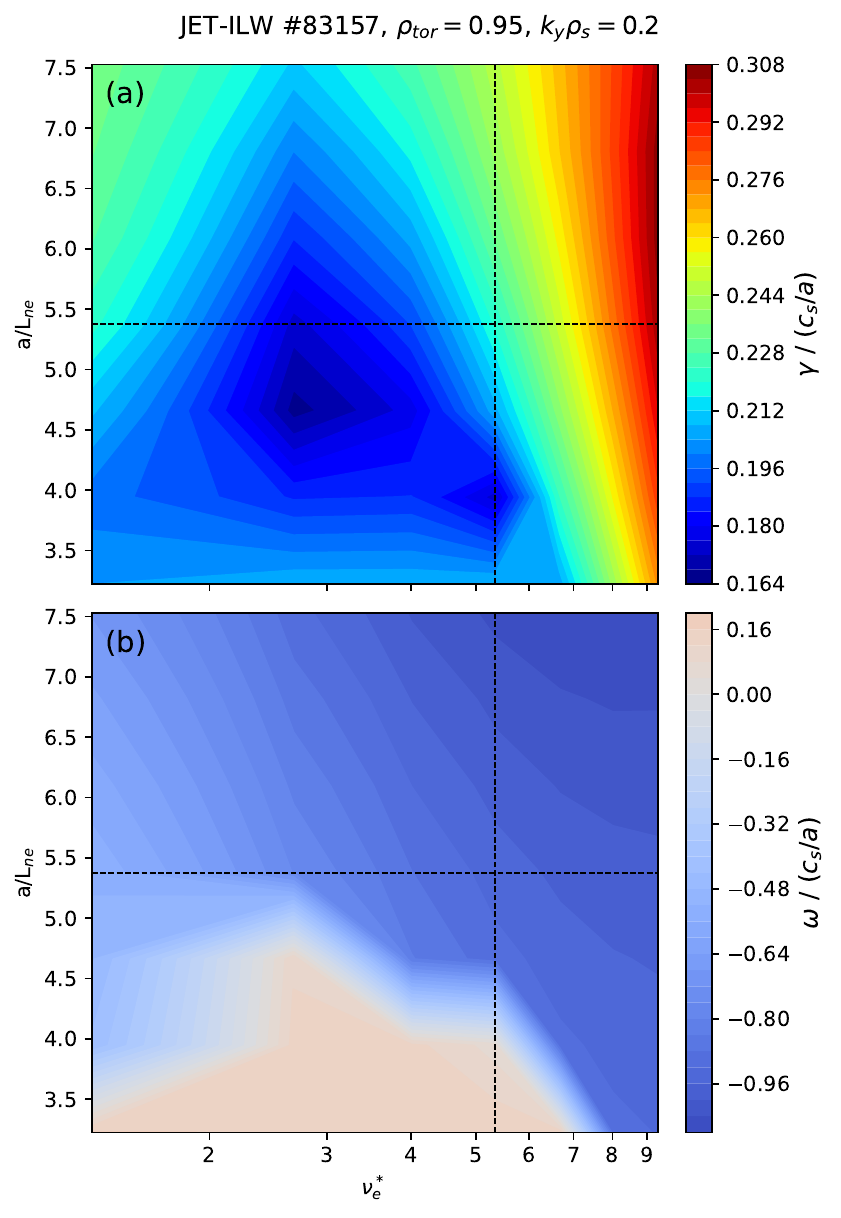}
   \caption{The linear (a) growth rate $\gamma$ and (b) frequency $\omega$ of the dominant mode at $k_y\rho_s=0.2$ as functions of both normalized electron collisionality $\nu_e^*$ and $a/L_{n_e}$ for \#83157 at $\rho_{\text{tor}}=0.95$.
   The nominal experimental values of both $\nu_e^*$ and $a/L_{n_e}$ are indicated as dashed lines.}
   \label{fig:omn-coll}
   \vspace*{-2em}
\end{figure}

For \#83157, the $P_{L-H}$ threshold is at its minimum as function of density, while Figure \ref{fig:col_scan}(c) showed it also had the lowest nominal growth rate.
Therefore, it is interesting to simultaneously investigate the sensitivity of the dominant modes to the electron collisionality $\nu_e^*$ and density gradient $a/L_{n_e}$ at $\rho_{\text{tor}}=0.95$ prior to H-mode entry.
The growth rates and mode frequencies of such a scan are shown in Figure \ref{fig:omn-coll}(a) and (b).
Ranges for both parameters were set by the propagated uncertainties on the respective quantities.
At experimental values of $a/L_{n_e}$ and $\nu_e^*$ (dashed cross), the dominant mode for this discharge is resistive.
Close to the nominal values, at $a/L_{n_e} = 4.75$ and $\nu_e^* = 2.9$, the global minimum of the growth rate is found.
This is akin to a triple point, around which all three branches seen in Figure \ref{fig:fetime} are present.
At $a/L_{n_e} < 4.75$, the $\ell$=1 ion mode gradually destabilizes for most values of $\nu_e$.
At $a/L_{n_e} > 4.75$ and $\nu_e^* < 2.9$, the growth rate slowly increases with decreasing collisionality, indicating the collisionless hybrid branch is being destabilized.
At $a/L_{n_e} > 4.75$ and $\nu_e^* > 2.9$, the growth rate increases with increasing collisionality, indicating the resistive drift-wave branch is being destabilized.

Overall, these results confirm previous work that suggested that the dominant instabilities in the L-mode pedestal-forming region, which must be stabilized to access H-mode, undergo a regime change as the density, and hence the collisionality, is increased \cite{Bourdelle2015,DeDominici2019}.
Especially at $\rho_{\text{tor}} = 0.95$, where the radial electric field is typically strongest in JET-ILW \cite{Silva2022}, the difference between the dominant-instability regimes on the low- and high-density branches is striking.
For the discharges closest to the minimum in $P_{L-H}$ vs density, the experimental collisionality also leads to growth rates of the dominant modes closest to their minima.

\begin{figure*}
   \centering
   \includegraphics[width=\textwidth]{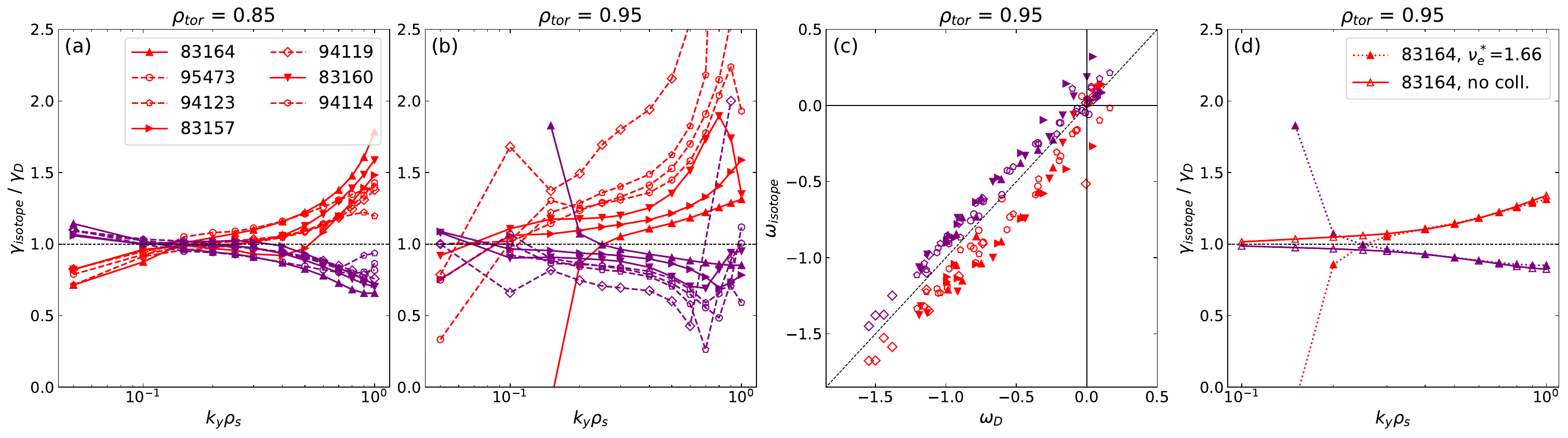}
   \caption{The effect of isotope mass on the linear spectra from GENE simulations with numerical equilibrium, where hydrogen (red) and tritium (purple) are compared to deuterium.
   The ratio of the linear growth rates for isotopes and deuterium $\gamma_{\text{isotope}}/\gamma_{D}$ as a function of $k_y$ for (a) $\rho_{\text{tor}}=0.85$ and (b) $\rho_{\text{tor}}=0.95$.
   (c) The frequency spectra of the isotopes $\omega_{\text{isotope}}$ for the mode branches shown in (b) against the frequency spectra for deuterium $\omega_D$.
   (d) $\gamma_{\text{isotope}}/\gamma_{D}$ as a function of $k_y$ for \#83164 at $\rho_{\text{tor}}=0.95$, both with and without collisions.}
   \label{fig:gamma_HDT}
\end{figure*}

\subsection{Effect of isotope mass}
\label{apdix:mass}
The experimental L-H power threshold $P_{L-H}$ has also been observed to depend on the mass of the main-ion species, with it being higher in hydrogen and lower in tritium discharges \cite{Maggi2018,Solano2022,Solano2023,Birkenmeier2023}.
At equal line-averaged density, the electron density and temperature profiles at the time of L-H transition were found to be very similar in JET-ILW (even equal within error bars) for plasmas with different hydrogen isotopes as main-ion species \cite{Maggi2018,Birkenmeier2023}. 
However, both the amount of fuelling required to attain the same line-averaged density and $P_{L-H}$ changed. 
Although the JET dataset considered here only includes deuterium discharges, in this context the sensitivity of the dominant linear instabilities observed in Sec.~\ref{sec:gene:lsa} to the main-ion isotope mass was investigated at all three radii.
For this purpose, only the ratio of the main-ion mass to the reference mass (deuterium) in the simulations was modified, while the electron-to-reference-mass ratio was kept fixed.

In Figure \ref{fig:gamma_HDT}(a) and (b), the ratio of the linear growth rates from simulations with the isotopes and with deuterium is plotted as a function of $k_y$ at $\rho_{\text{tor}}=0.85$ and $\rho_{\text{tor}}=0.95$. 
For the TEM/ITG mode branches dominant at $\rho_{\text{tor}}=0.85$ and 0.9 (not shown), the impact of the isotope mass on linear growth rate and frequencies is small, less than 10\% for $k_y\rho_s<0.5$, as shown by Figure \ref{fig:gamma_HDT}(a).
However, the hybrid modes dominant at $\rho_{\text{tor}}=0.95$ are quite sensitive to the isotope mass, with differences in linear growth rate of 10\%-50\%, as can be seen in Figure \ref{fig:gamma_HDT}(b).
This is consistent with Ref.~\cite{Bonanomi2019}, where it was demonstrated that such sensitivity originates from the contribution of passing electrons to the linear instabilities.
In the preceding section, we showed that of the three radii investigated, only for the nominal collisionality observed at $\rho_{\text{tor}}=0.95$ did the passing electron response become (to some extent) non-adiabatic.
At all three radii the mode frequencies are shifted by a small amount in either direction compared to those from simulations with deuterium, but no large changes in dominant mode branches are observed, as illustrated in Figure \ref{fig:gamma_HDT}(c).
Both the heat flux ratios and convective heat flux ratios change by less than 10\% compared to those from simulations with deuterium.
Therefore, the previous detailed characterization of the dominant mode characteristics is expected to hold for different hydrogen isotopes.

The discharge with the largest $a/L_{n_e}$ at $\rho_{\text{tor}}=0.95$, namely \#94119, also exhibits the largest change in growth rate in response to the change in isotope mass, see Figure \ref{fig:gamma_HDT}(b).
In fact, there appears to be a correlation between the difference in growth rate caused by the change in isotope mass and $a/L_{n_e}$.
When absolute differences in growth rate are compared, rather than ratios, this correlation remains.
This is in line with results from Ref.~\cite{Angioni2018}, which showed a dependence of the effect that isotope mass has on the linear growth rate on $a/L_{n_e}$, and this dependence is stronger at high collisionality.

For $k_y\rho_s < 0.15$, the trend of inverse scaling of the growth rate with the square root of the main ion isotope mass typically reverses.
When removing collisions from the simulations, this reversal of disappears, and the inverse scaling of the linear growth rate with the square root of the isotope mass remains intact for all $k_y$, as demonstrated in Figure \ref{fig:gamma_HDT}(d).
This indicates that at low $k_y$ the terms that scale with the thermal velocity $v_{th}$ in the linear gyrokinetic equation, which is proportional to the square root of the isotope mass in normalized units, are no longer dominant over collisions.

The impact of mixtures of two kinetic isotope species was also investigated in the simulations at $\rho_{\text{tor}}=0.95$.
Assuming equal ion-density gradients for both isotope species in the simulations, the linear growth rates were found to result in a linear combination of the isotope density fraction times the growth rate obtained in simulations with either isotope species in isolation.

The inverse scaling of the linear mode destabilization with the square root of the isotope mass is consistent with the experimentally observed trends in the L-H power threshold \cite{Solano2023} and how pedestal height \cite{Frassinetti2023} changes with isotope mass.
Experimentally, the line-averaged density at which the L-H power threshold is minimal $\bar{n}_{e,min(P_{L-H})}$ has been observed to be lower in JET-ILW for tritium than for deuterium \cite{Solano2023}.
A series of sensitivity scans in electron collisionality for various $k_y$ shows a slight shift in the minimum of the growth rate with respect to $\nu_e^*$ when comparing simulations with hydrogen or tritium to those with deuterium.
A detailed investigation of how this affects the nonlinear turbulent transport and the extent to which this plays a role in setting $P_{L-H}$ as a function of $\bar{n}_e$ for different isotopes is left for future studies.
It is also possible that differences in the plasma profiles and/or composition, \textit{e.g.~}due to mass-dependent sputtering of antenna and wall materials by the main-ion species, may be a contributing factor \cite{Solano2023,Callahan2023}. 

\subsection{Other sensitivities}
\label{sec:gene:other}
In addition to the collisionality and isotope mass, the sensitivity of the linear spectra to a number of other variables was tested.
A brief overview of these sensitivities is provided below, with further elaboration provided in the appendices.

\textbf{Electromagnetic effects:} Finite $\beta$ is weakly destabilizing to all the characterized (dominant) mode branches.
Setting $\beta=0$ in the simulations decreased growth rates typically by about 10\%, but did not result in any different mode branches becoming dominant.
Including parallel magnetic field fluctuations $\delta B_{||}$ was found to mostly have a negligible impact on growth rates and frequencies.
However, the heat flux ratios of the $\ell$=1 ion-direction mode are very sensitive to both $\beta$ and including $\delta B_{||}$.
At all three radii, the $\beta$-threshold for kinetic ballooning modes (KBM) was found to be at least an order of magnitude higher than the nominal experimental values.
More details are provided in Appendix \ref{apdix:beta}.

\textbf{Toroidal rotation:} Overall, the addition of toroidal rotation and parallel flow shear effects were found to have a minor impact on the described mode characteristics.
At $\rho_{\text{tor}}=0.95$, including only toroidal rotation was found to provide less than 10\% additional drive.
However, also adding the parallel velocity gradient to the simulations was found to (mostly) negate this increase.
No changes in the dominant mode branches were observed.
As mentioned previously, $E \times B$ shear was not included in these linear simulations.
For more details see Appendix \ref{apdix:rot}.

\textbf{Magnetic geometry:} The influence of the magnetic equilibria on the dominant low-$k_y$ modes at $\rho_{\text{tor}}=0.95$, as observed in Sec.~\ref{sec:gene:lsa:95}, prompted further sensitivity checks on the impact of magnetic geometry.
Using Miller geometry parameterizations of the local magnetic equilibria from MEGPy as a starting point, scans in elongation, triangularity, squareness and their related radial derivative quantities were made for \#83164 and \#95473.
In particular the elongation shear $s_{\kappa}$ and triangularity shear $s_{\delta}$ were found to be strong actuators of the linear growth rates and frequencies for the dominant linear instabilities at low $k_y$.
This indicates that the magnetic geometry can indeed influence the L-H power threshold directly, but further detailed study of these sensitivities is required.

\begin{figure}[!t]
   \centering
   \includegraphics[width=0.475\textwidth]{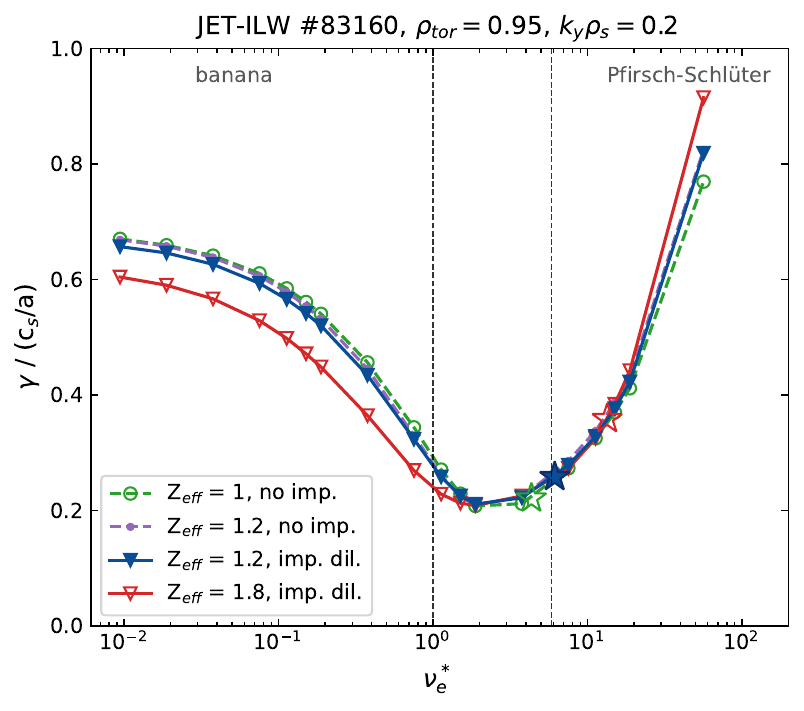}
   \caption{The effect of the effective charge Z$_{\text{eff}}$ on the linear growth rate $\gamma$ of the dominant mode at $k_y\rho_s=0.2$ as a function of normalized electron collisionality $\nu_e^*$ for \#83160 at $\rho_{\text{tor}}=0.95$ in GENE simulations with numerical equilibrium.
   Results at the nominal Z$_{\text{eff}}$ and with kinetic impurities (blue triangles) are compared against simulations with the nominal Z$_{\text{eff}}$ and with kinetic impurities (purple dots), Z$_{\text{eff}}=1$ (red circles) and an arbitrary increase in impurity content (density and density gradient increased by factors 4 and 2, respectively) with kinetic impurities (red triangles).}
   \vspace{-1em}
   \label{fig:coll_zeff}
\end{figure}
\textbf{Impurity composition:} The impact of including detailed impurity composition in the linear simulations was investigated.
Including impurities in these simulations has two effects: (1) the effective ion charge $Z_{\text{eff}}$ is increased, and (2) the main ion density (gradient), which is computed from quasi-neutrality, is diluted.
These two effects were investigated independently by performing scans of the nominal electron collisionality $\nu_e^*$ both with and without kinetic impurity species for different $k_y$, at different levels of impurity density.
Any additional effects that impurities can have on the plasma, such as radiative cooling, were not considered here.
In GENE, $Z_{\text{eff}}$ can be specified by including (kinetic) impurity species or, alternatively but not concurrently, by setting its value for use in calculations of collisions.
The two impurity species considered here are beryllium and, when included, a mass-weighted composite of the nickel and tungsten species.

Without any kinetic impurity species in the simulations, increasing $Z_{\text{eff}}$ independently of $\nu_e^*$ results in a (weak) stabilization of the collisionless ion-scale mode branches, typically $\leq$5\%, while it destabilizes the resistive mode branches by a similar amount.
The mode frequencies are affected more strongly than growth rates and show a shift towards values found at lower $\nu_e^*$ with nominal $Z_{\text{eff}}$.
Including the effect of dilution was found to have a stabilizing effect on the linear growth rates at all $\nu_e^*$ for the ion-scale instabilities for all three radii, regardless of mode propagation direction.
Combining these effects results in stabilization of the collisionless branches, while in the plateau and Pfirsch-Schl\"{u}ter regimes, the effects of $Z_{\text{eff}}$ and dilution were found to cancel.
This can result in a minor shift of the minimum in growth rate to lower $\nu_e^*$, by a factor of 2-3 at most for $Z_{\text{eff}} \leq 2$.
This is demonstrated in Figure \ref{fig:coll_zeff}, where scans of the dominant linear growth rate as function of normalized electron collisionality $\nu_e^*$ are plotted for $k_y\rho_s=0.2$ at $\rho_{\text{tor}}=0.95$ in \#83160.
Overall, it seems that excluding kinetic impurity species in gyrokinetic simulations at the low $Z_{\text{eff}}$ typical for metal-wall tokamaks is justified, as long as the nominal $Z_{\text{eff}}$ is included in the computation of $\nu_e^*$.

These findings are in agreement with Ref.~\cite{Callahan2023}, where it was shown that the power threshold on the low density branch was particularly sensitive to the main ion dilution caused by increased $Z_{\text{eff}}$, while this was not the case for the high density branch.
Furthermore, these findings also support the qualitative argument by Ref.~\cite{Bourdelle2014} that the low density branch of $P_{L-H}$ was not observed in JET with a carbon wall \cite{Maggi2014} due to the high $Z_{\text{eff}}$ resulting from the carbon impurities.

The stabilization due to impurities of the collisionless, electron-direction mode branches dominant in the banana regime is in agreement with the reduction in drive due to dilution typically seen for ITG modes.
This further strengthens the notion that these collisionless branches have a hybrid character, in this case mixing TEM and ITG.

\section{QUASILINEAR MODEL VERIFICATION}\label{sec:ver}
The dataset detailed in the previous section was used as a benchmark for the verification of the linear responses of QuaLiKiz and TGLF-SAT2 in the L-mode pedestal-forming region.
A full comparison of the nonlinear turbulent fluxes and the impact of $E \times B$ shear is considered outside the scope of this work and is left for future study.

First, the verification procedure is outlined.
Second, the setup of the quasilinear codes and the related model reduction benchmarks is described.
Third, the results of the comparison between QuaLiKiz and linear GENE simulations are presented.
Finally, the results of the comparison between TGLF-SAT2 and linear GENE are detailed.

\subsection{Verification procedure}
The following procedure was followed for the comparison between the quasilinear codes and the linear gyrokinetic simulations:
\begin{enumerate}
    \item A single routine, based on a workflow combining the EX2GK, MEGPy and GyroKit Python packages, was used to extract the local input quantities from the GPR profiles and magnetic equilibrium files for all linear GENE simulations.
    \item The effects of the model reductions employed by the quasilinear codes on the dominant linear instability characteristics in the L-mode pedestal-forming region were benchmarked by successively removing differences, as far as possible, in the setups of the linear GENE simulations and comparing the results to the dataset generated for Sec.~\ref{sec:gene}.
    \item The input files from the GENE simulations were used, as much as possible, to generate the corresponding input files for the respective quasilinear codes, eliminating potential errors introduced by separate data extractions.
    \item Switching between the frames of reference, both in units and magnetic equilibrium descriptions, used in the different codes also affected the local normalized binormal wavenumber, which is an input for all codes. Therefore, all simulations were performed at normalized wavenumbers for which the toroidal mode numbers corresponded to $k_y\rho_s \in [0.05,1]$ in the linear GENE simulations with numerical equilibria. For convenience, in the rest of this section all wavenumber dependent results are labelled $(k_y\rho_s)_{\text{num.~eq.}}$.
    \item A comparison was made between GENE and the quasilinear codes on the basis of the linear mode spectra (growth rate, frequency, heat flux ratio and convective heat flux ratio), the mode structures and, when relevant, input sensitivities.
\end{enumerate}

\subsection{Model setup}
\textbf{QuaLiKiz:} Version 2.8.2 of QuaLiKiz was used.
QuaLiKiz solves a reduced version of the electrostatic linear gyrokinetic dispersion relation, where the axisymmetric gyrokinetic problem is further simplified through the use of the lowest order ballooning transform.
A summary of all of its model reductions can be found in Table 2 of Ref.~\cite{Bourdelle2016}, while Ref.~\cite{Stephens2021a} provides a full derivation.
Version 2.8.2 introduced an updated Krook collision operator, that was tuned for improved agreement with TEM branches \cite{Stephens2021b}.

In comparison to the linear GENE simulations presented in Sec. \ref{sec:gene}, there are four main differences: QuaLiKiz (1) is electrostatic, (2) uses an $s$-$\alpha$ magnetic equilibrium model (circular geometry with a Shafranov shift), (3) uses a Krook collision operator and (4) does not include resistive physics in the dispersion relation.
The impact of the first two was benchmarked by repeating the wavenumber scans and several mode sensitivity studies in electrostatic linear GENE simulations with $s$-$\alpha$ equilibria.
There was no equivalent Krook collision operator available in GENE, thus it was not possible to separately asses its impact. 
Therefore, the Sugama operator was used in the GENE simulations. 
For the comparison with the reduced-fidelity linear GENE simulations, the QuaLiKiz simulations were run with default settings for version 2.8.2.

\textbf{TGLF:} The SAT2 version of TGLF was used.
TGLF solves the linear eigenvalue problem of a system of gyrofluid equations using a set of Hermite polynomial basis functions \cite{Staebler2007,Staebler2010}.
Since its inception, the TGLF saturation rule has undergone a series of evolutions, with the most recent iterations designated SAT2 \cite{Staebler2021a,Staebler2021b} and SAT3 \cite{Dudding2022}.
SAT2 represents the third saturation model of TGLF, which incorporates the poloidal variation of turbulence intensity. 
The linear solver was also modified between SAT1 and SAT2, in order to accommodate the updated spectral shift model in the radial direction \cite{Staebler2021b}, which was a requisite for SAT2.
Furthermore, for SAT2 the pitch-angle scattering collision model was also recalibrated to fit simulations with the gyrokinetic CGYRO code \cite{Staebler2021b}.
This model is only calibrated up to the banana-plateau boundary.

In comparison to the linear GENE simulations presented in in Sec. \ref{sec:gene}, there are three main differences: TGLF (1) solves a system of gyrofluid equations instead of gyrokinetic equations, (2) uses a (Turnbull-)Miller parameterization of the local magnetic equilibrium \cite{Miller1998,Turnbull1999} and (3) uses a pitch-angle scattering (PAS) collision operator.
The impact of the latter two was benchmarked by repeating the wavenumber scans and some mode sensitivity studies in linear GENE simulations with Miller equilibria, both with and without PAS collision operator (at $\rho_{\text{tor}}=0.95$).
For the comparison with the GENE simulations, the TGLF simulations were run with default settings for SAT2, with a custom $k_y$ grid (\texttt{kygrid\_model} = 0) of 40 gridpoints between $k_y\rho_s=$ 0.05 and 2 to ensure all the wavenumbers covered in the GENE scans could be compared directly.

\begin{figure}[!b]
    \centering
    \includegraphics[width=0.45\textwidth]{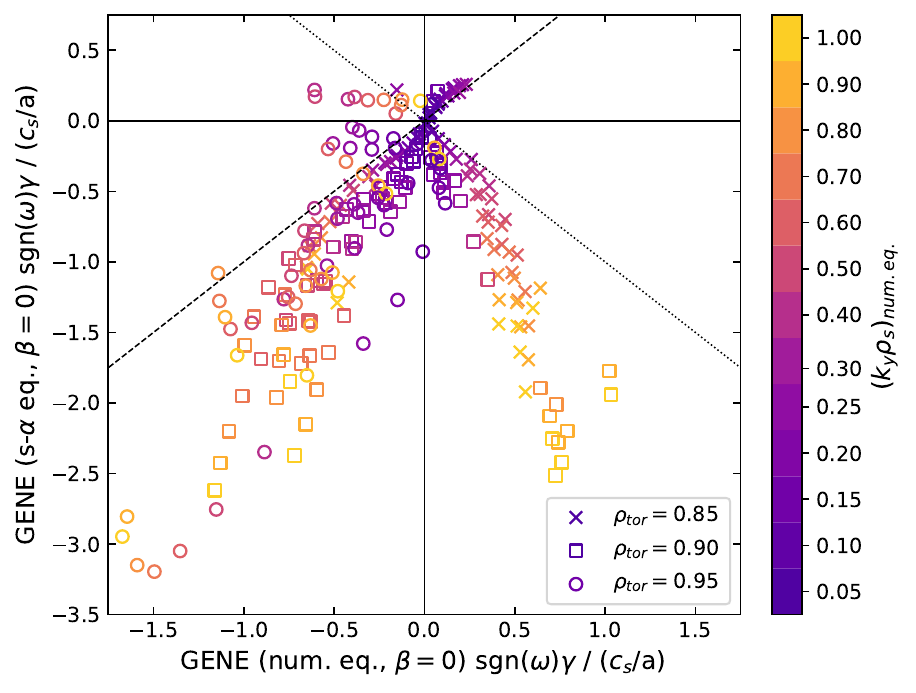}
    \caption{Comparison between the growth rates $\gamma$ times the sign of the corresponding mode frequencies $\omega$ from linear, electrostatic GENE simulations using $s$-$\alpha$ equilibria and numerical equilibria for all seven JET discharges at three radial positions (symbols) and ($k_y\rho_s)_{\text{num.~eq.}} \in [0.05,1]$ (colors). 
    Positive/negative values indicate modes propagating in the ion/electron diamagnetic drift direction.}
    \label{fig:efitvsalpha_linear}
\end{figure}
\begin{figure*}[!t]
    \centering
    \includegraphics[width=0.9\textwidth]{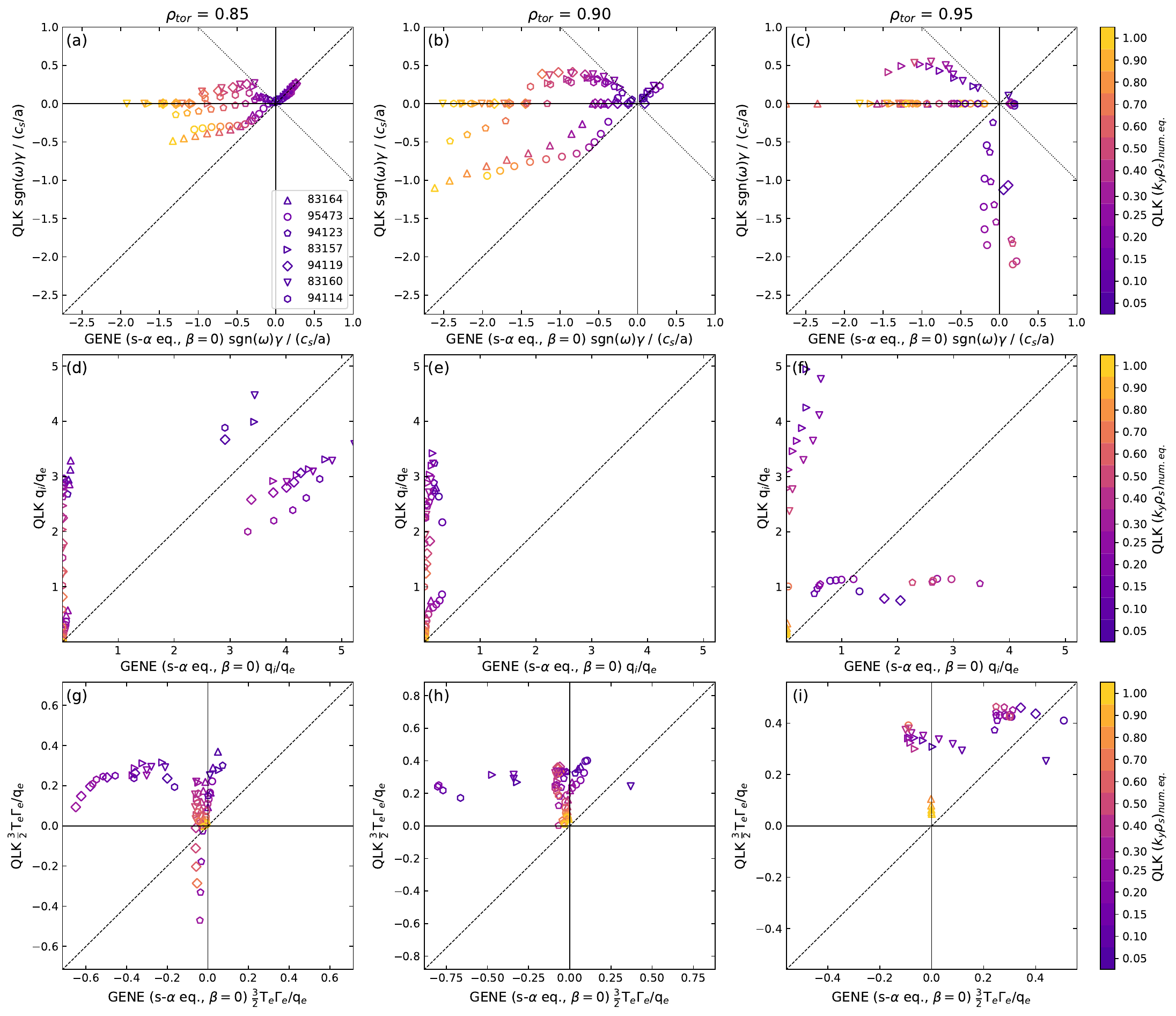}
    \caption{Comparison of the linear (a, b, c) growth rates $\gamma$ times sign of the mode frequencies $\omega$, (d, e, f) heat flux ratio q$_i$/q$_e$ and (g, h, i) convective heat flux ratio $\tfrac{3}{2}$T$_e \Gamma_e$/q$_e$ from QuaLiKiz (QLK) and linear, electrostatic GENE simulations using $s$-$\alpha$ equilibria and a Sugama collision operator for all seven JET discharges (symbols) at three radial positions (from left to right) and $(k_y\rho_s)_{\text{num.~eq.}} \in [0.05,1]$ (colors). 
    Positive/negative values indicate modes propagating in the ion/electron diamagnetic drift direction.}
    \label{fig:salphavsqlk_linear}
\end{figure*}
\begin{figure*}[!t]
    \centering
    \includegraphics[width=0.9\textwidth]{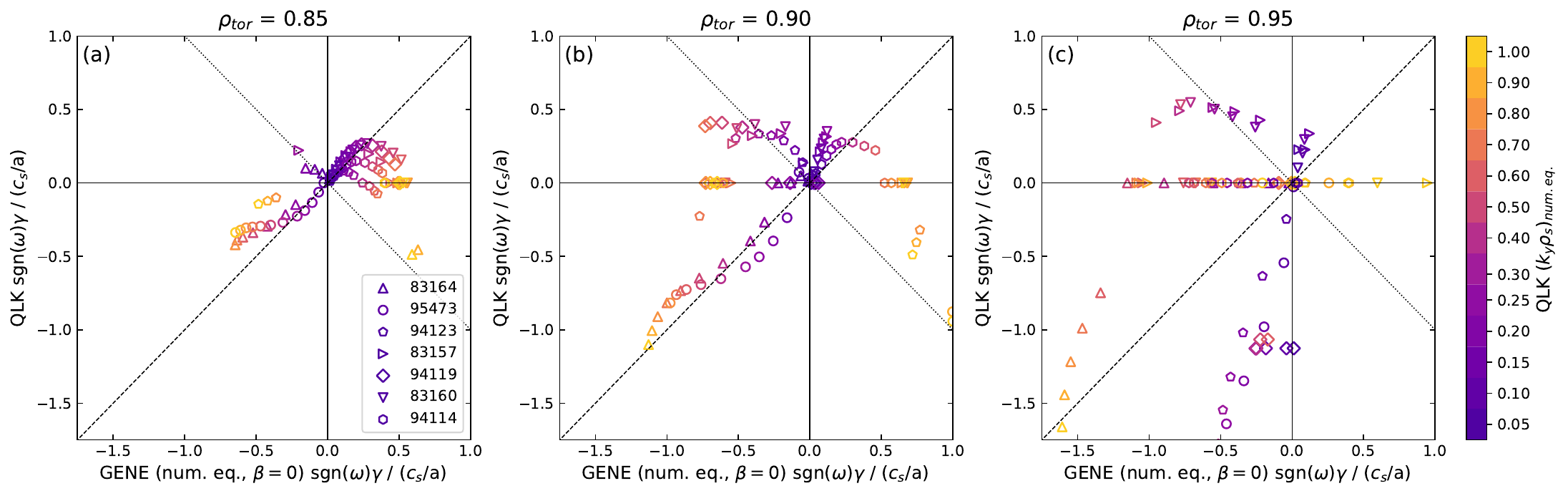}
    \caption{Comparison between the growth rates $\gamma$ times the sign of the corresponding mode frequencies $\omega$ from QuaLiKiz (QLK) and linear, electrostatic GENE simulations using numerical equilibria and a Sugama collision operator for all seven JET discharges (symbols) at three radial positions (from left to right) and $(k_y\rho_s)_{\text{num.~eq.}} \in [0.05,1]$ (colors).}
    \label{fig:numvsqlk_linear}
\end{figure*}

\subsection{Comparing QuaLiKiz against GENE}
\label{sec:qlk}
\subsubsection{Model reduction benchmark in GENE}
\label{sec:qlk:bench}
Turning off electromagnetic effects ($\beta=0$) in the linear GENE simulations did not have a large impact on the linear spectra and mode characteristics, as discussed in Sec.~\ref{sec:gene:other} and Appendix \ref{apdix:beta}.

Conversely, changing from numerical to $s$-$\alpha$ magnetic equilibria in the GENE simulations had a considerable impact on the linear spectra.
A comparison of the linear growth rates times the sign of the mode frequencies in electrostatic GENE simulations with both equilibrium descriptions is shown for all three radii and $(k_y\rho_s)_{\text{num.~eq.}} \in [0.05,1]$ in Figure \ref{fig:efitvsalpha_linear}.
The 1:1 diagonal (dashed) represents exact agreement in the linear growth rate and mode propagation direction, whereas the opposing diagonal (dotted) indicates modes with the correct growth rate but the (incorrect) opposite propagation direction.
At $\rho_{\text{tor}}=0.85$, marked by crosses in Figure \ref{fig:efitvsalpha_linear}, the dominance of TEM and ITG modes for the low- and high-density branch discharges, respectively, is qualitatively in agreement.
The linear growth rates agree mostly for $(k_y\rho_s)_{\text{num.~eq.}} \leq 0.4$, but the mode frequencies are larger by a factor of 3 in the simulations with $s$-$\alpha$ equilibria.
Heat flux ratios are larger by a factor 2, while convective heat flux ratios are typically smaller by a factor 2 in the $s$-$\alpha$ simulations.
However, for the high-density branch discharges, TEM-ETG modes become dominant over ITG at much lower wavenumbers in the simulations with $s$-$\alpha$ versus numerical equilibria ($(k_y\rho_s)_{\text{num.~eq.}} \geq 0.4$ vs $k_y\rho_s \geq 1$), respectively.
This is visible in the lower right quadrant in Figure \ref{fig:efitvsalpha_linear}.
At $\rho_{\text{tor}}=0.9$, marked by squares in Figure \ref{fig:efitvsalpha_linear}, even the qualitative agreement in dominant-mode characteristics starts to break down.
None of the dominant TEMs change propagation direction in the $s$-$\alpha$ simulations, indicating a lack of ubiquitous TEMs.
Instead, the dominant TEMs change to ETG modes at high wavenumbers.
There are also no odd-parity ion modes dominant at low wavenumbers.
These are mostly replaced by TEMs and, for two discharges, by standard ballooning-parity ITG.
For the correct mode branches linear growth rates are 30\%-50\% larger, while mode frequencies are again larger by a factor 3.
At $\rho_{\text{tor}}=0.95$, marked by circles in Figure \ref{fig:efitvsalpha_linear}, there are large discrepancies in all characteristics compared.
Although generally, TEM branches are dominant, and for a few discharges a shift in mode frequencies towards the ion direction occurs, the growth rates, mode frequencies and mode structures differ across all spectra.
This is not surprising, since at this radial location, replacing the strongly shaped flux-surfaces with circular geometry becomes a poor approximation.

Scans of the electron collisonality were repeated in GENE with $s$-$\alpha$ equilibria for the ion-scale instabilities.
These simulations showed that at experimental collisionalities, the linear spectra with $s$-$\alpha$ equilibria also include more dominant resistive mode branches than with shaped geometry, in particular at $\rho_{\text{tor}}=0.85$ and $\rho_{\text{tor}}=0.9$.

\subsubsection{Verification of the QuaLiKiz spectra}
\label{sec:qlk:veri}
The comparison between the linear growth rates times the sign of the mode frequencies, heat flux ratios and convective heat flux ratios from QuaLiKiz and the electrostatic GENE simulations with $s$-$\alpha$ equilibria is shown in Figure \ref{fig:salphavsqlk_linear} for all three radii and as a function of $k_y$.

For $k_y\rho_s \leq 0.4$ at $\rho_{\text{tor}} = 0.85$ and 0.9, the growth rates and frequencies for the TEM for the low-density branch discharges (\#83164, \#95473 and \#94123) and ITG for the high-density branch discharges respectively are in good agreement with the GENE $s$-$\alpha$ simulations, as shown in Figure \ref{fig:salphavsqlk_linear}(a) and (b).
The heat flux and convective heat flux ratios, however, mostly show large disagreement between the QuaLiKiz and GENE $s$-$\alpha$ simulations, see Figure \ref{fig:salphavsqlk_linear}(d), (e), (g) and (h).
Most of this comes from different mode branches being dominant in QuaLiKiz.
This is also reflected in the particle flux direction being opposite in a large fraction of the QuaLiKiz spectra.
For the mode branches that agree between QuaLiKiz and GENE, the heat flux ratios are $\sim$5$\times$ larger for TEM and $\sim$30\% smaller for ITG from QuaLiKiz than from the linear GENE simulations.

At higher wavenumbers in the low-density branch discharges, the TEM growth rates are too low in QuaLiKiz.
Furthermore, for most of the high-density branch discharges, QuaLiKiz predicts ITG dominance over the TEM-ETG modes observed to be dominant in the GENE $s$-$\alpha$ simulations.
Both of these differences are related to overdamping of TEM modes by the collision operator \cite{Stephens2021b}.
The recent re-tuning of the Krook collision operator in QuaLiKiz was performed for mid-radius plasma conditions, without consideration of the high-collisionality regime of the pedestal-forming region.

However, as shown in the previous section, the occurrence of TEM-ETG dominance for $k_y\rho_s \geq 0.4$ in the GENE $s$-$\alpha$ simulations was a departure from the results of the numerical equilibrium simulations.
In fact, comparing the linear growth rate spectra from QuaLiKiz against those from electrostatic GENE simulations with numerical equilibria at $\rho_{\text{tor}} = 0.85$, as shown in Figure \ref{fig:numvsqlk_linear}(a), the agreement in growth rates is much better for the low wavenumbers.
This may be indicative of the fact that the $s$-$\alpha$ geometry implementation in QuaLiKiz is not equivalent to the one in GENE, and has fewer approximations than the latter.
Nevertheless, the disagreement in mode frequencies is still 200\%.
At $\rho_{\text{tor}} = 0.9$ and 0.95, the comparison against GENE simulations with numerical equilibria does not show as much improvement.

\begin{figure}[!h]
    \centering
    \includegraphics[width=0.45\textwidth]{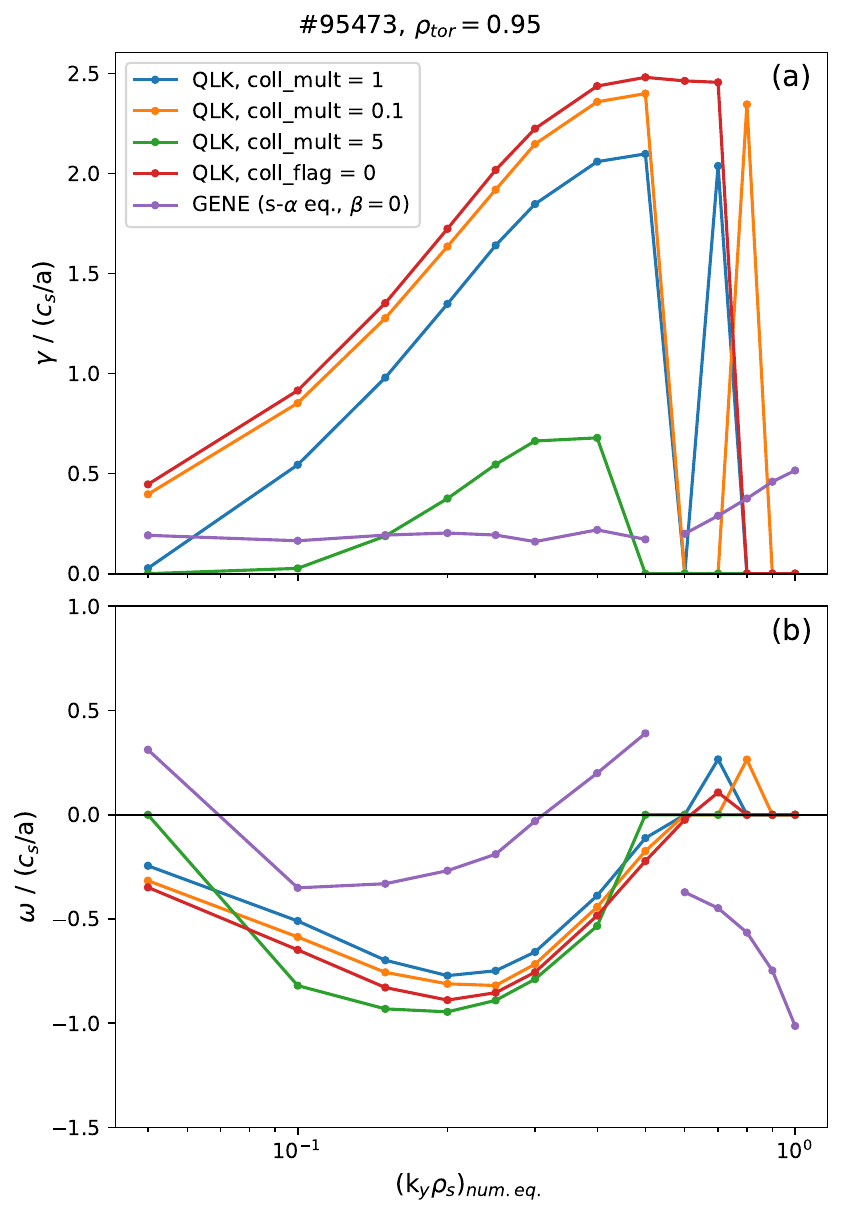}
    \caption{QuaLiKiz linear (a) growth rates $\gamma$ and (b) mode frequencies $\omega$ for \#95473 at $\rho_{\text{tor}}=0.95$. 
    Various settings for the Krook collision operator in QuaLiKiz (QLK) are compared against the linear spectra from GENE ($s$-$\alpha$ eq., $\beta=0$).}
    \label{fig:qlk_spectral_fix}
\end{figure}

As the nominal electron collisionalities shift into the plateau and Pfirsch-Schl\"{u}ter regimes at $\rho_{\text{tor}} = 0.95$, the scatter of the growth rates from QuaLiKiz becomes large, precluding any meaningful agreement with the GENE $s$-$\alpha$ simulations.
For a few of the low-triangularity, low-density branch discharges, TEM-like mode branches similar to those in the GENE simulations do become dominant in QuaLiKiz.
Their mode frequencies show similar trends, but their growth rates are about 10 times as large as the corresponding modes in the GENE $s$-$\alpha$ simulations, see for example Figure \ref{fig:qlk_spectral_fix}.
Gradient sensitivity scans in QuaLiKiz showed that increasing $a/L_{n_e}$ did not result in similar mode stabilization as in the GENE simulations, regardless of equilibrium type.
Comparing directly against the results from the electrostatic GENE simulations with numerical equilibria improves the mode frequency agreement, but the growth rates are still 3-5 times larger.

Moving from $\rho_{\text{tor}}=0.85$ to 0.95, there is an increasing range of wavenumbers that are linearly stable for QuaLiKiz.
Partly, this is a consequence of ETGs being artificially suppressed in QuaLiKiz from $k_y\rho_s \leq 2$.
Since the dominance of TEM-ETG modes for $(k_y\rho_s)_{\text{num.~eq.}} > 0.5$ in the GENE $s$-$\alpha$ simulations was a result of the circular magnetic geometry, this outcome may be beneficial for integrated modelling.
However, QuaLiKiz does not predict the UTEM instead, not even subdominantly.
At $\rho_{\text{tor}}=0.9$ and 0.95, there is also an increasing number of stable low wavenumbers.
These are in part caused by issues with the collision operator, as will be described in Sec. \ref{sec:qlk:coll}.

The Gaussian-shape, fluid-limit approximation of the mode structures that QuaLiKiz uses generally conforms to most of the ballooning mode structures of $\phi$ observed at $\rho_{\text{tor}}=0.85$ and even 0.9 in the GENE $s$-$\alpha$ simulations.
However, at $\rho_{\text{tor}}=0.9$ on the high-density branch and for all discharges at 0.95, unconventional mode structures appear in the GENE $s$-$\alpha$ simulations, similar to those observed in the GENE simulations with numerical equilibria, that do not agree with this approximation.
The mode structures with TEM characteristics were also more extended along the field lines in $s$-$\alpha$ than in numerical equilibria.
Subdominant high-order excitation mode branches ($\ell>0$) were also observed in the $s$-$\alpha$ simulations, and Ref.~\cite{Pueschel2019} has pointed out that the presence of such modes can add to the nonlinear fluxes.
However, QuaLiKiz almost never found subdominant modes, even with increased relative accuracy thresholds for the numerical solver.

\subsubsection{Effect of collisionality in QuaLiKiz}
\label{sec:qlk:coll}
In order to reproduce the electron collisionality sensitivity scans conducted in GENE, analogous scans with QuaLiKiz were performed. 
Given that QuaLiKiz does not take the collisionality as an input, scans of the collisionality multiplier were performed.
These scans did not reveal any evidence of dissipative instabilities when the collisionality was increased. 
Instead, at high collisionality, the linear spectra exhibited complete stabilization, as illustrated in Figure \ref{fig:qlk_spectral_fix}(a).
This was to be expected, as the current formulation of the linear dispersion relation in QuaLiKiz lacks resistive physics.

While scanning the collisionality in QuaLiKiz, it was observed that this resulted in an increase in the number of wavenumbers that returned stable solutions. This phenomenon is demonstrated in Figure \ref{fig:qlk_spectral_fix}. 
The cause of this was traced to the linear solver encountering a numerical issue in the calculation of the collision operator when $|\omega| \approx 0$.
When the collision operator was disabled (\texttt{coll\_flag} = 0), the aforementioned issue was resolved, and the solver returned a solution.
Regardless, these results show that QuaLiKiz cannot be expected to produce accurate predictions at $\rho_{\text{tor}}=0.95$, where the collisionality is substantial.

\subsection{Comparing TGLF-SAT2 against GENE}
\label{sec:tglf}
Following the assessment of the performance of QuaLiKiz, the focus is now turned to TGLF.
\subsubsection{Model reduction benchmark in GENE}
\label{sec:tglf:bench}
Changing from numerical equilibria to Turnbull-Miller parameterizations of the local equilibria in the GENE simulations generally had a negligible impact on the linear spectra and mode characteristics as described in Sec.~\ref{sec:gene}.
The plasma shapes of the discharges considered here have relatively small up-down asymmetry near the separatrix, and the optimized parameterizations from MEGPy can approximate the local numerical equilibria well.
For a few discharges, the exact wavenumbers at which the change of mode propagation direction to either UTEM or ETG occurred shifted slightly.
Furthermore, only at $\rho_{\text{tor}}=0.95$ do the linear growth rates exhibit a reduction -- of less than 20\% -- in growth rate for $k_y\rho_s \geq 0.5$, which is comparable to the results obtained in Ref.~\cite{Snoep2023}. 

The impact of utilizing a pitch-angle scattering collision operator instead of the Sugama operator has already been discussed in Sec.~\ref{sec:gene:coll}.
Deep in the banana regime, this change had a negligible impact.
Near the banana-plateau regime boundary, it resulted in deviations in the growth rates of up to 20\%.
In the plateau regime and near the plateau-Pfirsch-Schl\"{u}ter regime boundary, it resulted in changes in the growth rate of $\sim$35\% and affected the dominant mode characteristics.

Scans of the electron collisionality were also repeated for the ion-scale instabilities in the GENE simulations with Turnbull-Miller parameterizations and Sugama collision operator for a few cases.
The trends of the derivative of the growth rates with respect to the collisionality agreed qualitatively with the simulations with numerical equilibria.
 
\begin{figure*}
    \centering
    \includegraphics[width=\textwidth]{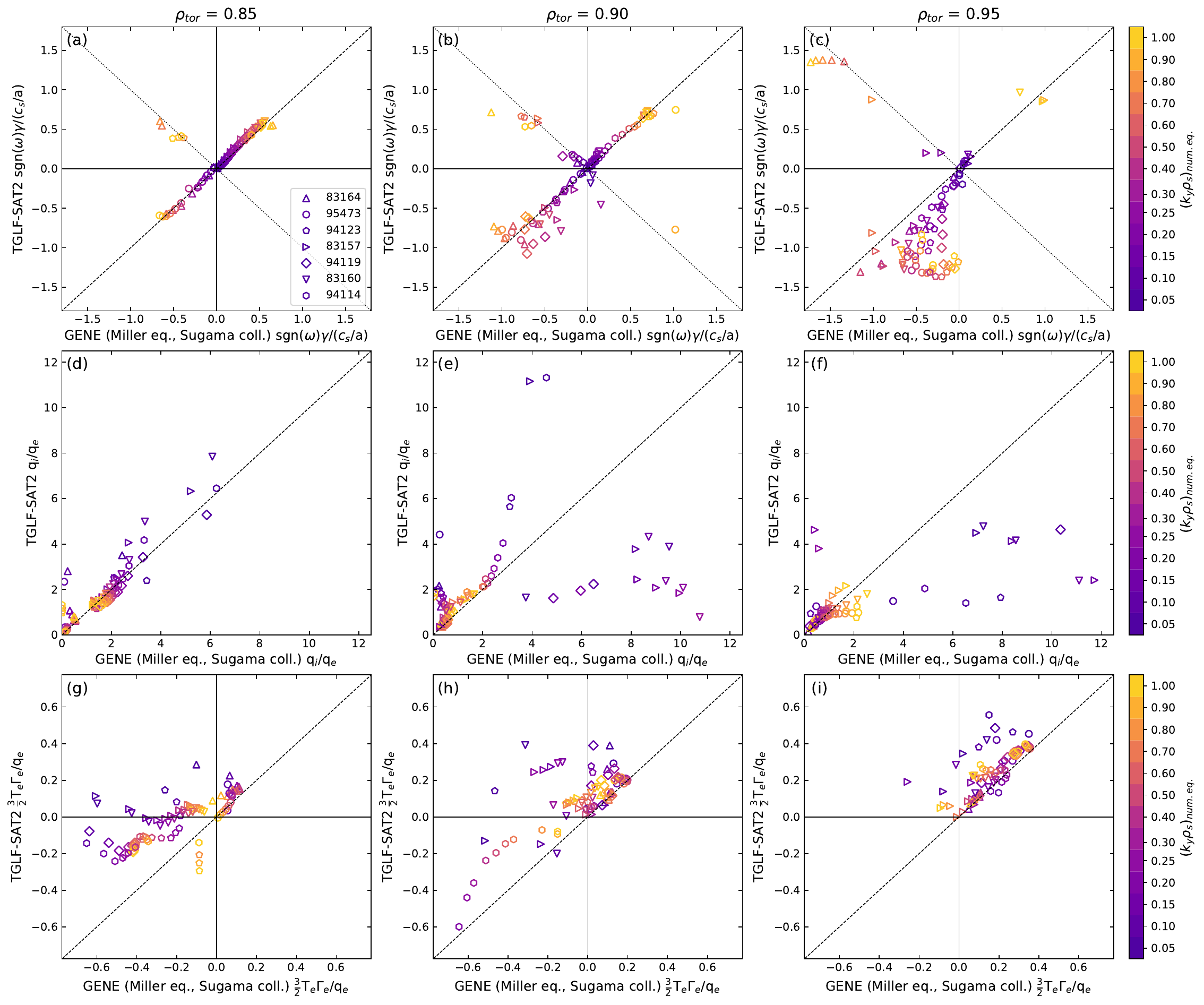}
    \caption{Comparison of the linear (a, b, c) growth rates $\gamma$ times sign of the mode frequencies $\omega$, (d, e, f) heat flux ratio q$_i$/q$_e$ and (g, h, i) convective heat flux ratio $\tfrac{3}{2}$T$_e \Gamma_e$/q$_e$ from TGLF-SAT2 against local linear, electromagnetic GENE simulations using Miller geometry for all seven JET-ILW discharges at three radial positions as a function of $k_y$. Positive/negative values indicate modes propagating in the ion/electron diamagnetic drift direction.}
    \label{fig:genevstglf-sat2}
\end{figure*}
\begin{figure*}[!t]
    \centering
    \includegraphics[width=0.9\textwidth]{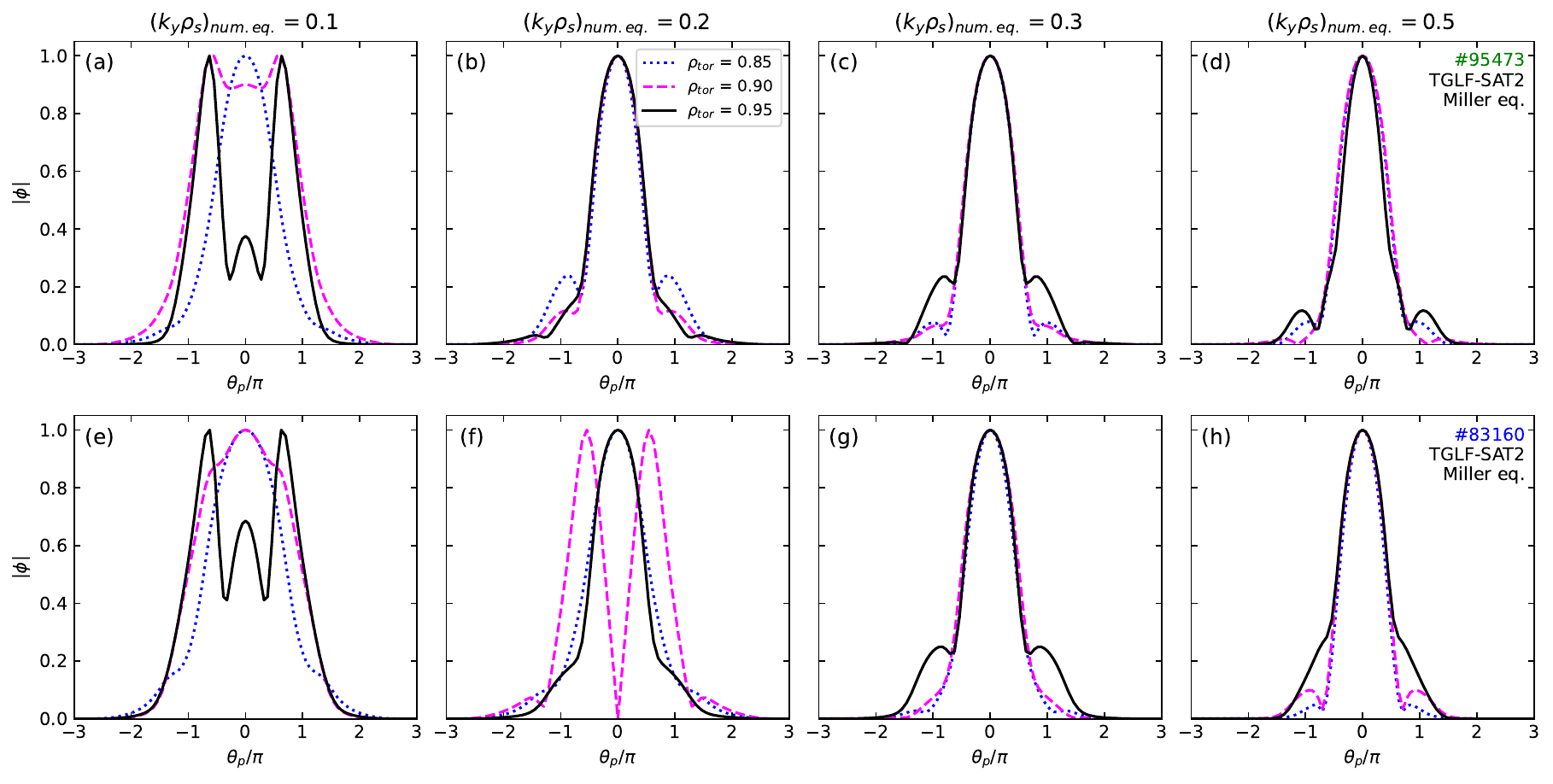}
    \caption{TGLF-SAT2 (Miller eq., electromagnetic) ballooning representation of the normalized electrostatic potential $\phi$ as a function of ballooning angle $\theta_{p}$, centered at the outboard midplane.
    Mode structures for $(k_y\rho_s)_{\text{num.~eq.}}$ = 0.1, 0.2, 0.3 and 0.5 are shown for \#95473 (a, b, c, d on top row) and \#83160 (e, f, g, h on bottom row) at $\rho_{\text{tor}}$ = 0.85 (blue dotted), 0.90 (purple dashed) and 0.95 (black solid).
    A truncated view of the simulated parallel domain ($\theta_{p} \leq \pm9\pi$) is shown for clarity.}
    \label{fig:phi_tglf}
 \end{figure*}

\subsubsection{Verification of the TGLF-SAT2 spectra}
\label{sec:tglf:veri}
In Figure \ref{fig:genevstglf-sat2}, a comparison of the linear growth rates times the sign of the mode frequencies, heat flux ratios and convective heat flux ratios from TGLF-SAT2 and the electromagnetic GENE simulations with Turnbull-Miller equilibrium parameterizations and Sugama collision operator for all three radii and as a function of $(k_y\rho_s)_{\text{num.~eq.}}$ is shown.

At $\rho_{\text{tor}}=0.85$, all the linear spectra show good agreement, as can be seen in Figure \ref{fig:genevstglf-sat2}(a), (d) and (g).
The shapes of the corresponding mode structures of $\phi$ from TGLF-SAT2 also agree reasonably well with those from GENE, as can be seen by comparing Figure \ref{fig:phi_tglf} with Figure \ref{fig:phi_gene}, although the TEMs are not as extended along the field lines.
For \#83164 and \#95473, the lowest $k_y$ in the spectra exhibit the wrong propagation direction, as TGLF-SAT2 predicts ITG instead of TEM.
As indicated by the colorbar, there are also a few outliers at high $k_y$ for two low-density branch discharges.
These are due to a shift between TGLF-SAT2 and GENE at which wavenumber a change in dominant mode propagation direction occurs in the spectrum, such as for TEM-UTEM and TEM-ETG branches.
The scatter of the heat flux and convective heat flux ratios is a bit larger than that of the growth rate spectra.
Generally, the direction of the particle fluxes from TGLF-SAT2 is correct, but for inward particle fluxes, the convective heat flux ratios are smaller in TGLF-SAT2.

Although the linear spectra at $\rho_{\text{tor}}=0.9$ generally still show agreement, the TGLF-SAT2 spectra exhibit more divergence from GENE than at $\rho_{\text{tor}}=0.85$, as can be seen in Figure \ref{fig:genevstglf-sat2}(b), (e) and (h).
TGLF-SAT2 correctly predicts the dominance of the $\ell$=1 ion modes at low wavenumbers for the high-density branch discharges, as illustrated in Figure \ref{fig:phi_tglf}(f), at the default number of Hermite basis functions.
However, this mode type is typically observed to be dominantly unstable for fewer wavenumbers in the TGLF-SAT2 spectra than in the GENE simulations, as can be seen by comparing Figure \ref{fig:phi_tglf}(e) with Figure \ref{fig:phi_gene}(e).
For $k_y\rho_s < 0.15$, TGLF-SAT2 tends to predict different dominant instabilities than GENE for all discharges.
This can be observed in Figure \ref{fig:genevstglf-sat2}(b) near the origin and by comparing the mode structures of $\phi$ for $k_y\rho_s=0.1$ in Figure \ref{fig:phi_tglf}(a) and (e) with Figure \ref{fig:phi_gene}(a) and (e).
For several low-triangularity discharges at $k_y\rho_s \leq 0.3$, the dominant modes in TGLF-SAT2 have (close to) the correct growth rate, but the wrong propagation direction.
For the low-density discharges this was due to predicting ITG over TEM, similar to $\rho_{\text{tor}}=0.85$, while for the high-density discharges this was due to a shift to a higher wavenumber of the transition from $\ell$=1 ion modes to TEM-UTEM.
Furthermore, for the high-density branch discharges, which are in the plateau regime at $\rho_{\text{tor}}=0.9$, TGLF-SAT2 overpredicts the linear growth rates for modes propagating in either direction by up to 30\% compared to GENE Miller with Sugama collision operator.
This is caused by the PAS collision model used in TGLF-SAT2.
The q$_i$/q$_e$ ratios show good agreement, except for the significantly higher ratios found in GENE for the $\ell$=1 ion-direction branches.
However, it is unknown if such modes were included in the databases of nonlinear gyrokinetic simulations involved in the tuning of the SAT2 saturation rule.
Preliminary nonlinear simulations with GENE of discharges for which this mode is linearly dominant indicated the heat flux ratios from TGLF-SAT2 provide reasonable predictions.

\begin{figure}[!h]
    \centering
    \includegraphics[width=0.475\textwidth]{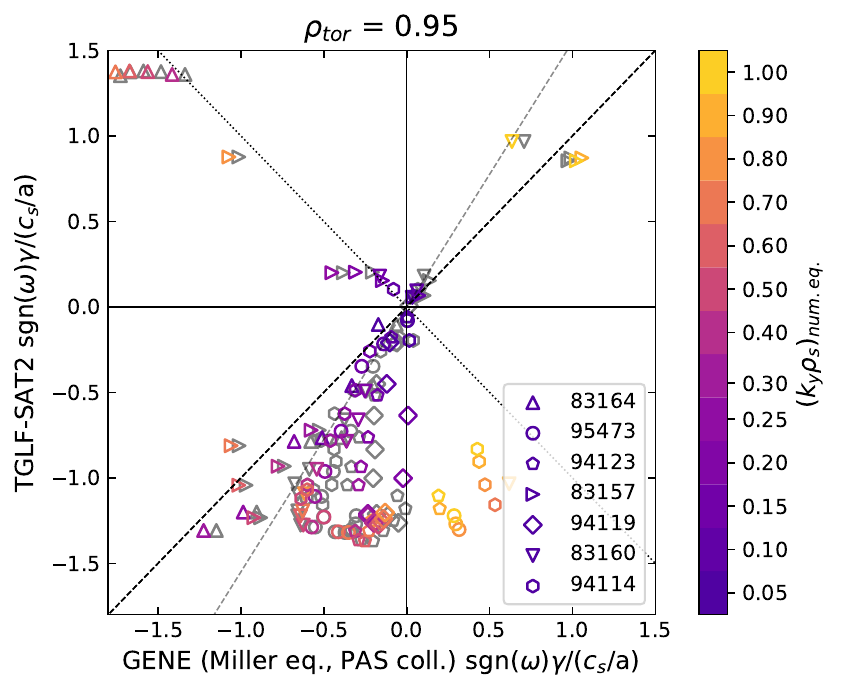}
    \caption{Comparison of the growth rate $\gamma$ times sign of the mode frequency $\omega$ from TGLF-SAT2 against the values from linear, electromagnetic GENE simulations using Miller geometry for all seven JET discharges at $\rho_{\text{tor}}=0.95$ and $(k_y\rho_s)_{\text{num.~eq.}} \in [0.05,1]$ with the pitch-angle scattering (PAS) collision operator (purple-yellow) and with the Sugama collision operator (gray). 
    TGLF-SAT2 uses a PAS collision model.}
    \label{fig:gene_pitchvstglf}
\end{figure}
\begin{figure*}
    \centering
    \includegraphics[width=0.44\textwidth]{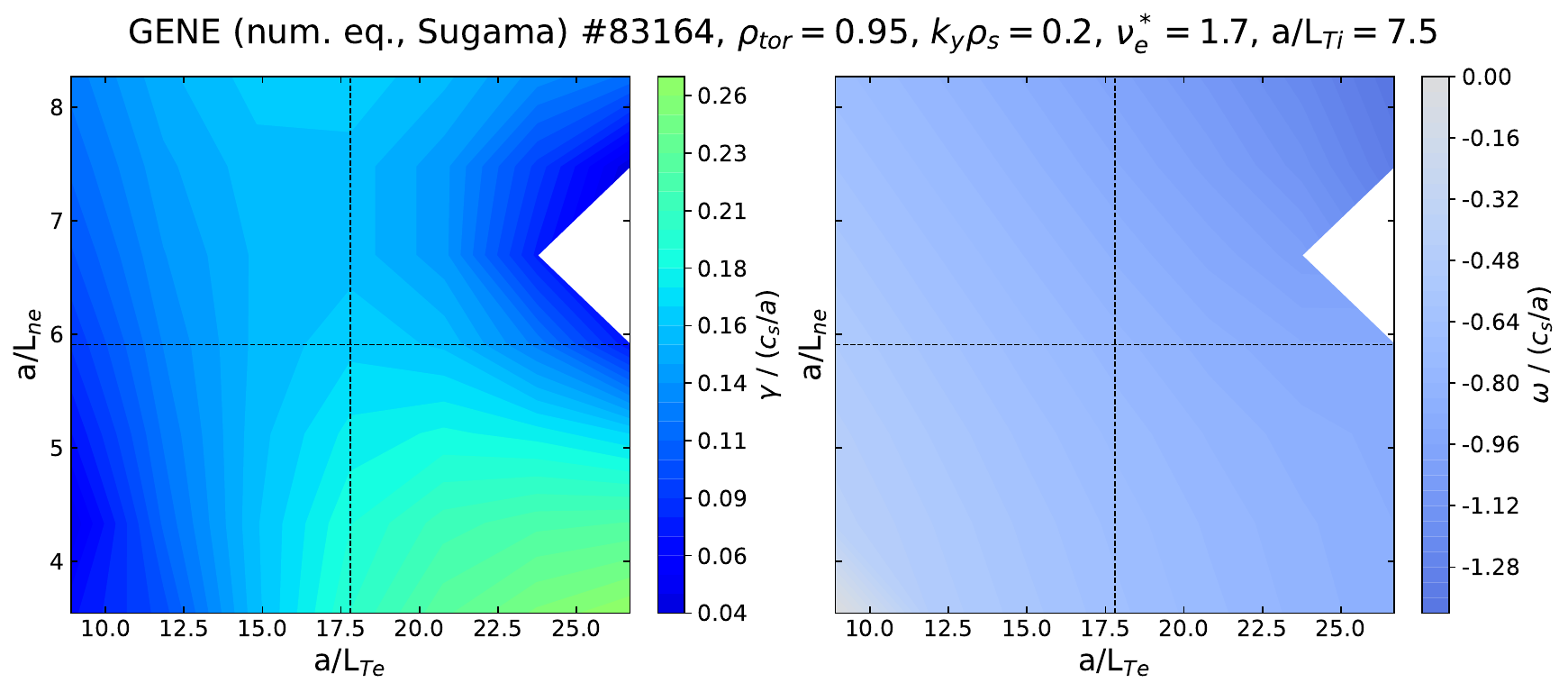}
    \includegraphics[width=0.44\textwidth]{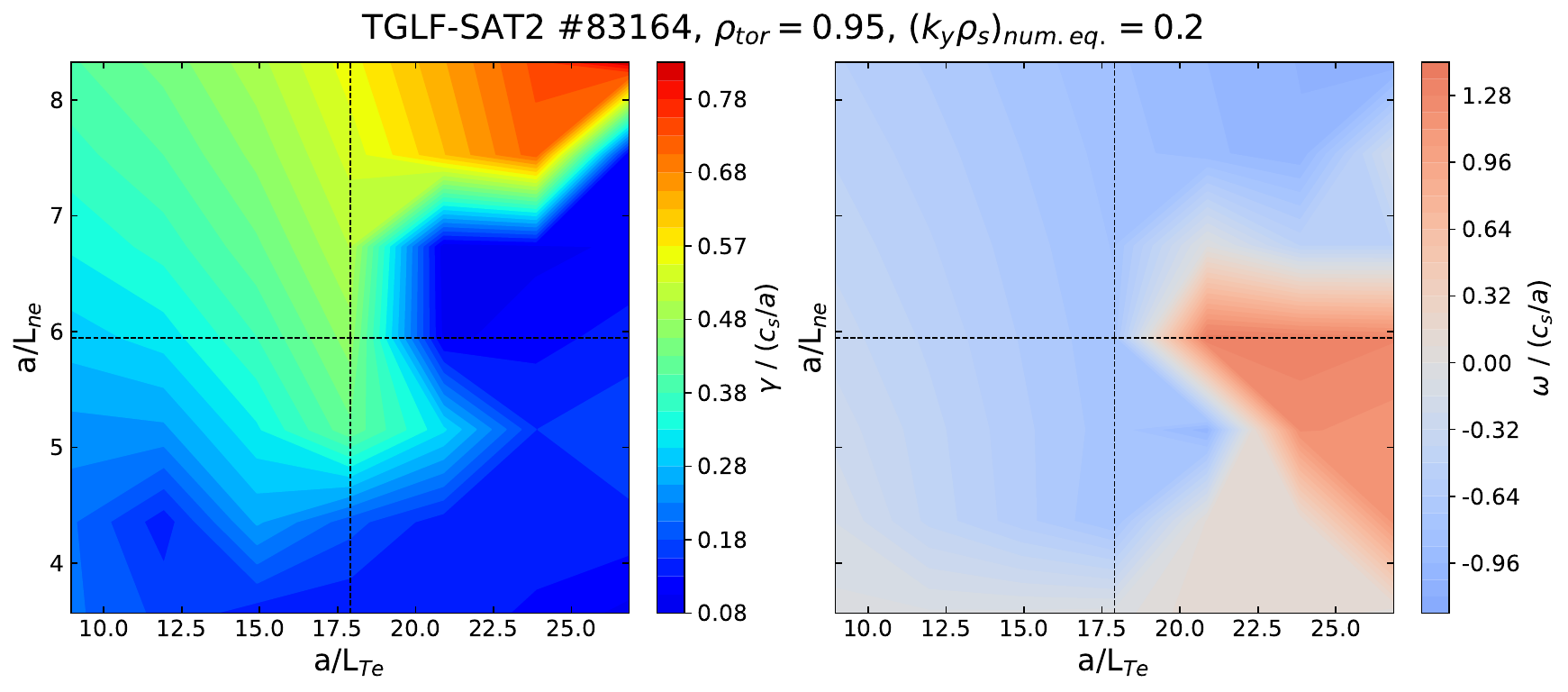}
    \includegraphics[width=0.44\textwidth]{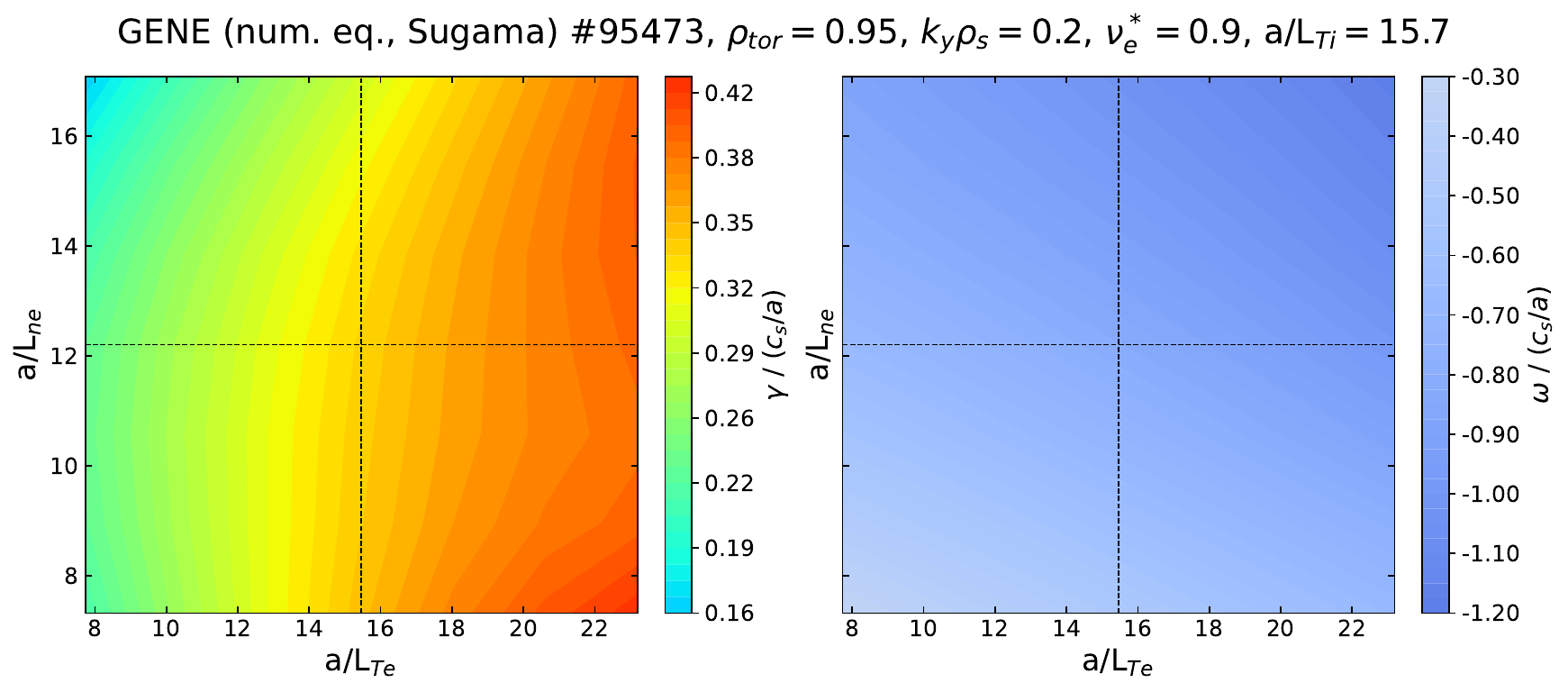}
    \includegraphics[width=0.44\textwidth]{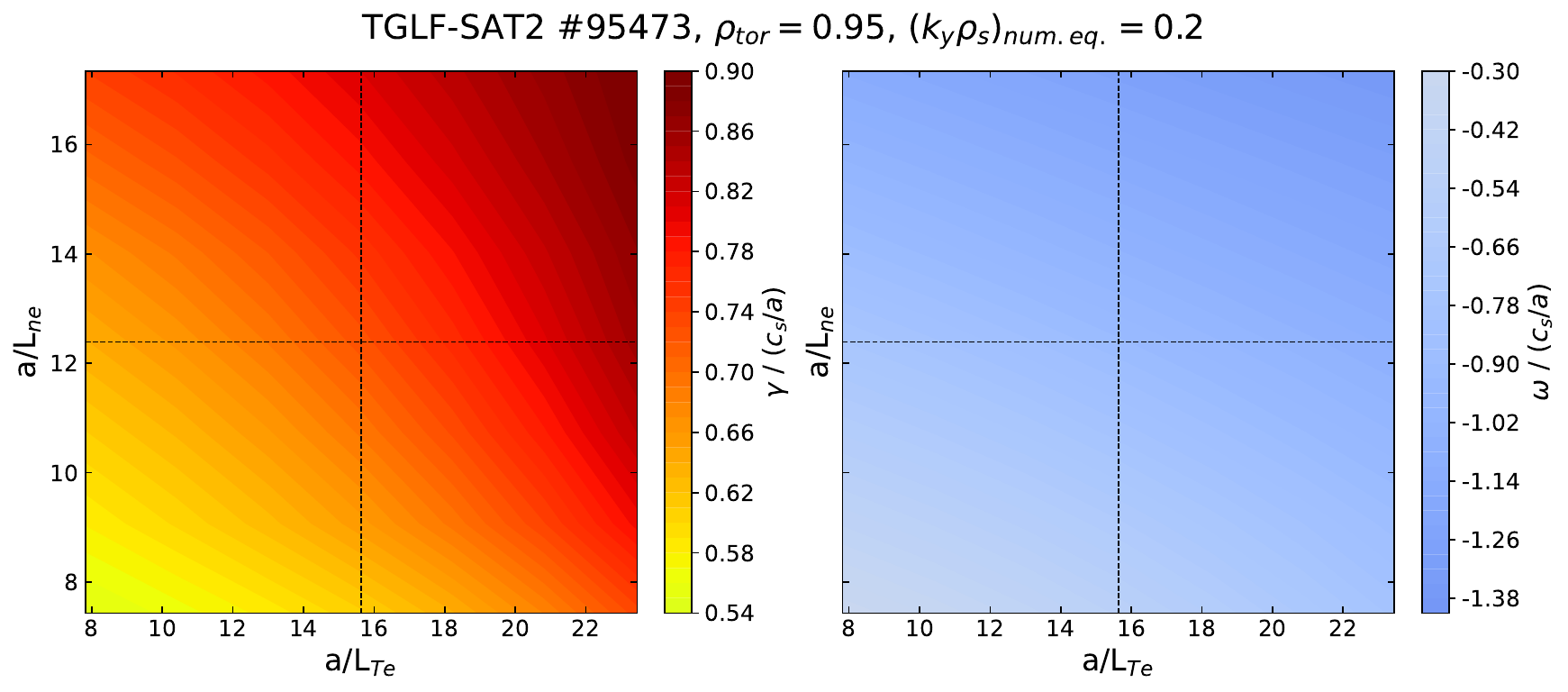}
    \includegraphics[width=0.44\textwidth]{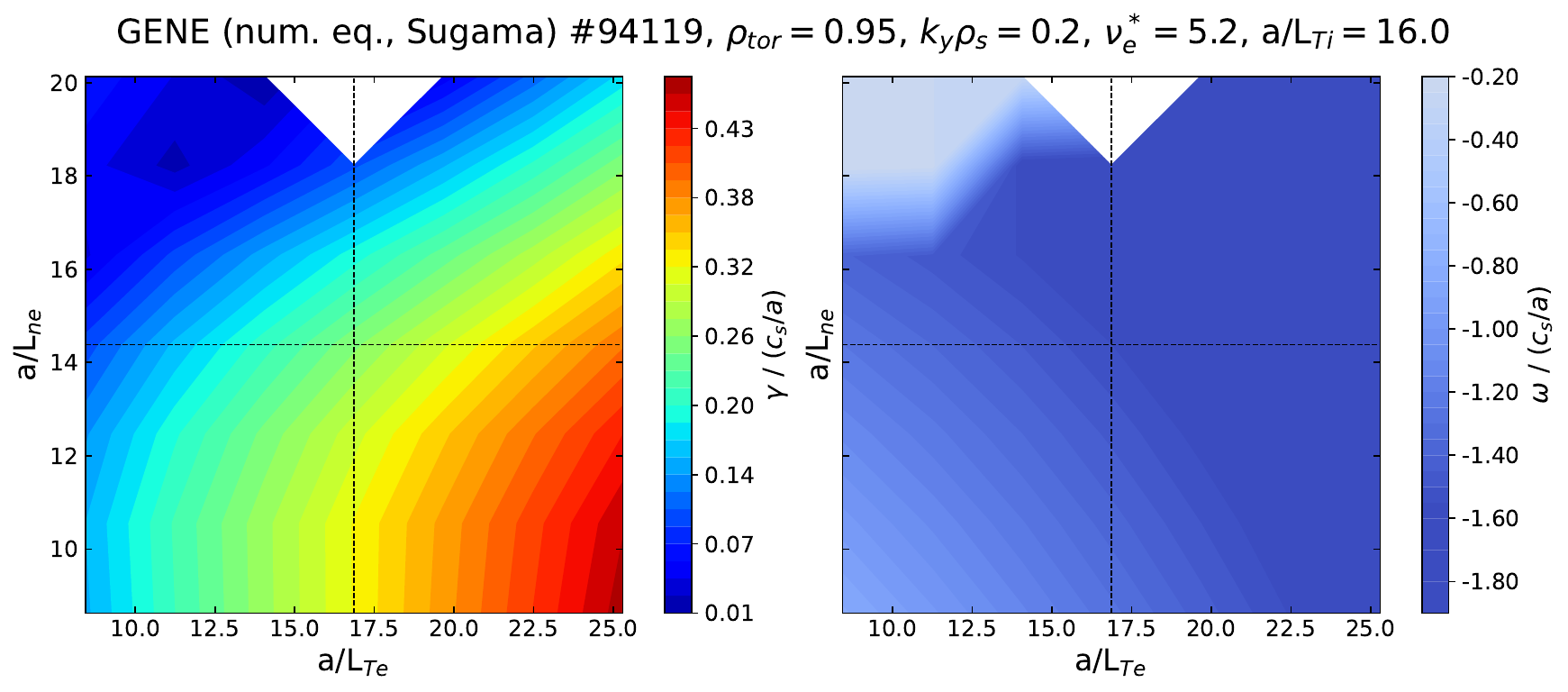}
    \includegraphics[width=0.44\textwidth]{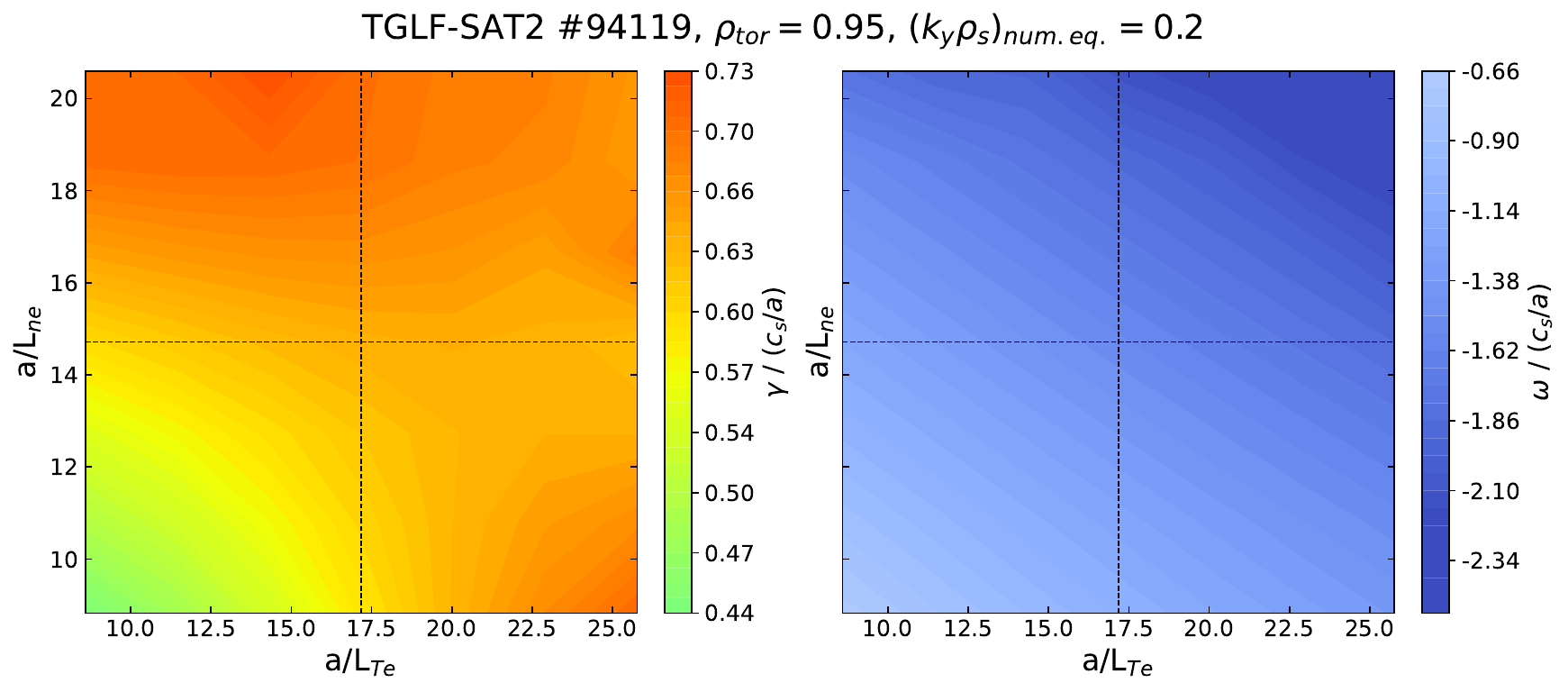}
    \includegraphics[width=0.44\textwidth]{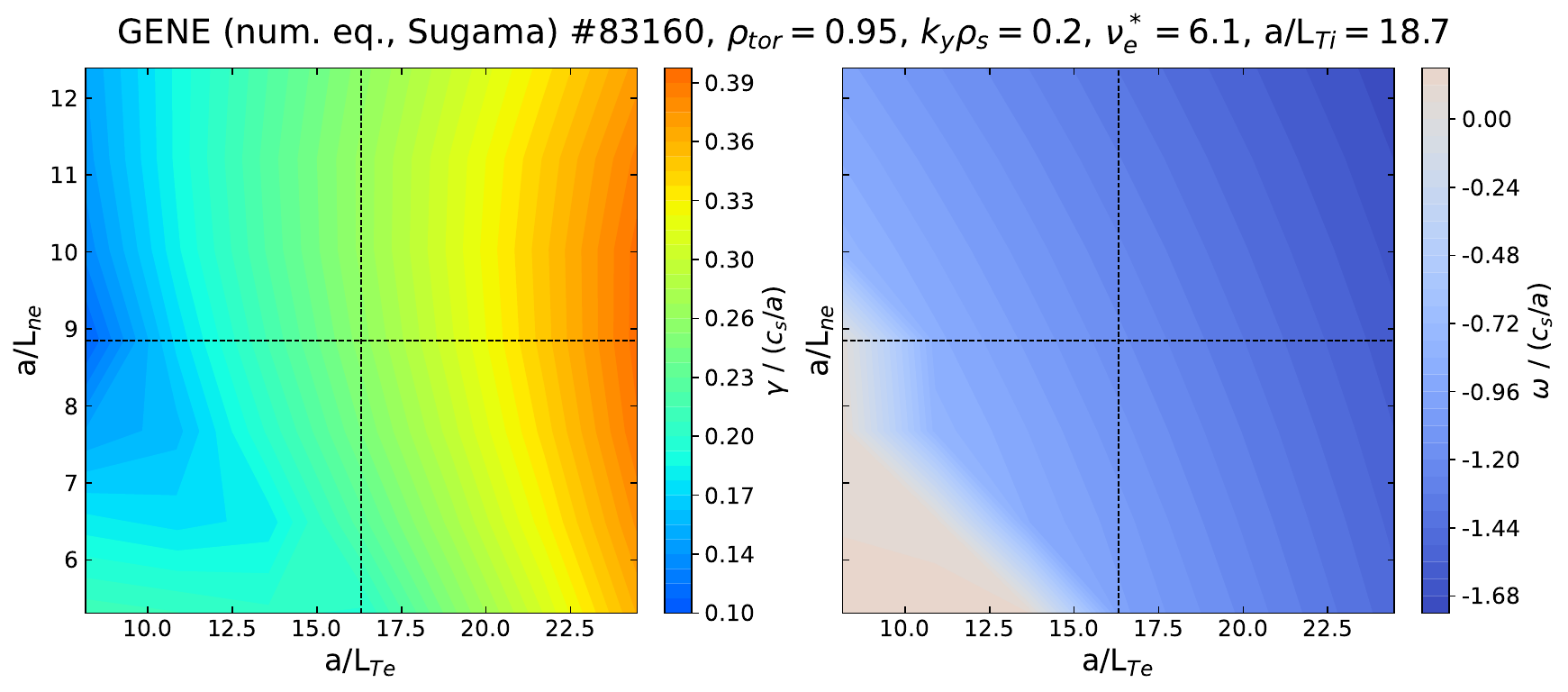}
    \includegraphics[width=0.44\textwidth]{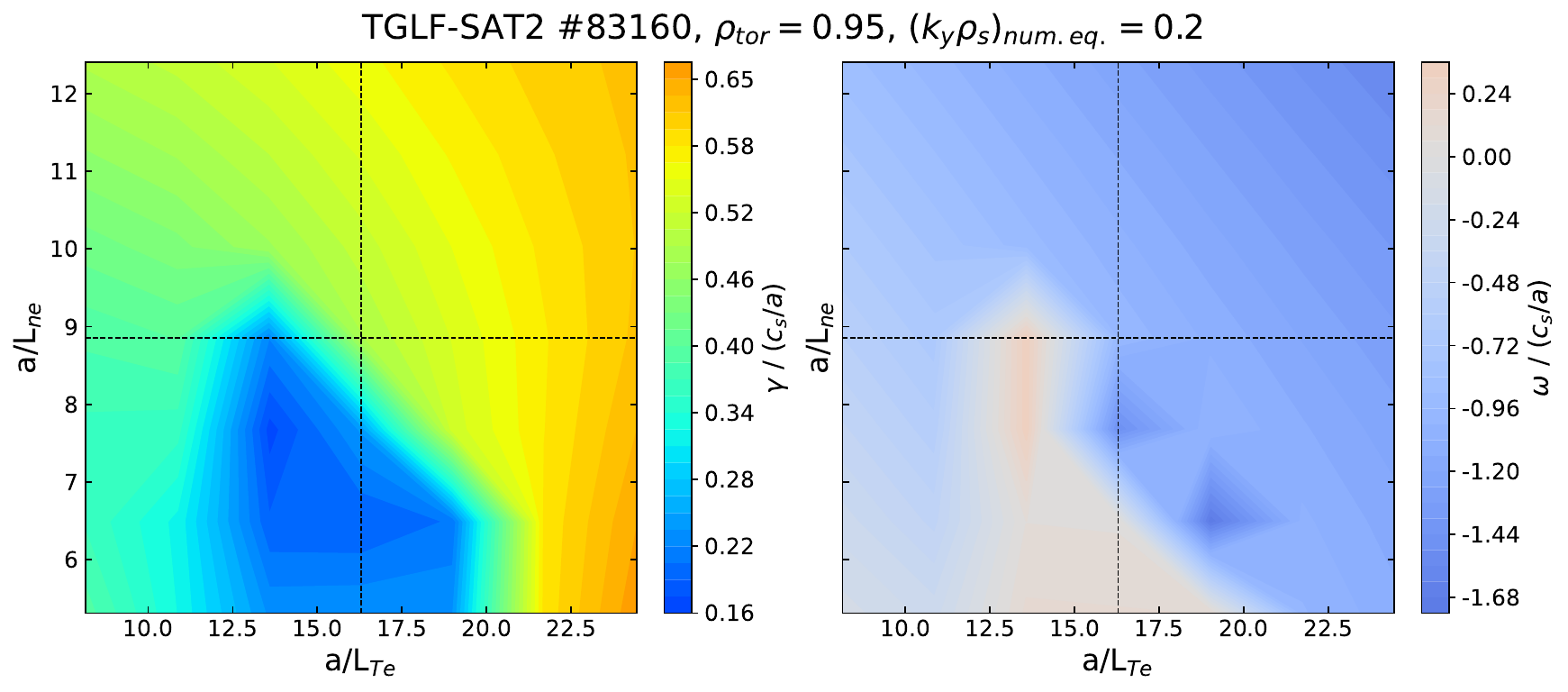}
    \caption{Comparison of the linear growth rate and frequency as a function of $a/L_{n_e}$ and $a/L_{T_e}$ between GENE (num.~eq., Sugama) in the left two columns and TGLF-SAT2 (Miller, PAS) in the right two columns. Separate color scales are used for the GENE and TGLF-SAT2 growth rates for clarity.}
    \label{fig:gradients}
    \vspace*{-1em}
\end{figure*}

At $\rho_{\text{tor}}=0.95$, TGLF-SAT2 consistently predicts larger growth rates than GENE Miller with Sugama collision operator, as can be seen in Figure \ref{fig:genevstglf-sat2}(c).
The growth rates for $(k_y\rho_s)_{\text{num.~eq.}} \leq 0.3$ from TGLF-SAT2 are 35\% larger than those from GENE for all discharges.
This is again due to the PAS collision model used in TGLF-SAT2.
To confirm this, the comparison at $\rho_{\text{tor}}=0.95$ was repeated against GENE Miller simulations with PAS collision operator, as shown in Figure \ref{fig:gene_pitchvstglf}.
For reference, the comparison against GENE Miller with Sugama from Figure \ref{fig:genevstglf-sat2}(c) is also shown in gray.
A thin gray dashed line indicates the 35\% error in linear growth rate typically observed in GENE between the Sugama and PAS collision operators in the plateau and Pfirsch-Schl\"{u}ter regimes, as shown in Sec.~\ref{sec:gene:coll} and Figure \ref{fig:coll_op}.
This line agrees well with the results of the comparison against GENE Miller with Sugama collision operator for $k_y\rho_s\leq 0.3$.
The comparison between TGLF-SAT2 and GENE Miller with PAS collision operator shows improved agreement for the same wavenumber range for most discharges.
However, it does not resolve all discrepancy at low $k_y$.

\begin{figure*}[!t]
    \centering
    \includegraphics[width=0.9\textwidth]{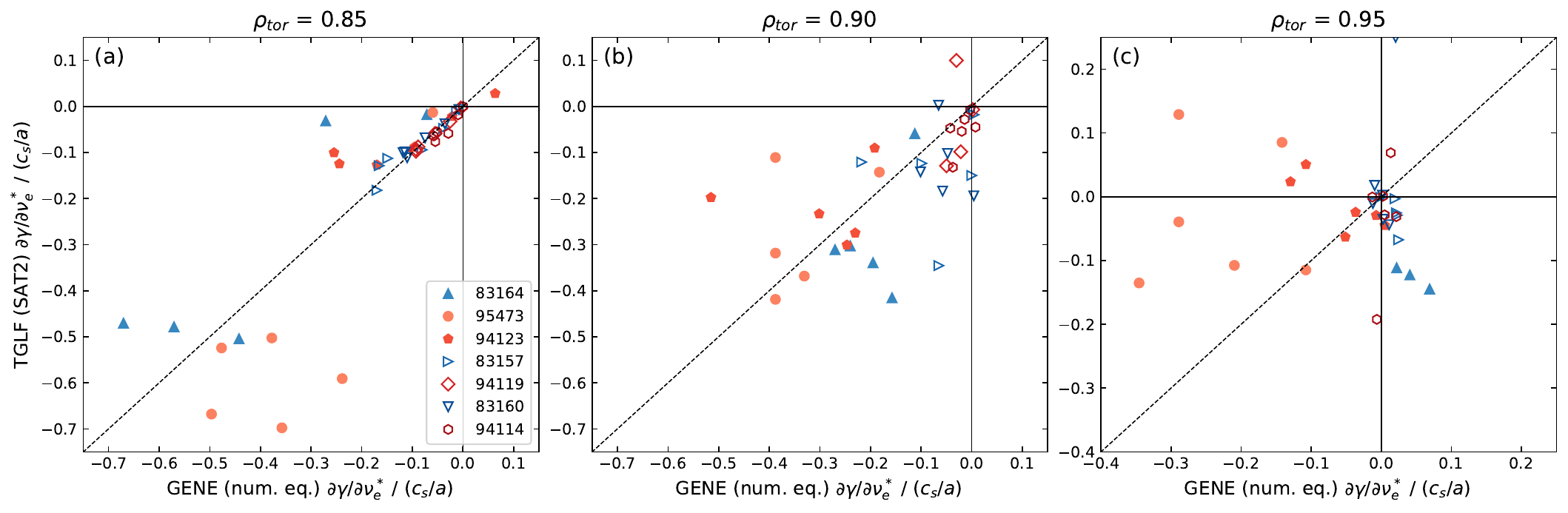}
    \caption{A comparison between TGLF-SAT2 (vertical) and GENE (horizontal) of the derivative of the linear growth rate with respect to the dimensionless electron collisionality $\partial \gamma/\partial\nu_e^*$ for $(k_y\rho_s)_{\text{num.~eq.}} \in [0.1,0.2,0.3,0.5,0.7,0.9]$ at (a) $\rho_{\text{tor}}$ = 0.85, (b) $\rho_{\text{tor}}$ = 0.9 and (c) $\rho_{\text{tor}}$ = 0.95.}
    \label{fig:dgammadnue_tglf}
\end{figure*}
\begin{figure}
    \centering
    \includegraphics[width=0.45\textwidth]{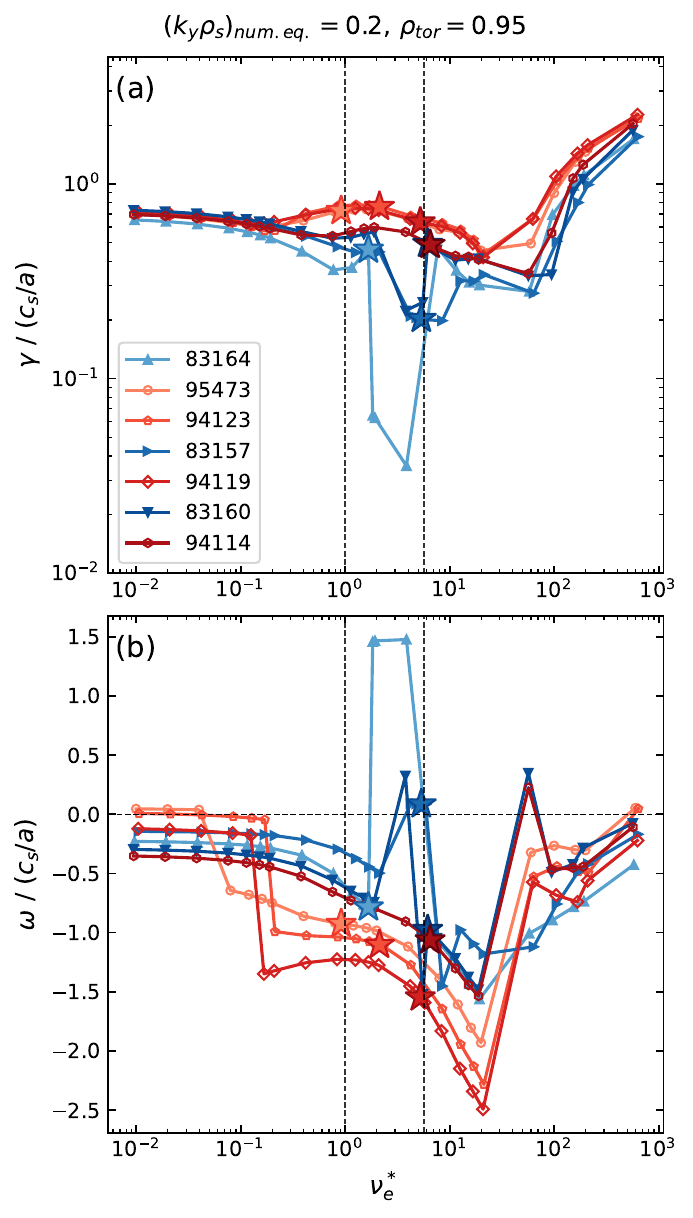}
    \caption{Linear (a) growth rates and (b) mode frequencies from TGLF-SAT2 as a function of $\nu_e^*$, for all seven discharges, for $(k_y\rho_s)_{\text{num.~eq.}} = 0.2$ at $\rho_{\text{tor}} = 0.95$.}
    \label{fig:gamma_coll_tglf}
\end{figure}
\begin{figure*}[!t]
    \centering
    \includegraphics[width=\textwidth]{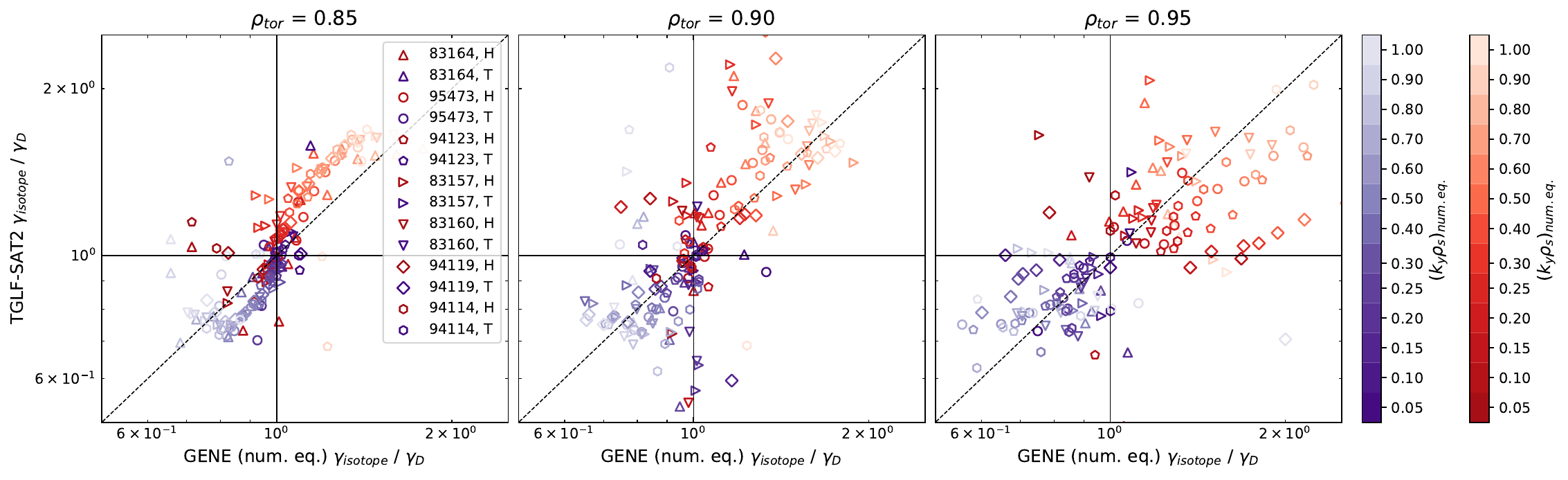}
    \caption{A comparison of the ratio of the linear growth rates for hydrogen (shades of red) and tritium (shades of purple) with deuterium $\gamma_{\text{isotope}}/\gamma_{D}$ between GENE (num.~eq., Sugama) and TGLF-SAT2 for $k_y\rho_s \in [0.05,1]$ at three different radii (from left to right).}
    \label{fig:gamma_HT_comp}
 \end{figure*}
Furthermore, two-dimensional sensitivity scans were performed to compare the gradient drives of the dominant instabilities at $(k_y\rho_s)_{\text{num.~eq.}} = 0.2$ at $\rho_{\text{tor}} = 0.95$ between TGLF-SAT2 and GENE (num.~eq., Sugama).
In Figure \ref{fig:gradients} scans of the growth rate and mode frequencies for $a/L_{n_e}$ and $a/L_{T_e}$ to $\pm$40\% and $\pm$50\% of their nominal values are shown for \#83164, \#95473, \#94119 and \#83160.
Even though the mode frequencies at nominal gradients are in relatively close agreement, the gradients sensitivities are not.
The sensitivity to $a/L_{T_e}$, increasing the growth rate with increasing $a/L_{T_e}$, is generally properly captured.
However, the non-monotonic sensitivity to $a/L_{n_e}$ is not reflected in TGLF-SAT2.

At $\rho_{\text{tor}}=0.95$, for $(k_y\rho_s)_{\text{num.~eq.}} > 0.3$ TGLF-SAT2 often predicts different dominant mode branches than GENE, based on mode frequencies and ballooning structures.
In GENE, the former show a shift in towards the ion diamagnetic drift direction and $\phi$ becomes increasingly split around the outboard midplane, as shown in Figure \ref{fig:phi_gene} in Sec.~\ref{sec:gene:lsa}.
However, in TGLF-SAT2 the ballooning structures are not showing any split, as can be seen in Figure \ref{fig:phi_tglf}(c), (d), (g) and (h), and both the mode growth rates and frequencies are larger than in GENE.
An increase in the number of Hermite polynomial basisfunctions (\texttt{nbasis\_max} $>$ 6) did not result in better agreement between the dominant modes in TGLF-SAT2 and GENE Miller.
Inspection of the first five subdominant mode branches in the TGLF-SAT2 simulations did not lead to better matches for the mode branches dominant in the GENE simulations.
The divergence is also not related to the type of collision operator, as can be seen in Figure \ref{fig:gene_pitchvstglf}, it remains when comparing against GENE Miller with PAS collisions.
Identifying the root cause of this discrepancy is left for future work.

In general, a larger amount of subdominant modes are unstable in TGLF-SAT2 than in the GENE simulations.
Although the growth rates for the subdominant modes follow similar trends to those of their dominant counterparts, both the mode frequencies and ballooning structures are mostly not in agreement with those from GENE.
Often, the ballooning structures are of higher $\ell$ values than in GENE, which suggests that higher harmonics are overpredicted in TGLF-SAT2.

A similar comparison of the linear spectra against GENE Miller was also made for TGLF-SAT1, which is occasionally still used in integrated modeling simulations.
For TGLF-SAT1, the linear dominant mode spectra were generally similar, but the growth rates and frequencies showed larger discrepancies at all three radii, in particular at low wavenumbers.
Furthermore, TGLF-SAT1 was predicting stability for the lowest $k_y$ in the simulations at $\rho_{\text{tor}}=0.9$ and 0.95.
Thus, TGLF-SAT2 constitutes an improvement over TGLF-SAT1 in the L-mode pedestal-forming region.
Finally, a comparison was also made with the recently developed SAT3 \cite{Dudding2022} saturation rule.
As SAT3 employs the same linear solver as SAT2, using it only affected the flux ratios.
The SAT3 flux ratios were found to be in strong agreement with the results from SAT2.
However, the fluxes themselves were typically two times larger for SAT3 than for SAT2, at all three radii.

With the geometry dependencies that were added to the saturation rule for TGLF-SAT2, it was observed that the specific local equilibrium Miller parameterization has a strong influence on the output fluxes.
In particular, the values of the radial derivatives of the flux-surface shaping coefficients were found to be important in the pedestal-forming region, see Appendix \ref{apdix:geo} for more details.
Sensitivity of the linear instabilities to elongation shear $s_{\kappa}$ and triangularity shear $s_{\delta}$, as was found in GENE, was not observed in TGLF-SAT2.

\subsubsection{Effect of collisionality in TGLF-SAT2}
\label{sec:tglf:coll}
To investigate if the dominant mode branches in TGLF-SAT2 have the correct characteristics as a function of the electron collisionality, a similar sensitivity study as described for GENE in Sec.~\ref{sec:gene:coll} was performed.

A comparison of the derivative of the linear growth rates with respect to $\nu_e^*$ between TGLF-SAT2 and GENE num.~eq.~with Sugama collision operator is shown in Figure \ref{fig:dgammadnue_tglf} for $(k_y\rho_s)_{\text{num.~eq.}} \in [0.1,0.2,0.3,0.5,0.7,0.9]$ and all three radii considered.
At $\rho_{\text{tor}}=0.85$ and 0.9, similar trends as in Figure \ref{fig:col_dgamma} in the derivative of the linear growth rates with respect to $\nu_e^*$ are observed.
Counterintuitively, at both of these radii the agreement is worst for the low-density branch discharges, although the scatter also increases for the high-density branch discharges at $\rho_{\text{tor}}=0.9$.
However, at $\rho_{\text{tor}}=0.95$ there are large discrepancies for all discharges.
Indeed, considering the linear growth rate and frequency spectra as a function of collisonality in TGLF-SAT2 at this radial location, as shown in Figure \ref{fig:gamma_coll_tglf} for $k_y\rho_s=0.2$, the transition from collisionless instabilities, that are stabilized by increasing collisionality, to resistive instabilities, which are destabilized by increasing collisionality, is less smooth than in GENE as shown in Figure \ref{fig:col_scan}(c).
As expected based on the results obtained in GENE with PAS collisions, there are no resistive mode branches at the plateau-Pfirsch-Schl\"{u}ter boundary in TGLF-SAT2.
This is due to the additional collisional processes that are missing in the PAS operator.
From the growth rate spectra in Figure \ref{fig:gamma_coll_tglf}, it is clear that multiple discharges are close to mode transitions at their experimental collision frequencies.
There are also several transitions to ion-direction branches for the high-triangularity discharges that are not present in the corresponding collisionality scans in GENE.
Furthermore, the low-triangularity discharges and \#83164 appear to be on dissipative mode branches, as their growth rates already increase with $\nu_e^*$ in the banana regime.
The collisionality also was found to affect the mode gradient sensitivities strongly.
However, even at artificially modified $\nu_e^*$ the non-monotonic $a/L_{n_e}$ was not recovered in TGLF-SAT2.

Overall these discrepancies can influence the trajectory of integrated modeling simulations.
Small variations in collisionality can trigger changes in the dominant instabilities that, based on the linear gyrokinetic simulations with GENE, are not expected to occur in the experiment.

\subsubsection{Effect of isotope mass in TGLF-SAT2}
\label{sec:tglf:mass}
Given recent work investigating the effect of isotope mass in experiments with the use of integrated modeling \cite{Birkenmeier2023}, it was verified to what degree TGLF-SAT2 captures the impact of isotope mass on the linear response correctly.
In Figure \ref{fig:gamma_HT_comp} a comparison between TGLF-SAT2 and GENE of the impact of isotope mass on the linear growth rate is shown.
The ratio of the growth rate for a given isotope, hydrogen in red and tritium in purple, to the growth rate for deuterium $\gamma_{\text{isotope}}/\gamma_D$ is plotted for TGLF-SAT2 against GENE simulations with numerical equilibrium and Sugama collision operator.
At all three radii, the impact of isotope mass largely correctly affects the linear growth rates.
However, the instabilities dominant at $\rho_{\text{tor}}$ = 0.95 in TGLF-SAT2 are less sensitive to the isotope mass than in the GENE simulations.
This may indicate that the increasingly non-adiabatic response of passing electrons is not captured entirely by TGLF-SAT2.

\section{SUMMARY AND DISCUSSION}
\label{sec:disc}
A detailed linear gyrokinetic stability analysis and mode characterization study was carried out in the pedestal-forming region of the L-mode phase just prior to H-mode entry for seven NBI-heated JET-ILW discharges.
The reported experimental sensitivities of the L-to-H transition threshold power $P_{L-H}$ to the line-averaged density, isotope mass, plasma shaping and dilution and increased effective ion charge due to plasma impurities were found to be reflected in the sensitivities of the dominant linear instabilities, in particular at $\rho_{\text{tor}}=0.95$.
A clear distinction between the ion-scale instabilities dominant in discharges on the low- and high-density branches for H-mode access was observed at all three radii considered in this work: at $\rho_{\text{tor}}=0.85$, these were TEM and ITG, respectively, at $\rho_{\text{tor}}=0.9$ TEM / Ubiquitous TEM (UTEM) and $\ell$=1 modes with drive from trapped ions and the ion-temperature gradient (resembling TITGs), respectively, and at $\rho_{\text{tor}}=0.95$, collisionless and resistive hybrid modes, respectively.
Both the electron collisionality $\nu_e^*$ and the ratio $L_{T_i} / L_{n_e}$ were identified as important factors in setting the dominant regime.

Corresponding to the differences in dominant modes, the quasilinear particle fluxes changed direction from outward for the low-density branch discharges to inward for the high-density branch discharges.
The dominance of the observed modes in both regimes was also found to be robust within the uncertainties of the density and temperature profile fits to the experimental data.
Furthermore, an inverse scaling of the linear mode destabilization with the square root of the isotope mass was observed, which is consistent with the experimentally observed trends in the L-H power threshold and pedestal height scaling. 
This effect was found to be enhanced by increased normalized density gradient.
It was also observed that the dominant linear instabilities exhibit a sensitivity to the radial derivatives of the plasma elongation and triangularity, but this requires further study.
Finally, pertaining to $\rho_{\text{tor}}=0.95$, the collisionless linear instabilities that are dominant on the low-density branch of the L-H transition were found to be significantly more sensitive to plasma dilution and increased effective ion charge due to impurities than the resistive instabilities that are dominant on the high-density branch.
Overall, this indicates that linear instability analysis can offer a path for the scoping and improvement of heuristic models that try to capture the approximate scaling of $P_{L-H}$.

The quasilinear turbulent transport models QuaLiKiz and TGLF-SAT2 were examined in the pedestal-forming region of L-mode.
First, the impact of the model reduction choices on the linear gyrokinetic spectra in the pedestal-forming region was investigated.
The $s$-$\alpha$ geometry employed by QuaLiKiz was found to be an inadequate description of the local magnetic equilibria in the pedestal-forming region.
This resulted in large changes to the linear spectra of the dominant instabilities.
Although several mode characteristics can be captured approximately, quantitative differences were found to be large.
The Miller geometry employed by TGLF was found to be sufficiently accurate for discharges with relatively small up-down asymmetry, such as those considered in this work.
The electrostatic gyrokinetic dispersion relation of QuaLiKiz was found to be accurate in comparison to GENE simulations with similar model reductions at $\rho_{\text{tor}}=0.85$.
Once the electron collisionality enters the plateau regime and plasma resistivity starts to play a role in the dynamics of the dominant instabilities for $\rho_{\text{tor}} > 0.9$, the physics included in QuaLiKiz starts to become insufficient.
Similarly, TGLF-SAT2 was found to agree well with GENE simulations in the banana regime, but once the electron collisionality entered the plateau regime and at high normalized gradients, discrepancies started to appear. 
Overall, QuaLiKiz can reliably be used up to $\rho_{\text{tor}}=0.85$ at most, as long as electron collisionality is in the banana regime.
TGLF-SAT2 is verified to apply in a larger radial range, but begins to break down at $\rho_{\text{tor}}=0.95$ at low-density and misses the correct hybrid modes with increased dominance of passing-electron dynamics; as a result it displays incorrect gradient-drive scalings and the flux ratios for $k_y\rho_s \in [0.1,0.5]$ at this radial location. At higher density and $\rho=0.95$, TGLF-SAT2 does capture the resistive mode branches at very high collisionalities correctly, but not the resistive drift waves near the plateau-Pfirsch-Schl\"{u}ter boundary. 
Nevertheless, the relative distribution of transport over the ion and electron channels for both heat and particle transport in TGLF agree well with quasilinear gyrokinetic values.
Recommendations for possible further model developments for QuaLiKiz and TGLF-SAT2, to extend their validity near the separatrix in L-mode, are proposed in Appendix \ref{apdix:rec}.

In the context of integrated modeling, when modeling L-modes for $\rho>0.85$, TGLF-SAT2 is recommended. 
Although TGLF-SAT2 predicts a number of mode characteristics accurately, the gradient and collisionality sensitivities at experimental collision frequencies are not in agreement with the high-fidelity GENE simulations.
This will likely affect the profile evolution in the pedestal-forming region in integrated modeling simulations.
Validating TGLF-SAT2 predicted profiles in this region using flux-driven simulations is challenging, in particular for the density profiles. 
Obtaining accurate heat transport may be possible, since the total flux is more important than flux ratio due to ion-electron heat exchange; however, capturing the direction of the predicted particle fluxes is challenging, especially in the presence of modes resembling TITGs, as well as UTEM or hybrid modes. 
When predicting density profiles, the interplay with a poorly constrained neutral particle source make the validation exercise even more challenging. 
Correctly capturing density-profile dynamics is important, since it will likely impact the dynamics of the pedestal build-up during the transition to H-mode. 

Towards fully predictive L-mode profiles, the accuracy of the linear response in reduced quasilinear models needs to be improved. 
One avenue for this could be a neural network regression on the linear response of high-fidelity gyrokinetic code such as GENE \cite{Citrin2023}.
Furthermore, analyses of databases of high-fidelity simulations can reveal new physics and aid in the further development of accurate reduced models; to this end, recent work has been conducted on composing gyrokinetic simulation databases \cite{Fuhr2023,Hatch2022}. 
The extensive simulation database used for the present publication will be made available and hence serve improving further model development. 
The validity of the quasilinear assumptions for L-mode edge turbulence is an active field of research.
Nonlinear simulations in Refs.~\cite{DeDominici2019} and \cite{Bonanomi2019} confirm that the quasilinear approximation is valid for L-mode edge turbulence, while more recent work is challenging this assertion \cite{Ashourvan2024}. 
Future research into these questions will be required to determine under what conditions precisely quasilinear modeling is no longer appropriate for such scenarios.

\appendix
\renewcommand\thesection{\Alph{section}}
\titleformat{\section}[block]{\bfseries}{\appendixname~\thesection}{1em}{}
\renewcommand{\appendixname}{APPENDIX}
\section{GPR PROFILE FITS}
\begin{figure*}
   \centering
   \includegraphics[width=0.675\textwidth]{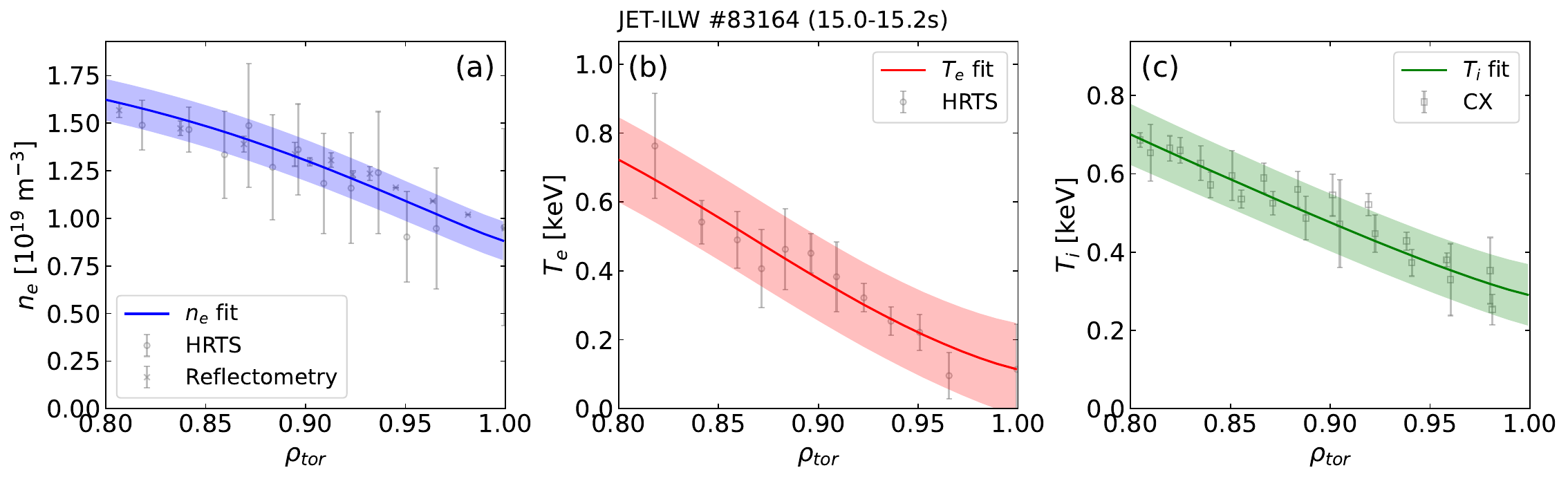}
   \includegraphics[width=0.675\textwidth]{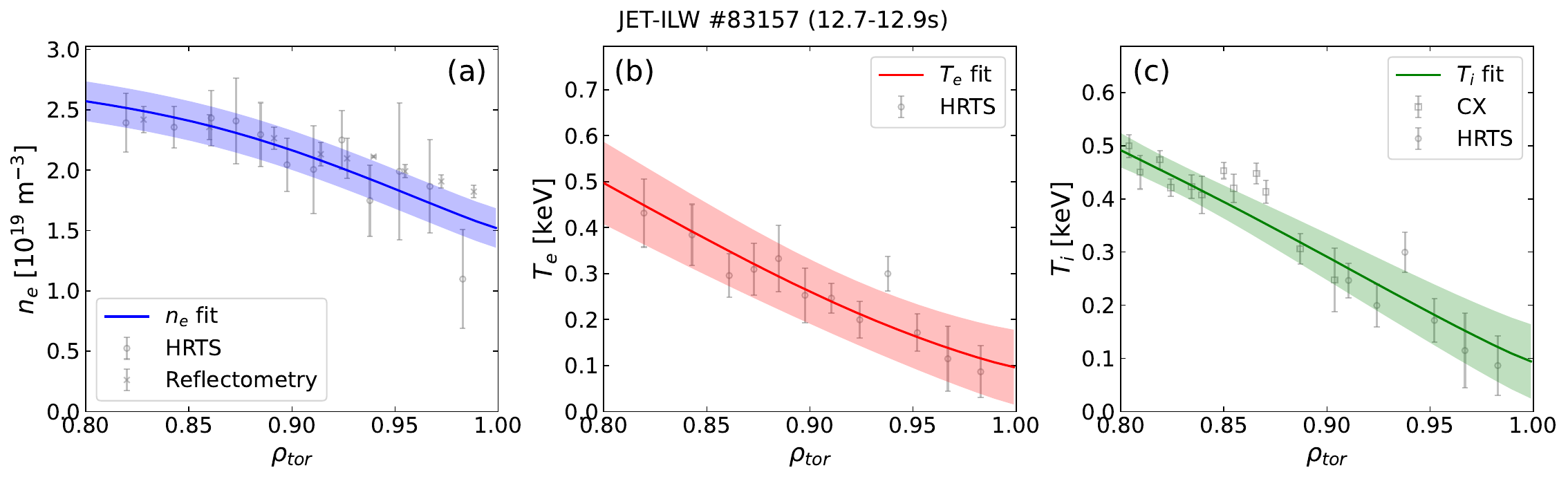}
   \includegraphics[width=0.675\textwidth]{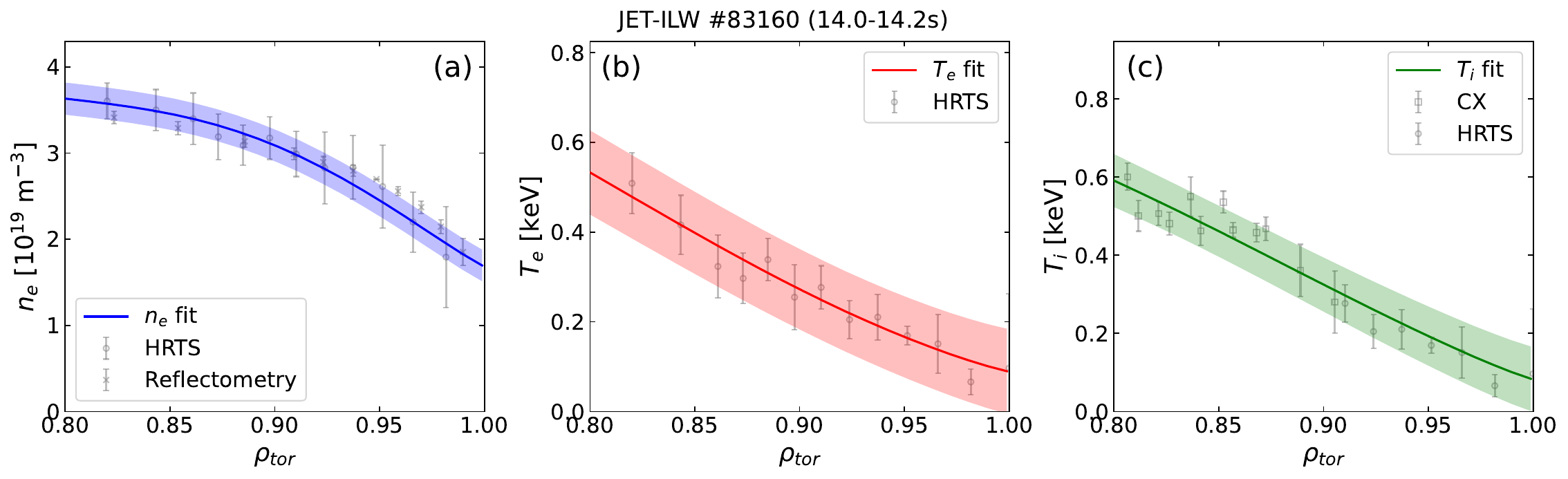}
   \includegraphics[width=0.675\textwidth]{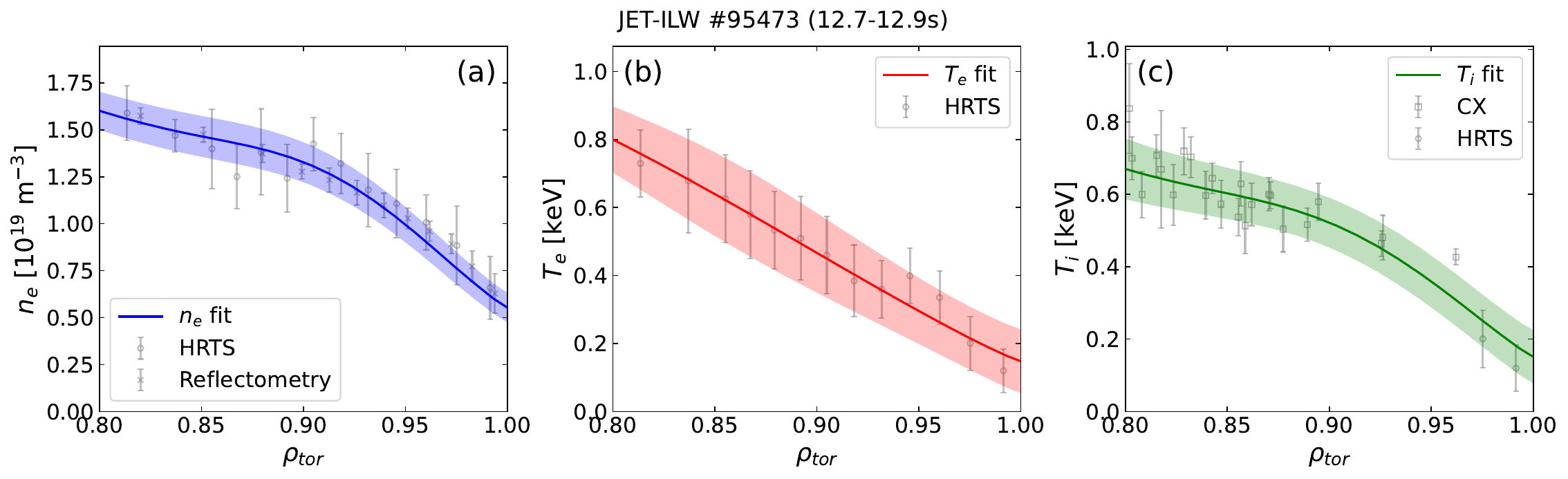}
   \includegraphics[width=0.675\textwidth]{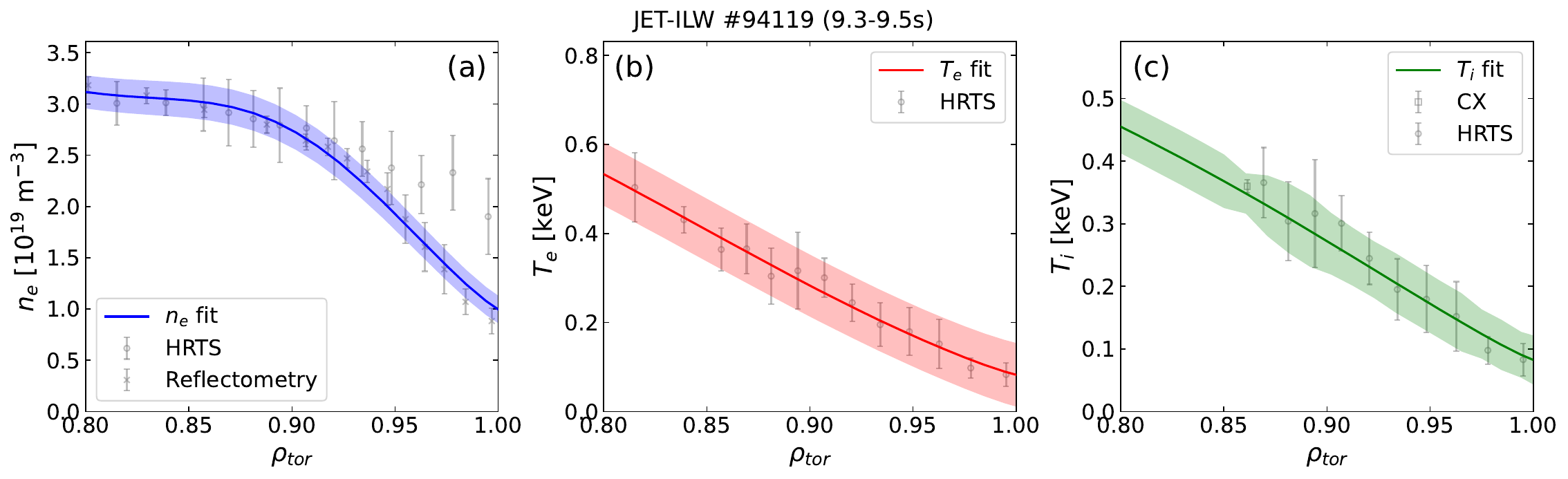}
   \includegraphics[width=0.675\textwidth]{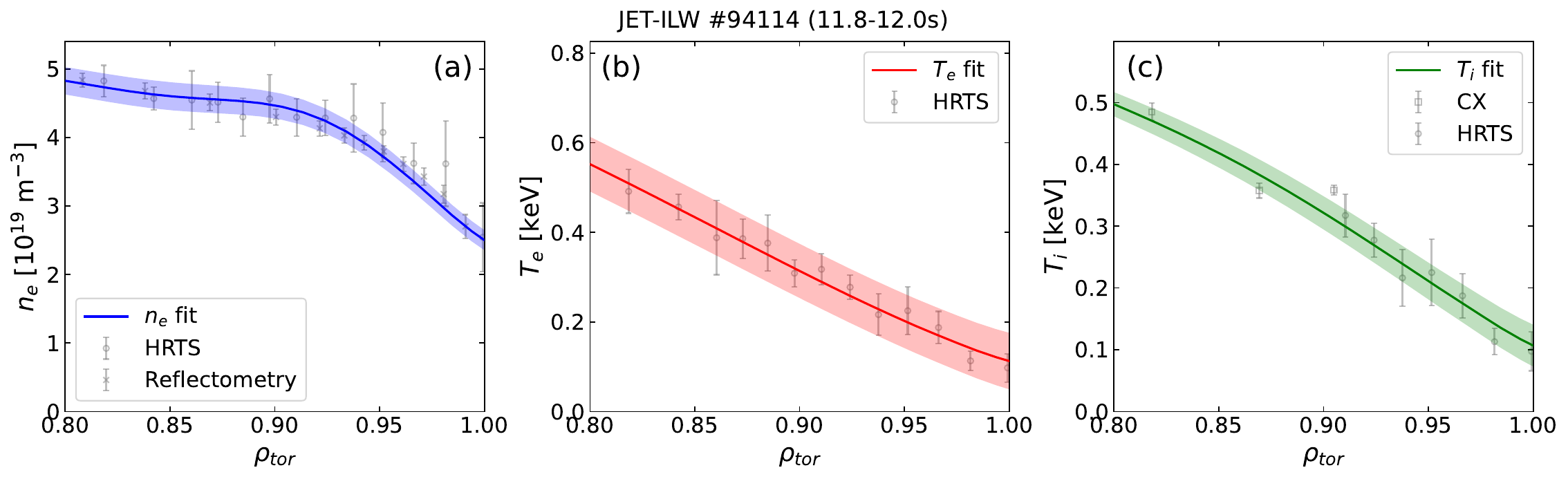}
   \caption{Gaussian process regression fits of time-averaged experimental measurements (grey markers) of the (a) electron density $n_e$ (blue), (b) electron temperature $T_e$ (red) and (c) ion temperature $T_i$ (green) near the edge of the plasma, $\rho_{\text{tor}}\in[0.8,1]$, for \#83164, \#83157, \#83160, \#95473, \#94119 and \#94114.}
   \label{fig:apdix:gpr_fits}
\end{figure*}

In Figure \ref{fig:apdix:gpr_fits} an overview of the Gaussian process regression fits of $n_e$, $T_e$ and $T_i$ in the pedestal-forming region for the remaining six JET-ILW discharges is provided.
\label{apdix:profiles}

\section{LINEAR GENE ELECTRON-SCALE SPECTRA}
\label{apdix:elecs}
\begin{figure*}[!h]
   \centering
   \includegraphics[width=\textwidth]{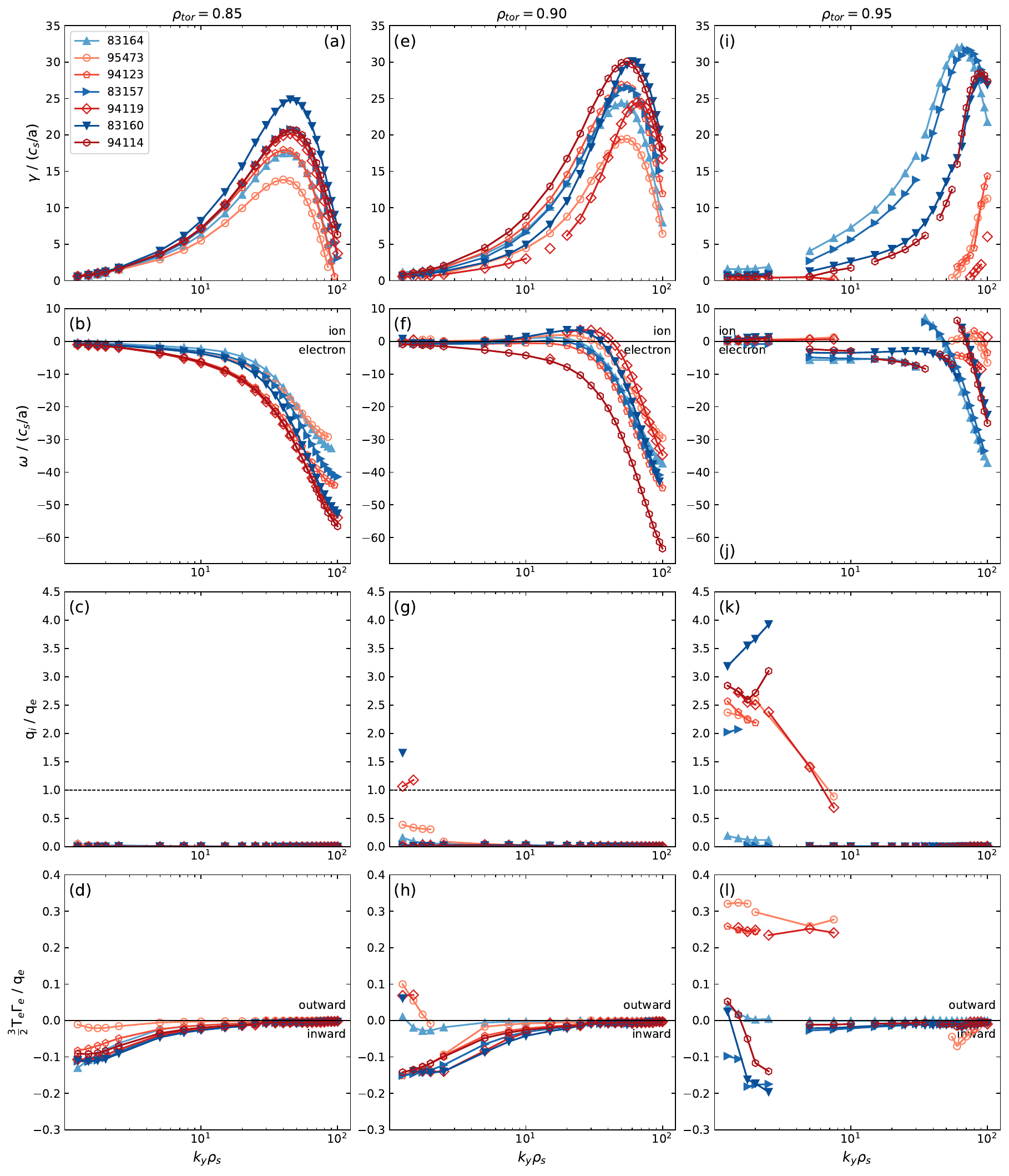}
   \caption{Linear \textsc{gene} electron-scale spectra as a function of binormal wavenumber $k_y\rho_s$ at $\rho_{\text{tor}}=$ 0.85, 0.9, 0.95.
   From top to bottom the linear (a) growth rate $\gamma$, (b) frequency $\omega$, (c) heat flux ratio q$_i$/q$_e$ and (d) convective heat flux ratio $\frac{3}{2}$T$_e\Gamma_e$/q$_e$ are shown for both high (closed symbols, shades of blue) and low (open symbols, shades of red) triangularity.
   Positive(/negative) frequencies indicate mode propagation in the ion(/electron) diamagnetic drift direction, while positive(/negative) convective heat flux ratio values indicate outward(/inward) flux directions.}
   \label{fig:ETG}
\end{figure*}

After the L-H transition a part of the remaining turbulent transport in H-mode is originates from electron temperature gradient (ETG) modes. 
Thus, to complete the picture of the dominant linear instabilities just prior to H-mode and how these might affect the transition, the intermediary and electron-scale spectra ($1 < k_y\rho_s \leq 100$) are shown in Figure \ref{fig:ETG}.

\subsection{Mode characteristics at $\rho_{\text{tor}}=0.85$}
On both density branches instabilities for $k_y\rho_s \geq 1$ gradually start to have more ETG mode characteristics.
Such modes only drive electron heat transport and are typically strongly ballooned around the outboard midplane, the latter of which was also observed for the dominant mode structures in our simulations.
Although Figure \ref{fig:ETG}(c) shows q$_i$/q$_e \approx 0$, non-zero electron particle fluxes remain for at least $k_y\rho_s \leq 30$ and trapped particles still play a role based on weight cross phase correlations.
Therefore, it is likely that the dominant electron-scale modes are TEM or TEM-ETG hybrids that gradually transition into pure ETG.

\subsection{Mode characteristics at $\rho_{\text{tor}}=0.9$}
The TEM-ETG regime seen at $\rho_{\text{tor}}=0.85$ continues here, with two notable differences.
Firstly, for four discharges, spread over both density branches, UTEM remains dominant for $k_y\rho_s \leq 3$, as evidenced by q$_i$/q$_e > 0$ and the change of particle flux directions in Figures \ref{fig:ETG}(g) and (h).
Secondly, for several discharges the propagation direction of the dominant modes gradually changes from the electron to ion and back to electron diamagnetic drift direction, as can be seen in Figure \ref{fig:ETG}(f).
These propagation direction changes are not accompanied by sudden changes in flux ratios or mode structure changes.
Therefore, these modes are ETG instability branches that also propagate in the ion direction, as previously observed by Pueschel \textit{et al.}~\cite{{Pueschel2020}}.

\subsection{Mode characteristics at $\rho_{\text{tor}}=0.95$}
As previously observed for the ion-scale instabilities at this radial location in Sec. \ref{sec:gene:lsa:95}, there are more pronounced differences between high and low triangularity discharges compared to previous radii.
For all discharges the UTEM-like branches that are dominant at the end of the ion-scale continue at the intermediary scale.
For the high triangularity discharges this extends up to $k_y\rho_s \leq 3$, while for the low triangularity discharges this extends all the way up to $k_y\rho_s = 10$, as can be seen in Figures \ref{fig:ETG}(k) and (l).
The exception among the low triangularity discharges is \#94114, which has $\sim$1.5$\times$ lower $a/L_{n_e}$ at this radial location compared to the other low triangularity discharges and whose spectra are similar to those of the high triangularity discharges.
This is correlated with \#94114 having about , as can be seen in Table \ref{tab:params}.

TEM-ETG hybrid instabilities, similar to those at $\rho_{\text{tor}}=0.9$, become dominant for the high triangularity discharges and \#94114 for $k_y\rho_s>3$, although the very small particle fluxes suggest ETG characteristics dominate.
As $k_y\rho_s$ increases several ETG mode branches that start propagating in the ion direction before quickly changing direction to the electron direction become dominant.
The low triangularity are completely stable for $3 < k_y \rho_s < 50$ and only at very large wavenumbers TEM-ETG hybrid branches become unstable, as can be seen by the non-zero outward particle fluxes in Figure \ref{fig:ETG}(l).
This clear difference between low and high triangularity discharges in their linear spectra shows that increased $a/L_{n_e}$ destabilizes the UTEM, while stabilizing the ETG modes, which was confirmed by sensitivity scans.
The increased stability of the electron-scale modes at low triangularity might be one of the reasons for the slightly lower $P_{L-H}$ required for the low triangularity discharges.

\subsection{Multi-scale effects}
The fraction $\gamma/k_y$ can be used as an indicator for when multiscale interactions between the ion and electron scale instabilities can become relevant.
At $\rho_{\text{tor}} = 0.85$ the values for $\gamma/k_y$ indicate there could be multiscale interaction, as $\gamma/k_y$ for the electron scale are about equal to or even larger than at ion scale \cite{Creely2019}.
Moving further out, $\gamma/k_y$ for the electron scale becomes half (at $\rho_{\text{tor}} = 0.9$) to a quarter (at $\rho_{\text{tor}} = 0.95$) of the values on the ion scale, thereby making it increasingly unlikely that there is any relevant multiscale interaction.
A further (nonlinear) investigation would be required to confirm this.

\section{FURTHER ION-SCALE SENSITIVITIES}
\subsection{Electromagnetic effects ($\beta$, $\delta B_{||}$)}
\label{apdix:beta}
\begin{figure}
   \centering
   \includegraphics[width=0.475\textwidth]{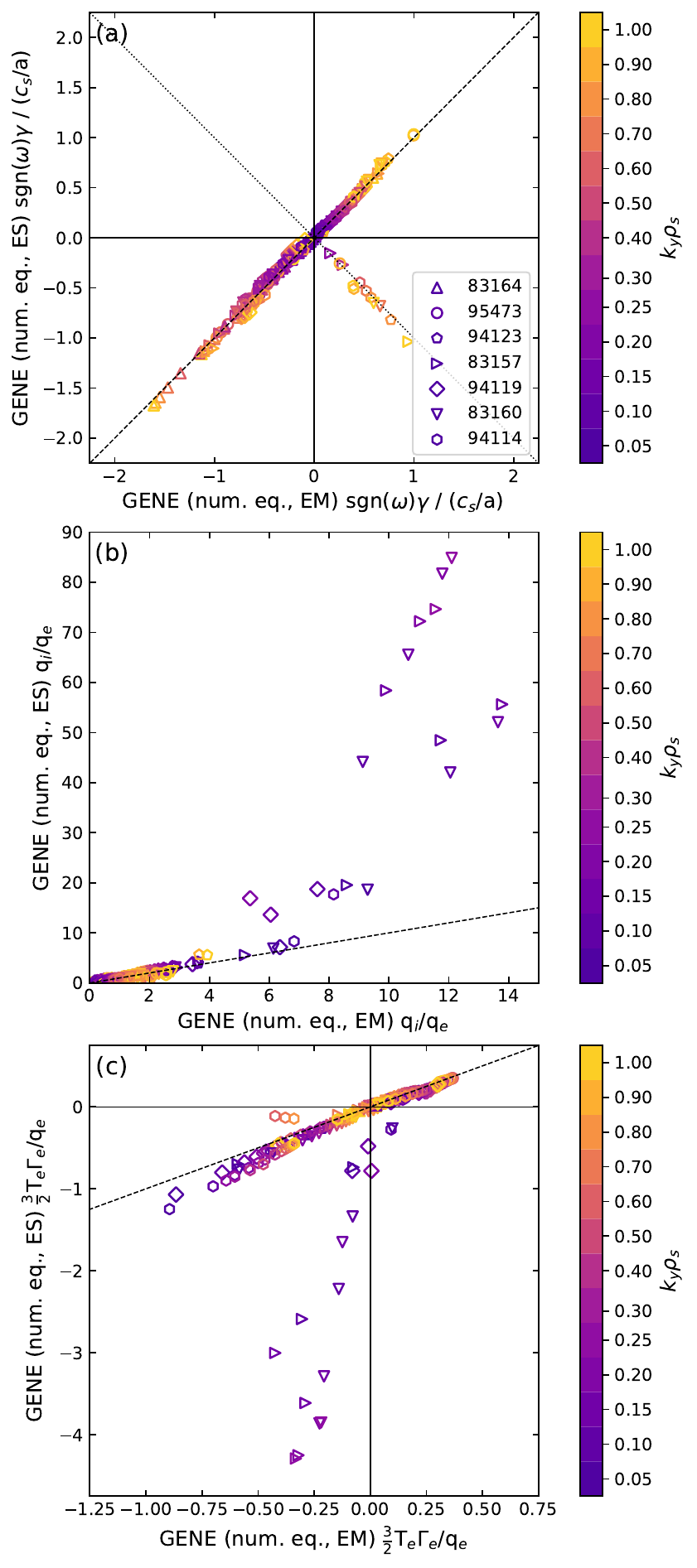}
   \caption{The impact of electromagnetic effects (finite $\beta$ and $\delta B_{||}$) on the linear (a) growth rate $\gamma$, (b) heat flux ratio q$_i$/q$_e$ and (c) convective heat flux ratio from linear \textsc{gene} simulations with a numerical equilibrium.
   The results from electrostatic (ES) simulations are on the vertical axis, while those from fully electromagnetic (EM) are on the horizontal axis.}
   \label{fig:beta=0}
\end{figure}
\begin{figure}[!t]
   \centering
   \includegraphics[width=0.475\textwidth]{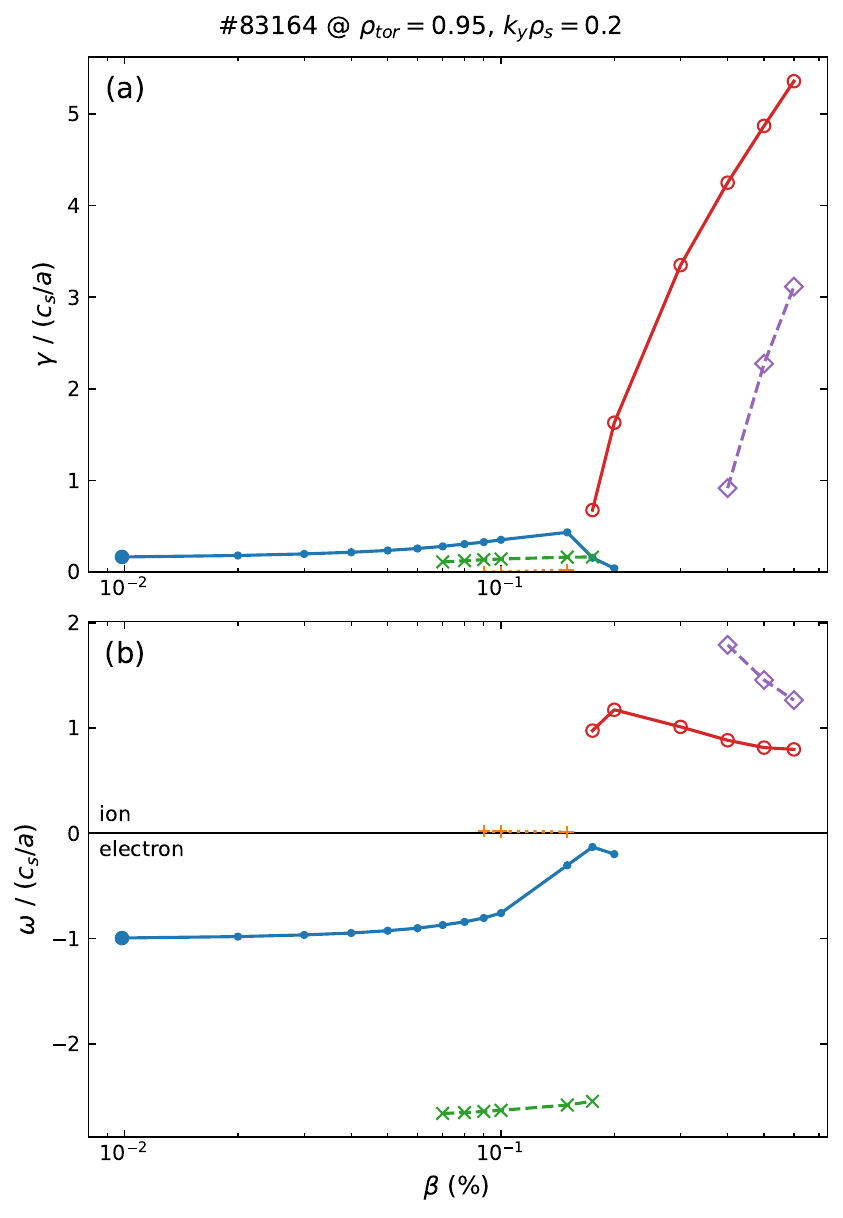}
   \caption{The linear (a) growth rate $\gamma$ and (b) frequency $\omega$ for the instabilities found at $k_y\rho_s=0.2$ for \#83164 at $\rho_{\text{tor}}=0.95$ as a function of $\beta$ in percent.
   The nominal values based on the fit of the experimental data (blue circle) are indicated on the left.
   Around the expected value of $\beta_{\text{crit}}$ both the dominant (blue and red solid) and subdominant (green, orange and purple dashed) instabilities are shown.}
   \label{fig:beta_scan}
\end{figure}
Previous characterization of the instabilities in the L-mode pedestal forming region emphasized the importance of electromagnetic effects \cite{Scott2005,Bonanomi2019}.
The impact of this on the linear instabilities was checked in three ways: (1) turning off electromagnetic effects entirely ($\beta_e = 0$), (2) turning off electromagnetic flutter (no $B_{||}$ fluctuations) and (3) $\beta$-scans at $\rho_{\text{tor}}=0.95$.
For the latter the plasma $\beta$ input in GENE was scanned with a self-consistent evaluation of the pressure term, but using the numerical equilibria at the nominal experimental $\beta$ value.

Electrostatic simulations showed that $\beta$ has a destabilizing effect on the dominant instabilities at low wavenumbers ($k_y\rho_s \leq 0.7$) and a stabilizing effect at high wavenumbers (for UM and ETG modes).
Growth rates typically change 5-10\% at $\rho_{\text{tor}} = 0.85$ and $0.9$ and between 10-30\% at $\rho_{\text{tor}}=0.95$, but this is hard to see in Figure \ref{fig:beta=0}.
For a few cases a change in mode propagation direction is shifted to slightly higher $k_y\rho_s$.
Additionally for the weakly resistive, $\ell$=1 ion-direction instabilities, dominant at low $k_y\rho_s$ on the high $\bar{n}_e$ branch, both the heat flux and convective heat flux ratios increase 2-8$\times$, as can be seen in Figures \ref{fig:beta=0}b and c.
This re-balancing of the heat fluxes of these odd-parity modes was found to be due to a relative reduction of the electrostatic electron heat flux q$_e$ by about an order of magnitude, not due to the removal of the electromagnetic contributions.
Thus, the electrostatic properties of these modes are strongly affected by electromagnetic effects.

Turning electromagnetic flutter, \textit{i.e.}~evolution of $\delta B_{||}$ fluctuations, on/off had a negligible effect on the reported growth rates ($\sim$1\%) and did not affect the linear heat flux ratios or any other linear mode characteristics.

Values for $\beta_{crit,MHD} = 0.6 \hat{s} / (q^2 (2 R/L_{n_e}+R/L_{T_e}+R/L_{T_i}))$ were used to estimate how close our discharges are to the onset of kinetic ballooning modes (KBMs) as function of $\beta$.
As shown in Table \ref{tab:beta_crit}, this fluid approximation of the KBM threshold is consistently about an order of magnitude higher than the nominal values of $\beta$ for all our discharges.
Scans in $\beta$ at $\rho_{\text{tor}}=0.95$, see Figure \ref{fig:beta_scan} for \#83164, shows that the actual kinetic threshold was typically $\sim$20\% lower than $\beta_{crit,MHD}$ at $k_y\rho_s=0.2$.
Linear GENE eigenvalue simulations around the KBM threshold showed the additional destabilization of subdominant modes up to and past the threshold.
At $\beta$ above the nominal experimental value a subdominant second-order excitation ($\ell$=1) of the dominant hybrid electron-direction mode found at $\rho_{\text{tor}}=0.95$ becomes unstable at about half the growth rate, as can be seen in Figure \ref{fig:beta_scan}.
Then, just prior to the KBM threshold, a very weak ITG mode becomes a second subdominant unstable mode.
After the threshold we see a subdominant $\ell$=1 excitation of the KBM, similar to \cite{Xie2018}.
Further investigation if and how such subdominant modes affect the nonlinear electromagnetic transport, which Ref.~ \cite{Bonanomi2019} found started increasing prior to reaching the KBM threshold, is beyond the scope of this work.

\begin{table}[!h]
   \setlength{\tabcolsep}{3.75pt}
   \setlength{\extrarowheight}{1pt}
   \begin{tabular}{c c c c c}
      \hline\hline\\[-2.25ex]
      $\rho_{\text{tor}}$ & & 0.85 & 0.9 & 0.95\\[1ex]
      \hline
      \#83164 & & 0.48 & 0.32 & 0.23 \\
      \#83157 & & 0.38 & 0.24 & 0.17 \\
      \#83160 & & 0.43 & 0.26 & 0.16 \\
      \#95473 & & 0.52 & 0.26 & 0.12 \\
      \#94123 & & 0.51 & 0.26 & 0.11 \\
      \#94119 & & 0.42 & 0.20 & 0.11 \\
      \#94114 & & 0.44 & 0.27 & 0.14 \\
      \hline\hline
   \end{tabular}
   \caption{Values for $\beta_{\text{crit},\text{MHD}} = 0.6 \hat{s} / (q^2 (2 R/L_{n_e}+R/L_{T_e}+R/L_{T_i}))$ in \% for the different discharges and radii considered here.}
   \label{tab:beta_crit}
\end{table}

\subsection{Effect of toroidal rotation}
\label{apdix:rot}
Toroidal rotation was excluded from the simulations reported up to this point to perform the linear stability analysis on an even keel, given that the rotation data was unavailable for two out of seven discharges.
It is however one of the contributing factors to the radial electric field $E_r$, which is thought to be an integral part of the L-H transition physics.
Therefore, as a sensitivity analysis we performed linear simulations at $\rho_{\text{tor}} = 0.95$ for the five discharges for which rotation profiles could be fitted.
The effects of toroidal rotation can be included in linear GENE simulations by setting the toroidal rotation velocity $\Omega_{\text{tor}}$ and the related parallel flow-shear rate (which is assumed to be purely toroidal rotation shear).
Including $E \times B$ shear in linear simulations has been observed to result in Floquet modes \cite{Dagnelie2019} and was therefore neglected here.
Adding the counter clock-wise toroidal rotation increased the linear growth rates at $\rho_{\text{tor}} = 0.95$ by less than 10\% and no major changes in mode characteristics were observed.
Adding the parallel flow shear on top of this nullified most of these gains in linear growth rate, in some cases even dropping below the values from the simulations without any rotation effects.
Overall, the addition of parallel flow-shear effects has limited impact on the linear stability spectra of these discharges.

\section{MODEL DEVELOPMENT RECOMMENDATIONS}
\label{apdix:rec}
\subsection{QuaLiKiz}
\begin{itemize}
    \item The collision model in QuaLiKiz needs to be re-tuned for TEM branches near the edge of the plasma. The value of the local normalized electron collisionality could serve as a switch to set the appropriate collision model in the code.
    \item The $s$-$\alpha$ geometry model should be replaced by a shaped flux-surface geometry, such as Miller or MXH, to be more accurate in predicting the correct dominant ion-scale mode branches close to the separatrix \cite{Arbon2021,Snoep2023}.
    \item A different (non-Gaussian) treatment of the eigenfunctions is required to properly capture the unconventional ballooning mode structures that occur in the steep gradient regimes.
    \item Further resistive/dissipative physics and a more advanced collision operator should only be added to the model after the geometry model is upgraded to include shaping, since in GENE simulations with $s$-$\alpha$ geometry the resistive mode branches were overrepresented compared to simulations with numerical equilibria.
    \item Resistive mode physics could be added to QuaLiKiz by extending the dispersion relation to include the parallel Ampere's law. Alternatively, QuaLiKiz could, at high collisionality, solve a separate dispersion relation based on the high-frequency expansion of the kinetic dispersion relation incorporating resistivity through Ohm's law \cite{Bourdelle2012}.
\end{itemize} 
\subsection{TGLF}
\begin{itemize}
    \item The collision operator model in TGLF needs to be extend with energy scattering and multi-species collisions to improve the response of the model past the banana-plateau boundary.
    \item Retuning the default number of parallel basis functions $n_z$ and the trapped fraction $\theta_{\text{trap}}$ in TGLF for the regime at $\rho_{\text{tor}}=0.95$ could also be attempted. 
    In Ref.~\cite{Patel2019} it was shown that this could result in 2$\times$ smaller RMS errors in the linear growth rates for electromagnetic instabilities in TGLF.
    \item The Miller geometry model should be upgraded to the MXH formulation, for strongly shaped and/or strongly up-down asymmetric plasmas.
    \item A comparison of the recent linear Gyro Fluid Solver (GFS) \cite{Staebler2023} against the linear gyrokinetic L-mode dataset presented in this work should be performed. 
    GFS uses general flux-surface geometry (MXH) and has implemented the mirror force to capture trapping effects instead of the bounce averaging approximation that TGLF uses.
\end{itemize}

\begin{figure}
   \centering
   \includegraphics[width=0.475\textwidth]{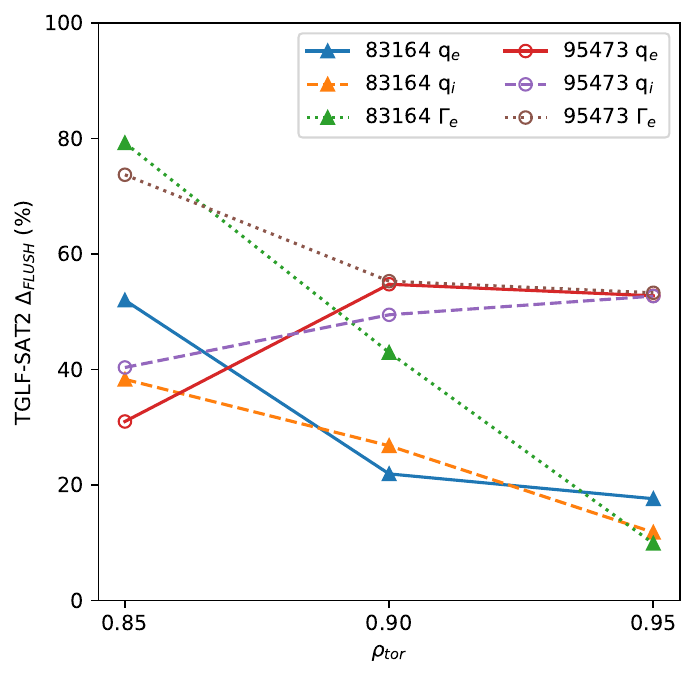}
   \caption{The relative difference in the TGLF-SAT2 heat and particle fluxes in percent when changing the shape parameters from FLUSH to MEGPy for $\rho_{\text{tor}} \in [0.85,0.9,0.95]$.}
   \label{fig:megpy}
\end{figure}
\section{IMPACT OF FLUX-SURFACE PARAMETERIZATION IN TGLF-SAT2}
\label{apdix:geo}
Near the edge of the plasma the shaping of the flux-surfaces typically varies more than in the core.
\mbox{(Turnbull-)}Miller parameterizations of the local equilibrium were generated for \#83164 and \#95473 with the FLUSH \cite{Pamela2015} and MEGPy packages.
The elongation and triangularity values found with both Miller parameterization methodologies were relatively close, see Table \ref{tab:mgeo}, however the radial derivatives of the shaping coefficients diverged significantly.
\begin{table*}[!t]
   \caption{Comparison of the shaping parameters for the local Miller equilibrium parameterizations from FLUSH and MEGPy at $\rho_{\text{tor}} \in [0.85,0.9,0.95]$ for JET-ILW \#83164 and \#95473.}
   \setlength{\tabcolsep}{3.75pt}
   \setlength{\extrarowheight}{1pt}
   \centering
   \begin{tabular}{c c c c c c c c c c c c}
      \hline\hline\\[-2.25ex]
      $\rho_{\text{tor}}$ & \# & \text{source} & $r$ & $R_0$ & $\partial_r R_0$ & $\kappa$ & s$_{\kappa}$ & $\delta$ & s$_{\delta}$ & $\zeta$ & s$_{\zeta}$ \\[1ex]
      \hline
      \multirow{4}{*}{0.85}
         & \multirow{2}{*}{\textcolor{pblue}{83164}}
          & MEGPy & 0.82 & 2.90 & -0.12 & 1.46 & 0.49 & 0.19 & 0.70 & -0.02 & 0.02\\
          & & FLUSH & 0.81 & 2.90 & -0.12 & 1.45 & 0.55 & 0.20 & -0.01 & 0.00 & 0.00\\
         & \multirow{2}{*}{\textcolor{pred}{95473}}
          & MEGPy & 0.84 & 2.89 & -0.14 & 1.43 & 0.40 & 0.13 & 0.39 & -0.02 & -0.03\\
          & & FLUSH & 0.84 & 2.89 & -0.14 & 1.45 & 0.52 & 0.11 & 0.33 & 0.00 & 0.00 \\
      \hline
         \multirow{4}{*}{0.9}
            & \multirow{2}{*}{\textcolor{pblue}{83164}}
            & MEGPy & 0.86 & 2.90 & -0.13 & 1.50 & 0.66 & 0.23 & 0.93 & -0.01 & -0.04 \\
            & & FLUSH & 0.85 & 2.89 & -0.13 & 1.49 & 0.77 & 0.23 & 0.64 & 0.00 & 0.00\\
            & \multirow{2}{*}{\textcolor{pred}{95473}}
            & MEGPy & 0.88 & 2.88 & -0.16 & 1.46 & 0.51 & 0.15 & 0.50 & -0.02 & -0.04\\
            & & FLUSH & 0.87 & 2.88 & -0.15 & 1.49 & 0.73 & 0.14 & 0.60 & 0.00 & 0.00\\
      \hline
         \multirow{4}{*}{0.95}
            & \multirow{2}{*}{\textcolor{pblue}{83164}}
               & MEGPy & 0.89 & 2.89 & -0.14 & 1.54 & 0.95 & 0.26 & 1.05 & -0.02 & -0.20\\
               & & FLUSH & 0.89 & 2.89 & -0.14 & 1.55 & 1.20 & 0.28 & 1.57 & 0.00 & 0.00 \\
            & \multirow{2}{*}{\textcolor{pred}{95473}}
               & MEGPy & 0.91 & 2.88 & -0.17 & 1.50 & 0.68 & 0.17 & 0.71 & -0.02 & -0.12 \\
               & & FLUSH & 0.91 & 2.87 & -0.13 & 1.55 & 1.40 & 0.16 & 1.27 & 0.00 & 0.00\\
      \hline\hline
   \end{tabular}
   \label{tab:mgeo}
\end{table*}
This was found to result in significant differences in the absolute heat and particle fluxes output by TGLF-SAT2 at fixed nominal gradient values, as can be seen in Figure \ref{fig:megpy}.
At $\rho_{\text{tor}}=0.85$ the relative difference between TGLF-SAT2 using the parameterization from FLUSH and using the parameterization from MEGPy is between 40\%-80  \%, while at 0.95 the relative difference are slightly lower at 20\%-60\%.
The area under the linear growth rate spectra only differed by $\sim$8\% at all radii, so this is not due to the impact of the shaping on the linear response of TGLF-SAT2.
However, SAT2 added the dependence of the quasilinear intensity spectrum on the plasma shape, in which the Jacobian of the flux-surface is found.
The Jacobians resulting from the FLUSH parameterization had a subtantially larger error with respect to the numerical equilibrium than those from the MEGPy parameterizations.
Repeating this with SAT0 and SAT3, which don't include a similar dependence of the quasilinear intensity spectrum on the plasma shaping, negligible differences between the output fluxes for both parameterization methodologies were found.

\section*{DATA AVAILABILITY STATEMENT}
The data that support the findings of this study are available upon reasonable request from the authors.

\section*{ACKNOWLEDGEMENTS} \label{sec:ack}
The authors are grateful to our colleagues in the TSVV1 and TSVV11 EUROfusion E-TASC projects, in particular Clemente Angioni, Nicola Bonanomi, Hugo de Blank, Francis Casson, Tobias G\"{o}rler, Florian K\"{o}chl and Alessandro di Siena, for various discussions and suggestions.
We would also like to thank Gary Staebler for his insights on TGLF.

This work has been carried out within the framework of the EUROfusion Consortium, funded by the European Union via the Euratom Research and Training Programme (Grant Agreement No 101052200 — EUROfusion). Views and opinions expressed are however those of the author(s) only and do not necessarily reflect those of the European Union or the European Commission. Neither the European Union nor the European Commission can be held responsible for them.

DIFFER is a part of the institutes organization of NWO.

This research was supported in part by the grants FIS2017-85252-R and PID2021-127727OB-I00, funded by the Spanish MCIN/AEI/10.13039/501100011033 and by “ERDF A way of making Europe”.

Work supported, in part, by the US DOE under Contract No. DE-AC05-00OR22725 with UT-Battelle, LLC.

\end{document}